\pdfoutput=1
\documentclass[useAMS,usenatbib,twocolumn]{mn2e}
\usepackage{epsfig,amsmath,amssymb,bm}

\def\bea{\begin{eqnarray}}
\def\eea{\end{eqnarray}}
\def\be{\begin{equation}}
\def\ee{\end{equation}}
\def\simlt{\ \raise -2.truept\hbox{\rlap{\hbox{$\sim$}}\raise5.truept   %
\hbox{$<$}\ }}
\def\simgt{\ \raise -2.truept\hbox{\rlap{\hbox{$\sim$}}\raise5.truept   %
\hbox{$>$}\ }}    
\usepackage{aas_macros}

\title{Local Group dSph radio survey with ATCA (I): Observations and background sources}

\author[Regis, Richter, Colafrancesco, Massardi, de Blok, Profumo, Orford]{{\normalsize Marco Regis$^{1,2}$, Laura Richter$^3$, Sergio Colafrancesco$^4$, Marcella Massardi$^{5}$, W.J.G. de Blok$^{6,7,8}$, Stefano Profumo$^{9,10}$ and  Nicola Orford$^4$}\\
$^1$Dipartimento di Fisica, Universit\`{a} di Torino, via P. Giuria 1, I--10125 Torino, Italy\\
$^2$Istituto Nazionale di Fisica Nucleare, Sezione di Torino, via P. Giuria 1, I--10125 Torino, Italy\\
$^3$SKA South Africa, 3rd Floor, The Park, Park Road, Pinelands, 7405, South Africa\\
$^4$School of Physics, University of the Witwatersrand, Johannesburg, South Africa\\
$^{5}$INAF - Istituto di Radioastronomia, Via Gobetti 101, I-40129, Bologna, Italy.\\
$^6$Netherlands Institute for Radio Astronomy (ASTRON), Postbus 2, 7990 AA Dwingeloo, The Netherlands\\
$^7$Astrophysics, Cosmology and Gravity Centre, Department of Astronomy, University of Cape Town, Private Bag X3, \\Rondebosch 7701, South Africa\\
$^8$Kapteyn Astronomical Institute, University of Groningen, PO Box 800, 9700 AV, Groningen, The Netherlands\\
$^9$Department of Physics, University of California, 1156 High St., Santa Cruz, CA 95064, USA\\
$^{10}$Santa Cruz Institute for Particle Physics, Santa Cruz, CA 95064, USA\\
E-mail:regis@to.infn.it,laura@ska.ac.za}

\begin{document}

\date{}

\maketitle

\begin{abstract}
Dwarf spheroidal (dSph) galaxies are key objects in near-field cosmology, especially in connection to the study of galaxy formation and evolution at small scales. In addition, dSphs are optimal targets to investigate the nature of dark matter. However, while we begin to have deep optical photometric observations of the stellar population in these objects, little is known so far about their diffuse emission at any observing frequency, and hence on thermal and non-thermal plasma possibly residing within dSphs.
In this paper, we present deep radio observations of six local dSphs performed with the Australia Telescope Compact Array at 16 cm wavelength. 
We mosaiced a region of radius of about one degree around three ``classical" dSphs, Carina, Fornax, and Sculptor, and of about half of degree around three ``ultra-faint" dSphs, BootesII, Segue2, and Hercules. The rms noise level is below 0.05 mJy for all the maps.
The restoring beams FWHM ranged from $4.2\times2.5$ arcseconds to $30.0\times2.1$ arcseconds in the most elongated case. 
A catalogue including the 1392 sources detected in the six dSph fields is reported.
The main properties of the background sources are discussed, with positions and fluxes of brightest objects compared with the FIRST, NVSS, and SUMSS observations of the same fields.
The observed population of radio emitters in these fields is dominated by synchrotron sources. We compute the associated source number counts at 2 GHz down to fluxes of 0.25 mJy, which prove to be in agreement with AGN count models.
\end{abstract}


\section{Introduction}
\label{sec:intro}

Dwarf galaxies are the most common type of galaxy throughout the Universe and are interesting objects for many different reasons.
First of all, they dominate -- by number -- the total galaxy population.
Dwarf galaxies are also our closest neighbors, allowing us to collect data of the highest quality available about galaxies besides our own. Their structure, chemical composition and kinematics pose important challenges to our theoretical understanding of galaxy formation (see, for example, \citet{Mateo:1998wg} and \citet{McConnachie:2012vd} for reviews). Finally, dwarf spheroidal (dSph) galaxies have been recognized as key probes for the presence and the nature of dark matter (DM).
They are the most DM dominated objects discovered in the local Universe.
Their stellar population (spread on scales $\sim 100$ pc) have central velocity dispersion $>5$ km/s, which lead to an inferred dynamical mass of $\sim10^7\,M_{\odot}$, and imply very large mass-to-light ratios, up to $(10^3-10^4)\, M_{\odot}/L_{\odot}$.

On the other hand, very little is known about them, partially because these objects are small and dimly lit.

Recent searches for dwarf galaxies in the Sloan Digital Sky Survey data have more than doubled the number of known dSph satellite galaxies of the Milky Way (MW), and have revealed a population of ultra-faint galaxies, less luminous than any galaxy previously known~\citep{Willman:2005cd,Belokurov:2006ph,McConnachie:2008ji}. 
In the last decade, twenty five new dwarf galaxy companions of the Milky Way and M31 have been discovered.
The SDSS analysis and survey completeness studies suggest that their detection is complete only within a $\sim 50$ kpc radius from us \citep{Koposov:2009ru,Walsh:2008qn}. Applying luminosity bias corrections, \citet{Tollerud:2008ze} found that a few hundreds of these extremely faint MW satellites should be discovered at larger distances and in different angular regions of the sky. 

Spectroscopic studies have revealed that the recently discovered dSphs are the faintest (the most extreme ultra-faint dwarfs have luminosities smaller than the average globular cluster $L_V \sim 10^3-10^4 L_{\odot}$), most dark matter dominated \citep[see, for example,][]{Strigari:2008ib}, and most metal poor galaxies in the Universe (with mean stellar metallicity $\langle [Fe/H]\rangle \sim-2$ \citep[see, for example,][]{Tolstoy:2009jb}). 

dSphs are unique probes for testing structure formation models at small scales and early times.
They have challenged the standard cold DM cosmological paradigm (with, for example, the so-called ``missing satellite problem'' \citep{Klypin:1999uc,Moore:1999nt}), demanding a deeper understanding of the efficiency of small DM halos at forming stars, and of the dSph star formation feedbacks and chemical enrichment.
Forthcoming instruments and deep dSph searches will in fact extensively scrutinize the low-luminosity threshold of galaxy formation~\citep{Bullock:2009au}.

dSph galaxies have been also recognized as optimal laboratories for indirect DM searches \citep{Colafrancesco:2006he}. Due to their proximity, high DM content, and low level of astrophysical backgrounds, dSphs are widely considered to be among the most promising targets for detecting the diffuse electromagnetic radiation possibly induced by DM annihilations or decays.\\
A recent attempt in this direction making use of single dish radio observations was performed by \citet{Spekkens:2013ik} and \citet{Natarajan:2013dsa} with the Green Bank Telescope.
The field of view (FoV) of Draco, Ursa Major II, Coma Berenices, and Willman I were mapped at 1.4 GHz with a resolution of 10 arcmin and a sensitivity of 7 mJy/beam (after discrete source subtraction).
No significant emission was detected from the dSphs, with 95\% C.L. bounds being about two orders of magnitude above the expected flux for a reference model of synchrotron emission induced by annihilation of DM particles with 100 GeV mass. For a more extended discussion, see \citet{Natarajan:2013dsa}.

Studies of possible truly diffuse emission in dSph are also important for assessing the amount of thermal and non-thermal plasma in those structures, as well as the presence of large-scale magnetic fields, for which very little information is available up to date.
Full use of dSphs as DM laboratories will require synergy between large-area photometric surveys, deep spectroscopic and astrometric follow-ups, and subsequent observations at multifrequency with telescopes operating from radio to gamma-rays for DM indirect detection \citep[see, for example,][for an outlook of future perspectives]{Bullock:2009au}.

Our project moves along this context. 
We present here deep mosaic radio observations of a sample of six local dSphs. Data have been collected making use of the Australia Telescope Compact Array (ATCA) observing at 16~cm wavelength.
Three ``classical" dSphs (CDS), i.e., Carina, Fornax, and Sculptor, and three ``ultra-faint" dSphs (UDS), i.e., BootesII, Segue2, and Hercules, were observed. In this paper (Paper I), we present the small-scale sources detected in the six fields of investigation.
Our experimental setup is specifically designed to seek a diffuse radio continuum signal from particle DM (on the scale of a few arcminutes). However, the experiment also allows detection of radio emission on scales of a few arcsec to about 15 arcmin, with a sensitivity of approximately 50 $\mu$Jy at 2 GHz. The results concerning the diffuse emission are presented in \citet{Paper2} (hereafter Paper II) and \citet{Paper3} (hereafter Paper III). 

Searches for point-like radio emissions in dSphs are essential for understanding the star formation and evolution in dSphs.
The knowledge of background sources is crucial as well, in the identification of any kind of dSph diffuse emission (including the signal from DM annihilations). 
Indeed, one of the major issues that needs to be addressed in this context is the contamination of maps by both unresolved and truly diffuse radio background sources.
The arcsec scale spatial resolution of the employed ATCA telescope configuration allows to distinguish between background point-source contributions and diffuse emission (for the latter, we refer to scales of the order of the dSph size, which is about a few arcmin).

We produced a deep search for background radio sources in the dSph fields of view, which add up to about 8 square degrees of the sky (covered by means of a mosaic strategy).
The average synthesised beams ranged from $4.2\times2.5$ arcseconds, $-6.4$ degrees in major and minor axes FWHM and position angle respectively (Carina), to $30.0\times2.1$ arcseconds, $1.2$ degrees in the most elongated case (Bootes), see Table~\ref{tab:beams}. 
The low level of Galactic contamination towards the six selected objects and the good spatial resolution and sensitivity of our ATCA observations allowed to reach
an rms noise value $\lesssim 50$ $\mu$Jy in all the dSph maps. Our radio sample is thus complete at $5 \sigma$ confidence level (hereafter C.L.) down to $\approx 250$ $\mu$Jy, in terms of peak flux density.
This sensitivity level allowed us to  extract a total of 1392 radio sources. This number is sufficiently large to derive precise source number counts at 2 GHz.

The outline of the paper is the following: in Section~\ref{sec:obs}, we describe the dSph observations performed with the ATCA. The process of data reduction is presented and discussed in Section~\ref{sec:red}. We describe the procedure adopted for source extraction and building of the source catalog in Section~\ref{sec:cat}. We compare the results of our source catalog with previous radio surveys in Section~\ref{sec:comp} and we derive the radio source number counts in Section~\ref{sec:num}. We finally discuss results and draw conclusions in Section~\ref{sec:concl}.

\section{Observations}
\label{sec:obs}
The observations presented in this paper were performed during July/August 2011 with the six 22-m diameter ATCA antennae operating at 16 cm wavelength.
As mentioned in the Introduction, three CDS (Carina, Fornax, and Sculptor), and three UDS (BootesII, Hercules, and Segue2) were observed. The project was allocated a total of 123 hours of observing time.

The spectral setup included the simultaneous observation of a 2~GHz-wide band centered at 2100 MHz with a 1~MHz spectral resolution for continuum observations (recording all four polarization signals), and of a $32\times 64$~MHz-channel band centered at 1932 MHz to observe the 1420~MHz HI spectral line. The latter configuration allowed the use of ``zoom bands'' and the spectral resolution for the line observations is 32 kHz. In the present paper, we will consider the results obtained from the analysis of the continuum band only, while the HI-line emission will be presented in further papers from our collaboration.

The observations of Carina, BootesII, Segue2, and part of Hercules were conducted with the hybrid array configuration H214 with maximum baseline of 214~m for the 5 antennas in the core of the array, while for Fornax, Sculptor, and the second part of Hercules, the hybrid configuration H168 with maximum baseline of 168~m was used. A sixth antenna located at about 4.5 km from the core of the array can be used to earn sensitivity on the smaller angular scales. In these configurations, the primary beam ranges between 42' at 1.1 GHz and 15' at 3.1 GHz. The synthesized beams are $\sim3.5$' and $\sim1$' at the extreme of the frequency band if we do not include the long baselines involving antenna 6, while it is $\sim12''$ and $\sim4''$ if we do include it. 

The mapping of the three CDS required a 19 field-mosaic with a total on-source integration time of about 1 hour/field. For BootesII and Hercules, a 7 field-mosaic with an on-source integration time of about 2 hours/field was chosen, while Segue2, due to its smaller size, was imaged with a 3 field-mosaic with about 4 hours/field of integration time (with the purpose of maximizing the sensitivity).
More precisely, a total of 16.5, 15.0, 17.0, 13.0, 10.9, 9.6 hours were spent on-source for Carina, Fornax, Sculptor, BootesII, Hercules, Segue2, respectively.

The pointing grid pattern for the mosaic has been chosen to be an hexagonal grid following Nyquist sampling of the primary beam, 
which leads to a spacing of $\sim15$ arcmin, and an image of nearly 1 degree of radius for CDS and about half-degree for UDS (except for Segue2, where observations involved a row made by three pointings, as mentioned).

For the short-baselines of interest, the time taken for a baseline to rotate to a completely independent point is about 50 minutes.
In order to ensure a good UV plane coverage, we set the time to perform a cycle over all mosaic pointing centres to half of such time, and, more precisely, the dwell time has been $\sim26\,{\rm min}/(n_p+1)$, where $n_p$ is the number of mosaic panels.

The nominal rms sensitivity in each panel for the actual observing time is 36, 38, 35, 25, 28, 20 $\mu$Jy for Carina, Fornax, Sculptor, BootesII, Hercules, Segue2, respectively.
It has been computed by means of the ATCA sensitivity calculator\footnote{http://www.narrabri.atnf.csiro.au/cgi-bin/obstools/atsen8.pl} and assuming no flagging and natural weighting.

For typical spectral indices of synchrotron sources (as found in our catalogue, see next Sections), the average observing frequency is $\langle \nu \rangle\simeq2$ GHz.

\section{Data reduction}
\label{sec:red}

The data were reduced using the {\it Miriad} data reduction package~\citep{Sault:95}. 
We followed the standard reduction procedure recommended in the {\it Miriad} user guide\footnote{http://www.atnf.csiro.au/computing/software/miriad} to calculate and apply the instrumental bandpass, gain, phase, and flux density calibration. 

A total of about 26 hours were devoted to setup and calibration. PKS1934-638, observed twice a day, was used as flux density and bandpass calibrator throughout. 
Secondary calibrators were taken from the ATCA calibrators catalogue as bright sources close to the target galaxies: PKS~0647-475 (Carina), PKS~0237-233 (Fornax), PKS~0022-423 (Sculptor), PKS~0823-500 (BootesII), PKS~0215+015 (Segue2), and PKS~1705+018 (Hercules).
The secondary calibrators were used also as gain calibrators.
The calibration was performed for four frequency bins equally spaced across the band, to account for gain changes across the wide CABB frequency band.

The data were considerably contaminated by radio frequency interference. Bad data were identified through a combination of hand-made flagging and the automated flagging routines provided by {\it Miriad}. Approximately one third of the data were flagged and removed from each data set.

The data were imaged using the {\it Miriad} task MFCLEAN, an implementation of the multi-frequency CLEAN algorithm developed by \citet{Sault:94}. For each target a four-iteration self-calibration was performed for each mosaic panel. The final images were cleaned to a sensitivity cutoff of about three times the nominal rms sensitivity adjusted for flagged data, assuming that one third of each data set is flagged.

Due to a correlator bug during the time of the observations, all of the mosaic panels for each observation were correlated at the position of the first panel. This was corrected for in the image plane, by correcting the position information in the image headers.

\begin{figure}
\centering
 \includegraphics[width=0.5\textwidth]{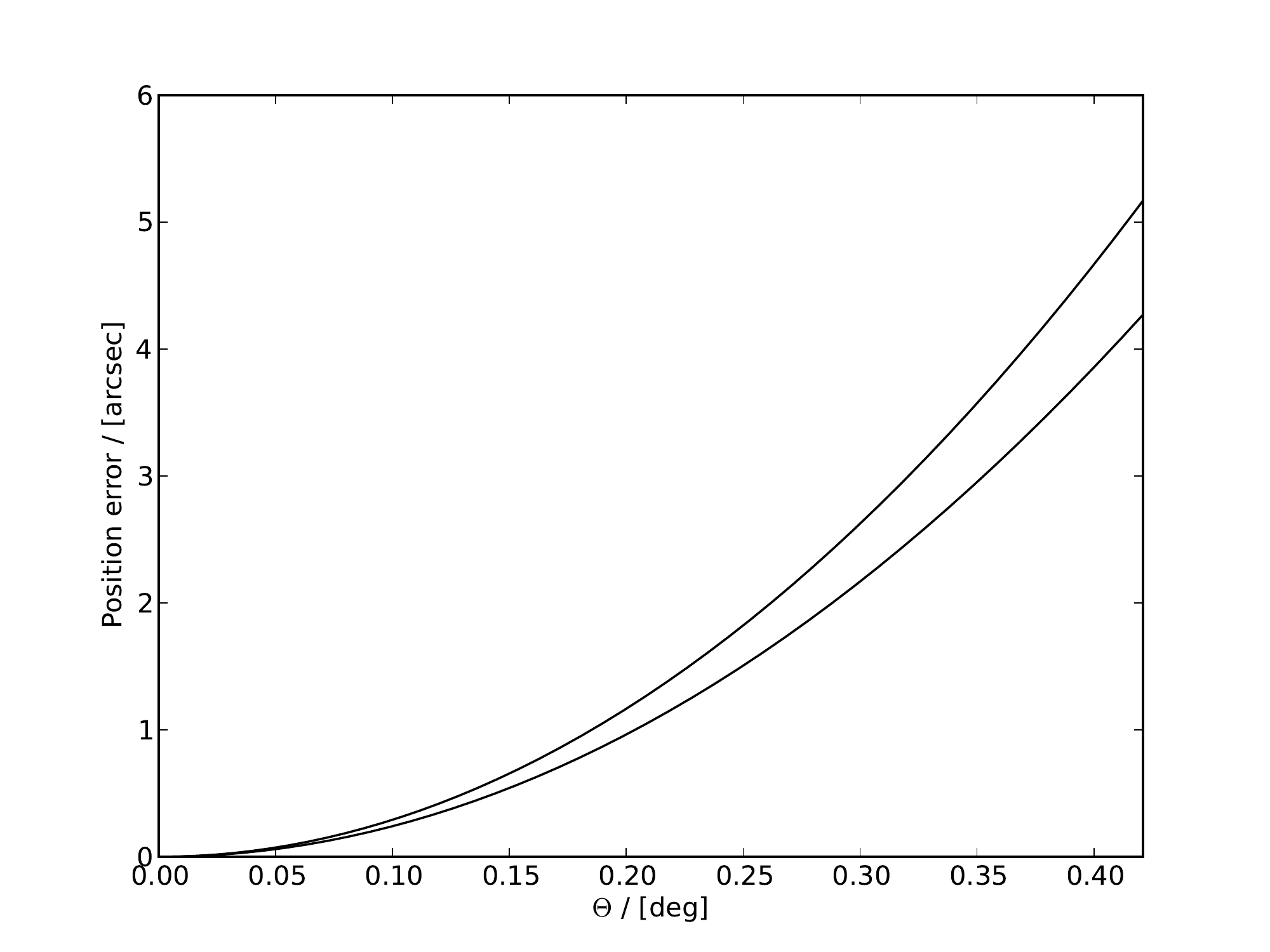}
\caption{Position error of Eq.~\ref{eq:poserr} as a function of distance $\Theta$ from the phase centre. The two curves are for zenith angles of, respectively, 50 (lower) and 68 (upper) degrees (which are the extrema of the range covered in the Segue2 observations).}
\label{fig:zenith}
\end{figure}

The images displayed significant w-term effects. These include a systematic offset in source position across the fields, increasing with increasing distance from the phase centre. For a co-planar array, this position shift is approximately~\citep{Cotton:99}

\be
\mbox{position error} \simeq \frac{\Theta^2}{2 \times 2.06 \times 10^5} \sin{z}                    
\label{eq:poserr}
\ee

where $\Theta$ is the distance of a source from the phase centre, $\Theta \equiv \sqrt{l^2 + m^2}$, and $z$ is the zenith angle. Amongst our target source list, Segue2 is most distant from the ATCA latitude, with the zenith angle ranging from 50 to 68 degrees. Fig.~\ref{fig:zenith} shows Eq.~\ref{eq:poserr} for these zenith angles, plotted against a range of $\Theta$ up to the cutoff point of the ATCA beam model (25.21 arcmin). The mean Segue2 restoring beam (including long baselines) is $17 \times 1.9$ arcseconds. Towards the edge of the ATCA beam the position errors are an appreciable fraction of this beam width, so that when the individual images were mosaiced together bright sources incorrectly appeared to have multiple slightly offset components. 
The arcs often associated with wide field effects are not present in these images, as the change in position offset across the zenith angle range is not significant compared to the restoring beam dimensions. Even at the cutoff point, where the position offset is largest, the difference between the offset for the two zenith angles is less than an arcsecond.

\begin{figure}
\centering
   \includegraphics[width=0.5\textwidth]{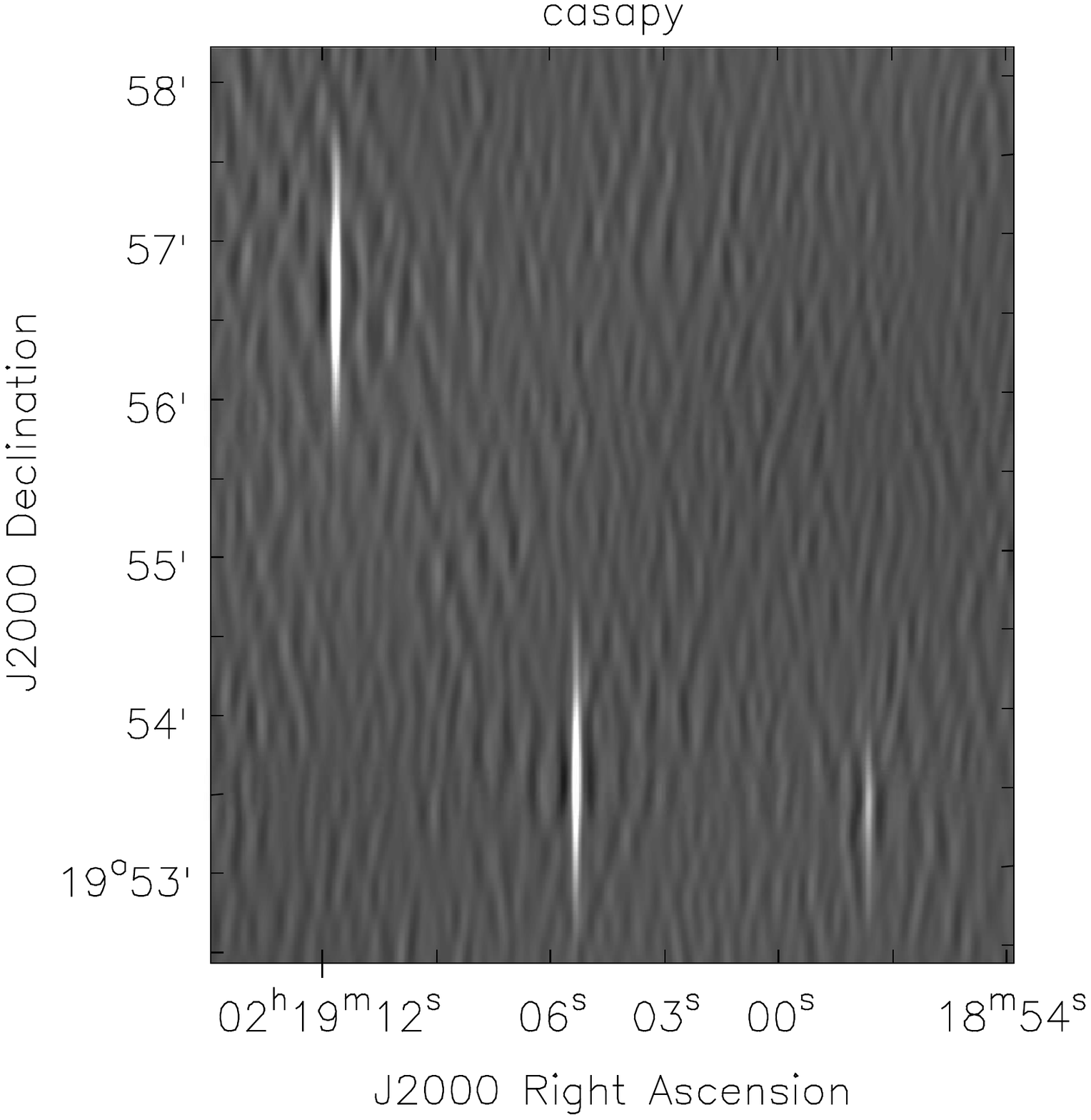}\\
\vspace{-3.5cm}
   \includegraphics[width=0.5\textwidth]{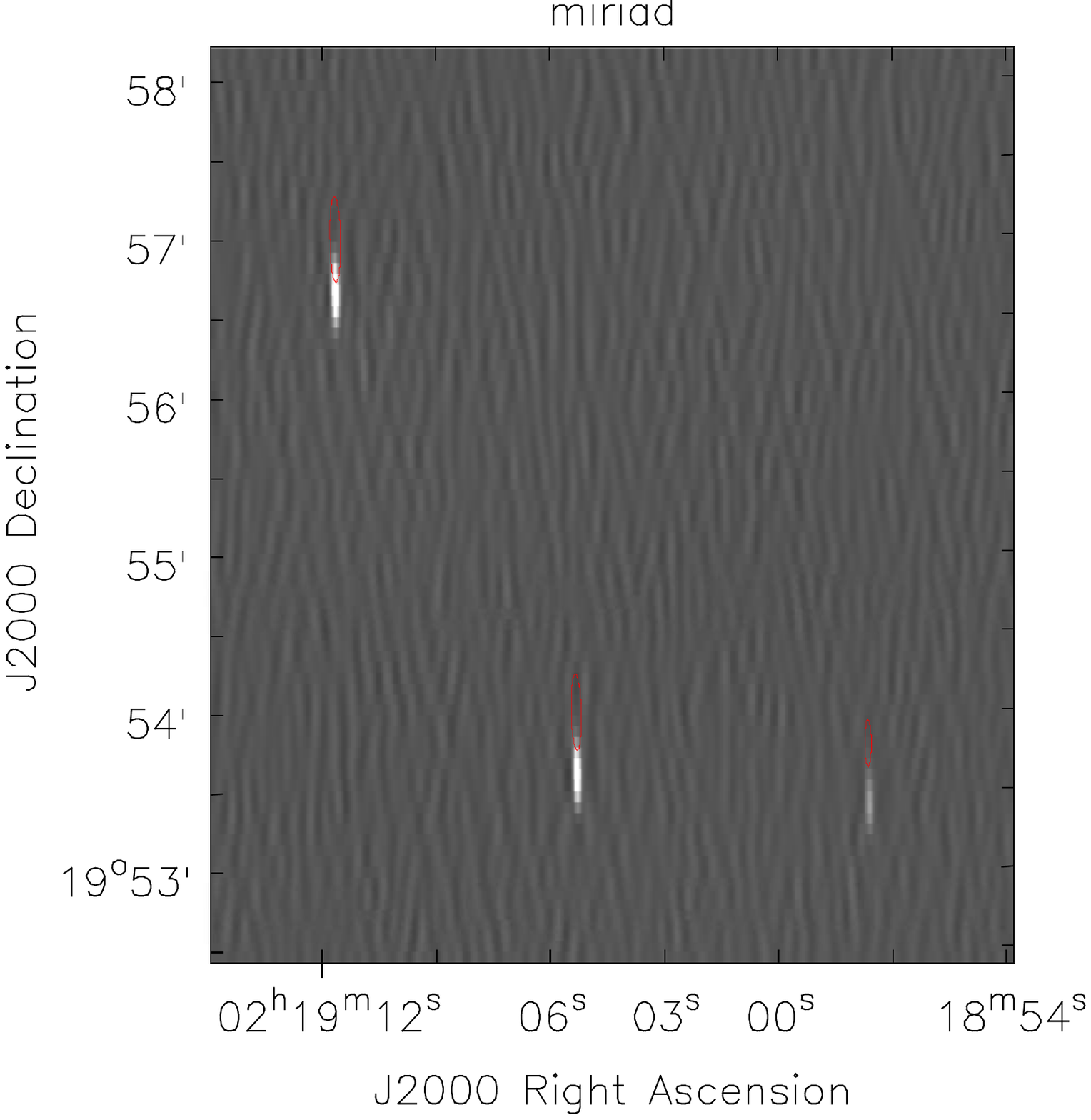}
\vspace{-4.5cm}
\caption{Upper: Image of a central region of the Segue2 FoV obtained with casapy CLEAN. Lower: Same of upper panel, but performing the imaging by means of the {\it Miriad} MFCLEAN algorithm and enforcing NCP projection. Above each source in the {\it Miriad} image is a contour showing the position of sources with SIN projection.}
\label{fig:NCP}
\end{figure}

In order to address this issue, the wide-field imaging capabilities of the CLEAN task in the casapy software package were investigated.
We focused on the most problematic data set (in terms of w-term effects), which is the Segue2 one. 
A three-iteration self-calibration was performed, imaging to the same final flux as the final {\it Miriad} images, and using two Taylor terms to correspond to the {\it Miriad} MFCLEAN.
The use of 256 wprojection planes was found to correct the source positions. However, the {\it Miriad} imager was still preferable, for a number of reasons: i) The casapy CLEAN fit for the restoring beam was poor, in some cases not converging at all; ii) The noise floor of the casapy images was found to be 25\% to 50\% higher than in the corresponding {\it Miriad} images; and iii) The casapy CLEAN with 256 wprojplanes was prohibitively slow. We attribute the higher noise floor of the casapy images to the (relatively incomplete) UV-coverage of the ATCA data, making the image-plane clean of the {\it Miriad} package more suitable. 

We show an example of these findings in Fig.~\ref{fig:NCP}. The two images compare a region of the central Segue2 panel imaged using the {\it Miriad} MFCLEAN and casapy CLEAN. In both images the greyscale range is [-0.1 mJy, 1 mJy]. The contours on the {\it Miriad} image indicate the positions of the sources (before the NCP projection was enforced). The presence of larger noise and larger restoring beam in the casapy case are clearly visible. The rms noise calculated in the source-free top right region of each image is 44 $\mu$Jy for the casapy image and 29 $\mu$Jy for the {\it Miriad} image.

The {\it Miriad} imager was therefore still preferred. The w-term imaging problem was solved by enforcing NCP projection. 
Indeed, the long baselines (which are the ones between the sixth antenna and each of the five antennas of the core) form approximately an East-West array.
In this case, the NCP projection reduces the imaging problem to a two-dimensional Fourier Transform, without the need for w-term approximations.

However, the ATCA was not used in purely E-W mode for these observations, and short baselines include antennas on the northern spur. 
The w-component of the visibility data from these baselines is not removed by enforcing the NCP projection, leading to artifacts in the images including data from these baselines, predominantly for the Segue and Hercules fields, which were most distant from the ATCA latitude. 
On the other hand, if the long baselines data are not included, the restoring beam size becomes larger than 1 arcmin. 
This is sufficiently large that the position offsets between baselines (of order a few arcsec) are not significant.

We proceeded producing two maps for each target.\footnote{Maps and source catalogue presented in this project can be retrieved at http://personalpages.to.infn.it/$\sim$regis/c2499.html.}
The data were first imaged with a Briggs robustness parameter of -1 \citep{Briggs:thesis} leading to an high resolution map where short baselines are down-weighted, and the offset issues are solved by enforcing the NCP projection.
Table~\ref{tab:beams} lists the average restoring beam parameters over all mosaic panels for each field.

\begin{table}
\centering
\begin{tabular}{|ccc|}
\hline
FoV & FWHM & Position Angle \\ 
      & [arcsec $\times$ arcsec] & [deg] \\
\hline 
Carina     & $4.2 \times 2.5$ &  $-6.4$ \\
Fornax    & $7.7 \times 2.2$  & $4.0$ \\
Scultpor  & $8.0 \times 2.2$ & $-0.8$ \\
BootesII   & $30.0 \times 2.1$ & $1.2$ \\
Hercules  & $28.5 \times 1.9$ & $-1.3$ \\
Segue2    & $17.1 \times 1.9$ & $1.5$ \\
\hline
NVSS  & $45 \times 45$ & \\
SUMSS & $45 \times 45\cos{\delta}$ \\
FIRST  & $5.4 \times 5.4$ & \\
\hline
\end{tabular}
\caption{Average restoring beam parameters across all mosaic panels for each field of view, for the robust -1 maps with no Gaussian taper.
The angular resolutions of the NVSS \citep{Condon:1998iy}, SUMSS~\citep{Mauch:2003zh} and FIRST~\citep{FIRST} surveys are included
for comparison. The rms sensitivity of the three surveys amounts to approximately 0.45 mJy/beam, 1 mJy/beam and 0.14 mJy/beam, respectively. For the rms of the observations presented here, see Fig.~\ref{fig:rms}. }
\label{tab:beams}
\end{table} 

A second set of maps was then generated, by imaging again with the same robustness parameter, but applying a Gaussian taper of 15 arcseconds to the data before Fourier inversion.
The beam becomes sufficiently large that the w-term corrections do not show up in the images. 
The un-tapered robust -1 maps were used to determine the source positions and to provide the lowest off-source image rms noise.
However, they are not sensitive to scales above a few tens of arcsec (since they basically rely on long baselines data) and can underestimate the flux of extended sources.
The tapered images were used in conjunction with the un-tapered ones to determine the total source fluxes, as we will describe below.

\begin{figure*}
\centering
\hspace{-15mm}
 \begin{minipage}[htb]{7.5cm}
   \includegraphics[width=0.85\textwidth,angle=-90]{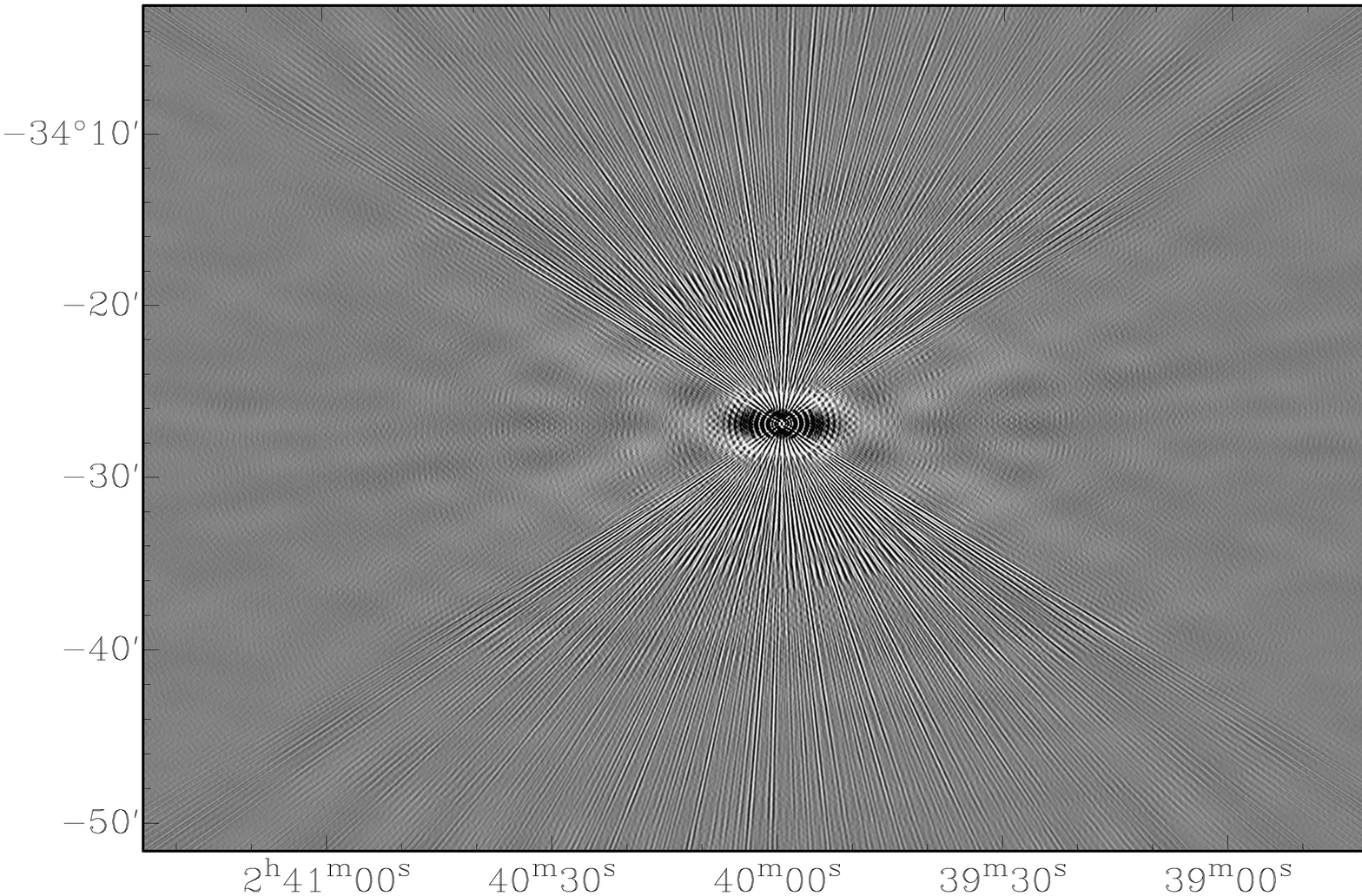}
 \end{minipage}
\hspace{10mm}
 \begin{minipage}[htb]{7.5cm}
   \centering
   \includegraphics[width=0.85\textwidth,angle=-90]{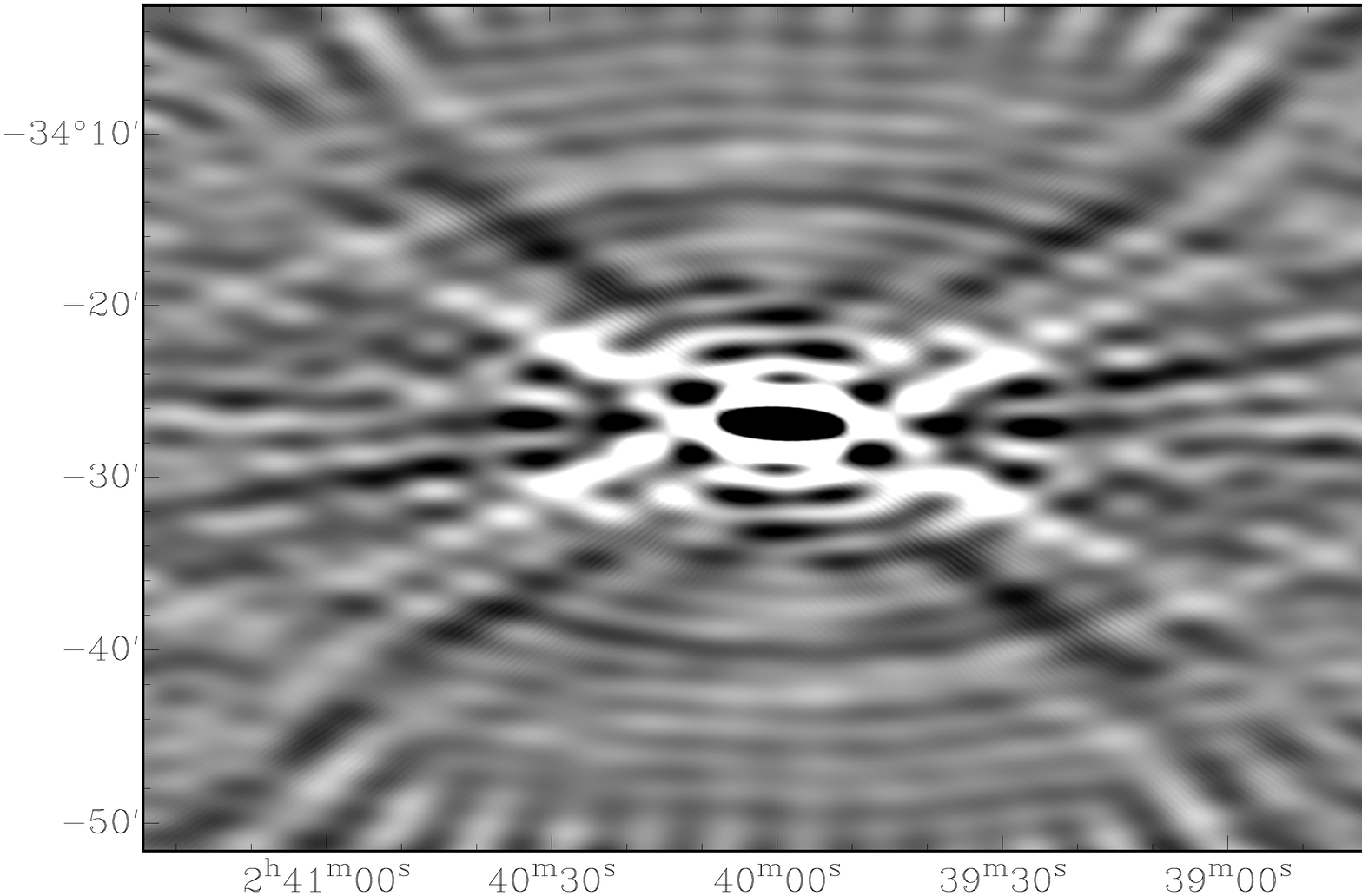}
 \end{minipage}\\ 
\hspace{-15mm}
 \begin{minipage}[htb]{7.5cm}
   \includegraphics[width=0.85\textwidth,angle=-90]{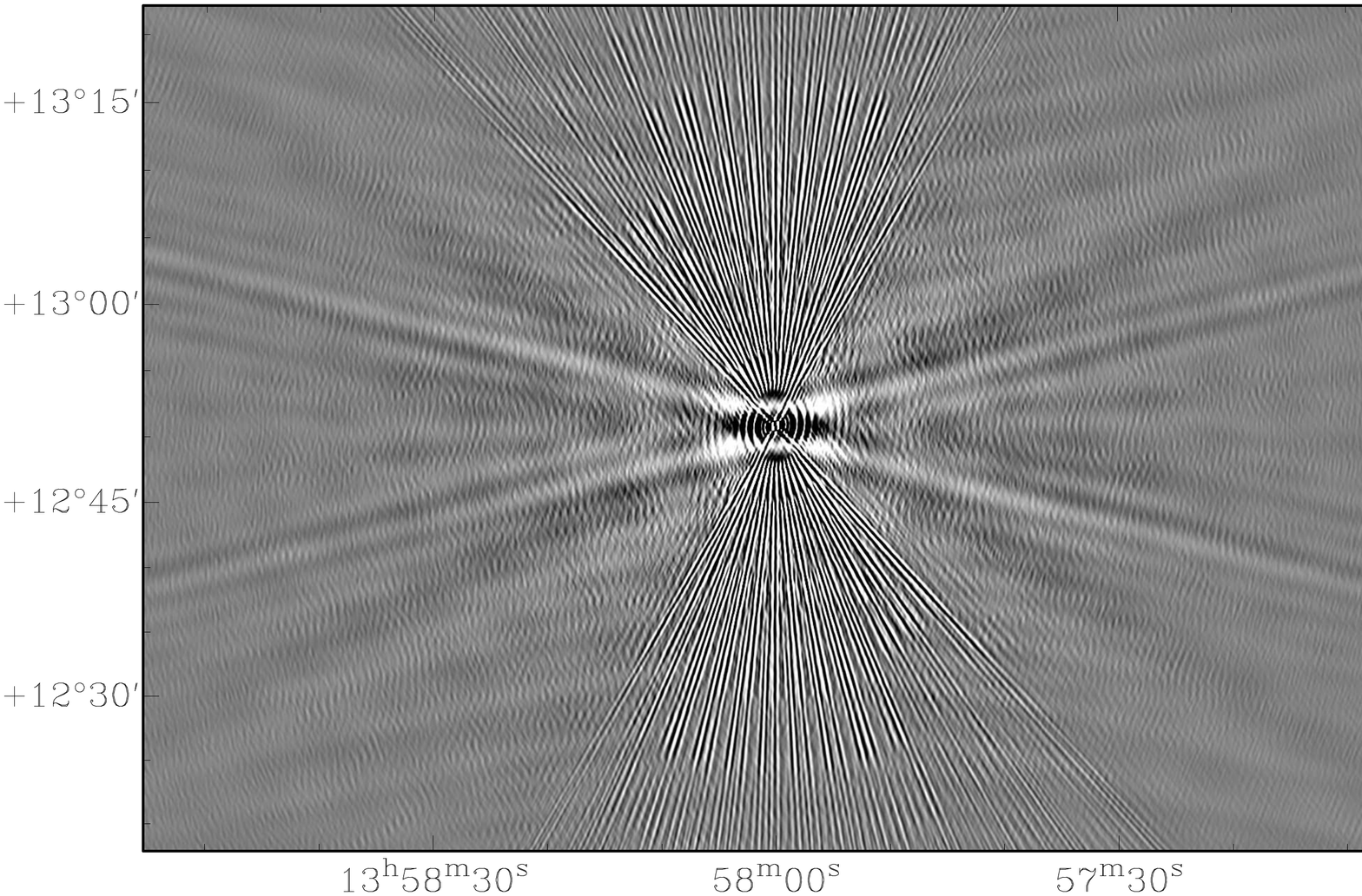}
 \end{minipage}
\hspace{+10mm}
 \begin{minipage}[htb]{7.5cm}
   \centering
   \includegraphics[width=0.85\textwidth,angle=-90]{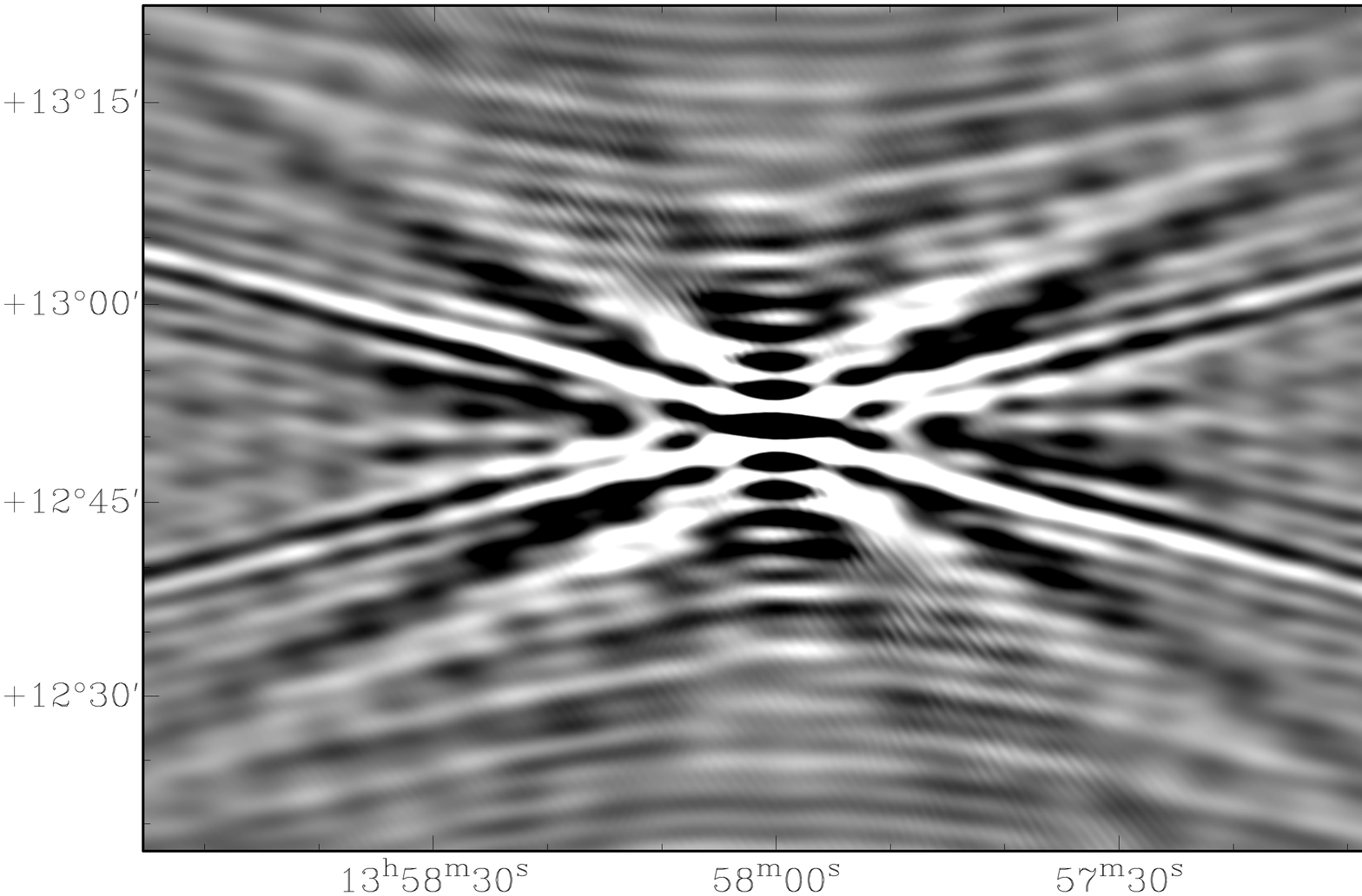}
 \end{minipage}
\caption{{\bf Beams.} Plots of the central region of the synthesized beam for the Fornax (upper panels) and BootesII (lower panels) fields.
Left panels: Imaging with robustness parameter -1, with greyscale range $\{-0.01,0.01\}$.
Right panels: Imaging with robustness parameter -1 and Gaussian taper of 15 arcsec, with greyscale range $\{-0.03,0.03\}$.}
\label{fig:beams}
\end{figure*}

As described earlier, the H168 and H214 ATCA configurations used for these observations had a compact core and a single 4.5~km baseline. This creates a synthesized beam with a central peak on the scale of the resolution provided by the 4.5~km baseline, and a plateau on the scale of the compact core resolution. Examples of this are shown in Fig.~\ref{fig:beams}, for the first pointing of a CDS (Fornax) and UDS (Bootes) field. A consequence of the beam shape is that the noise in the final images is correlated on the scale of the 4.5~km resolution (a few arcseconds) and the core resolution (arcminute).

The complex beam shape is echoed in deconvolution artifacts in the images. These artifacts are accounted for in subsequent processing through the use of a variable noise background in the source detection algorithm, as discussed in Section~\ref{sec:cat}. The noise background is higher in localized regions around stronger sources, where deconvolution errors are most extreme \citep[see, for example,][]{Mauch:2003zh}.

\begin{figure*}
\centering
\hspace{-10mm}
 \begin{minipage}[htb]{7.5cm}
   \includegraphics[width=0.85\textwidth,angle=-90]{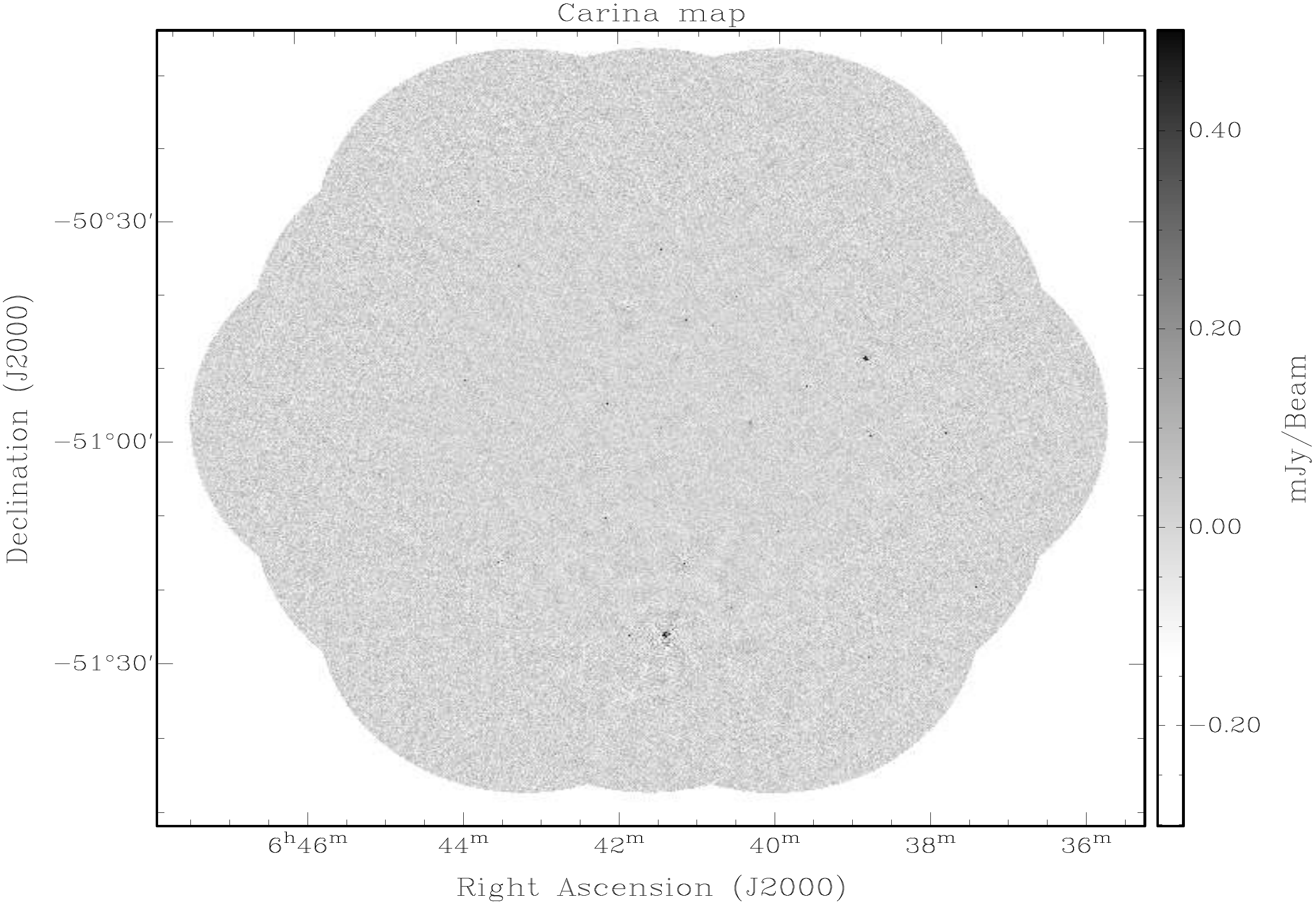}
 \end{minipage}
\hspace{25mm}
 \begin{minipage}[htb]{7.5cm}
   \centering
   \includegraphics[width=0.85\textwidth,angle=-90]{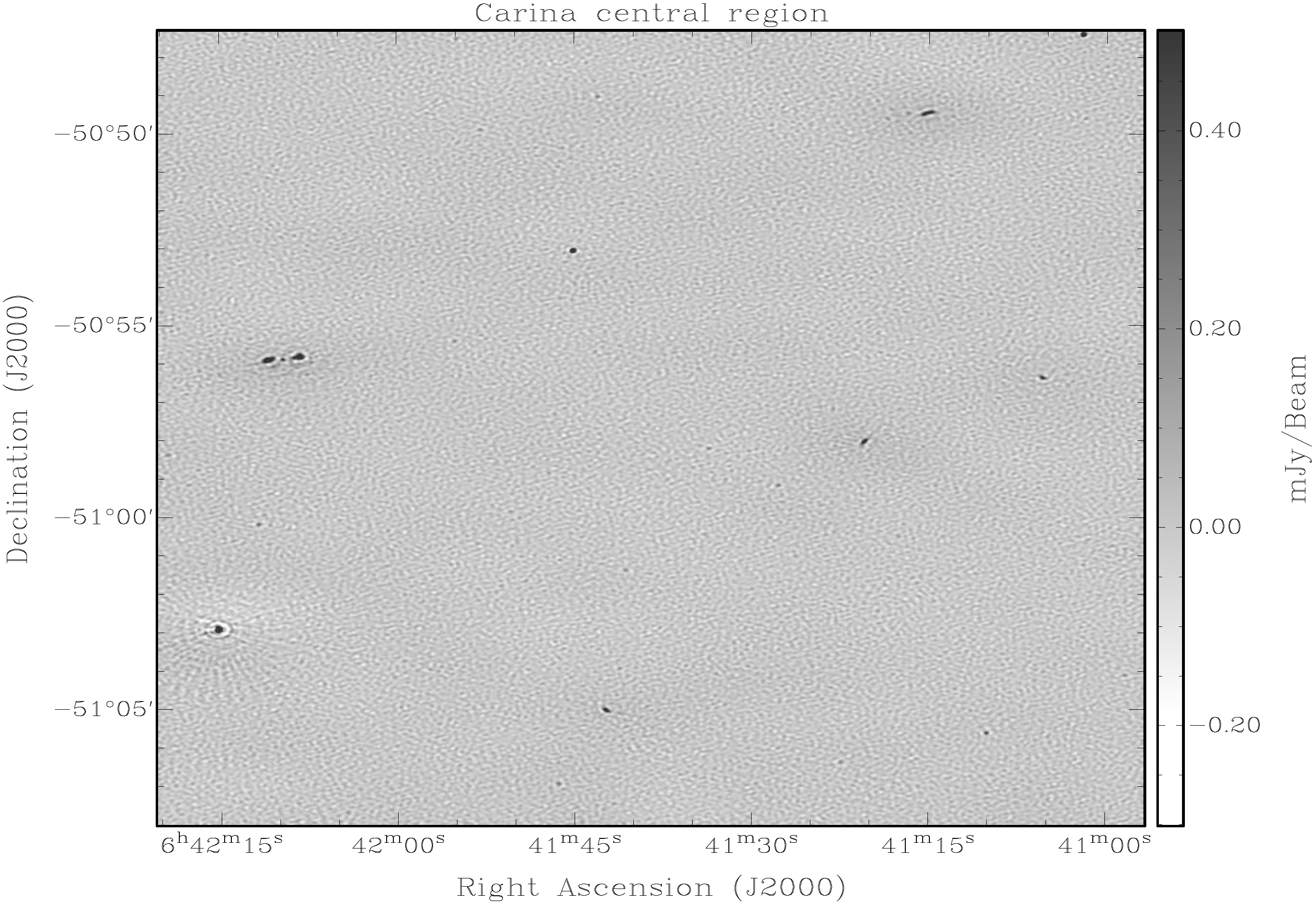}
 \end{minipage}\\ 
\vspace{5mm}
\hspace{-10mm}
 \begin{minipage}[htb]{7.5cm}
   \includegraphics[width=0.85\textwidth,angle=-90]{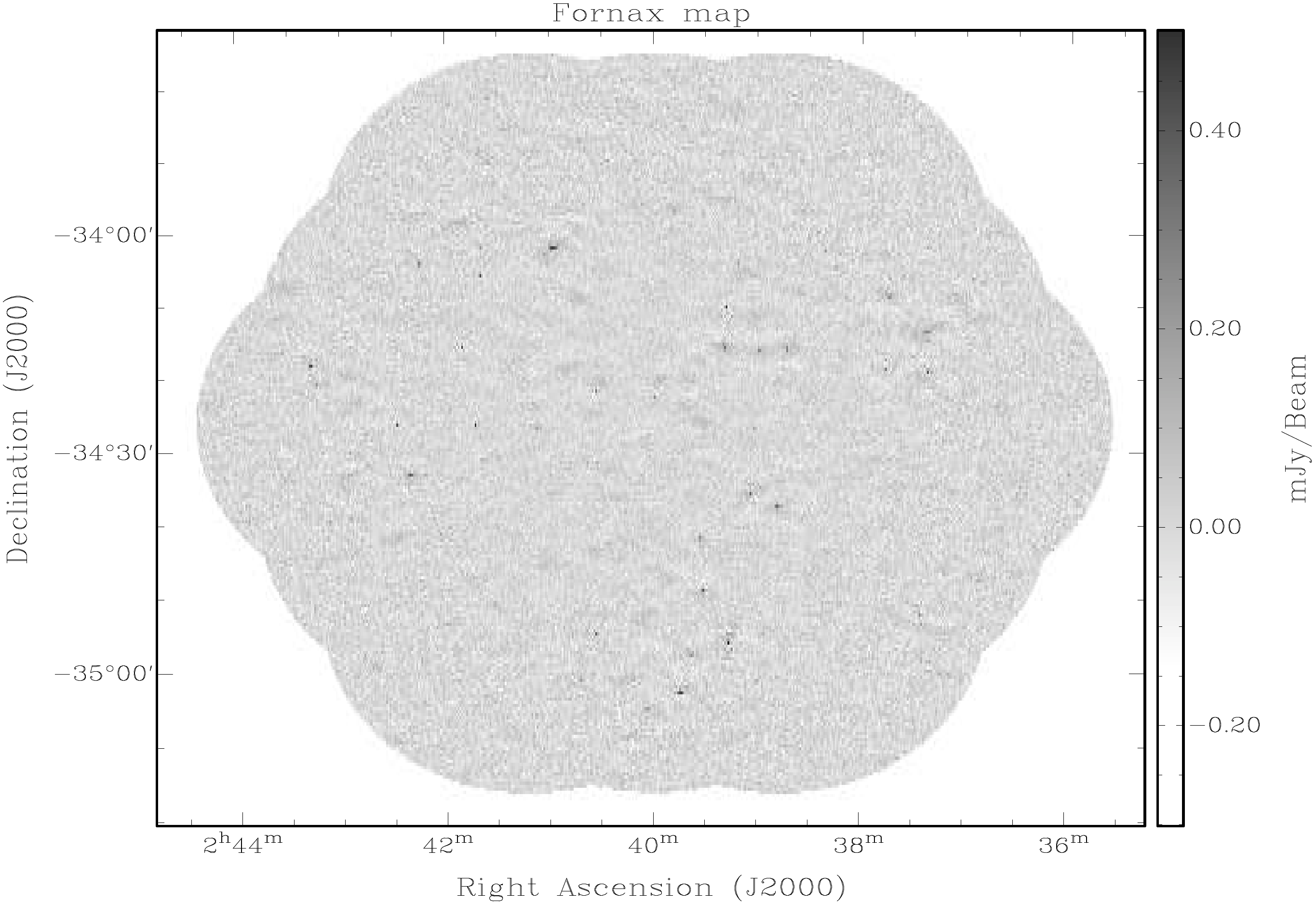}
 \end{minipage}
\hspace{+25mm}
 \begin{minipage}[htb]{7.5cm}
   \centering
   \includegraphics[width=0.85\textwidth,angle=-90]{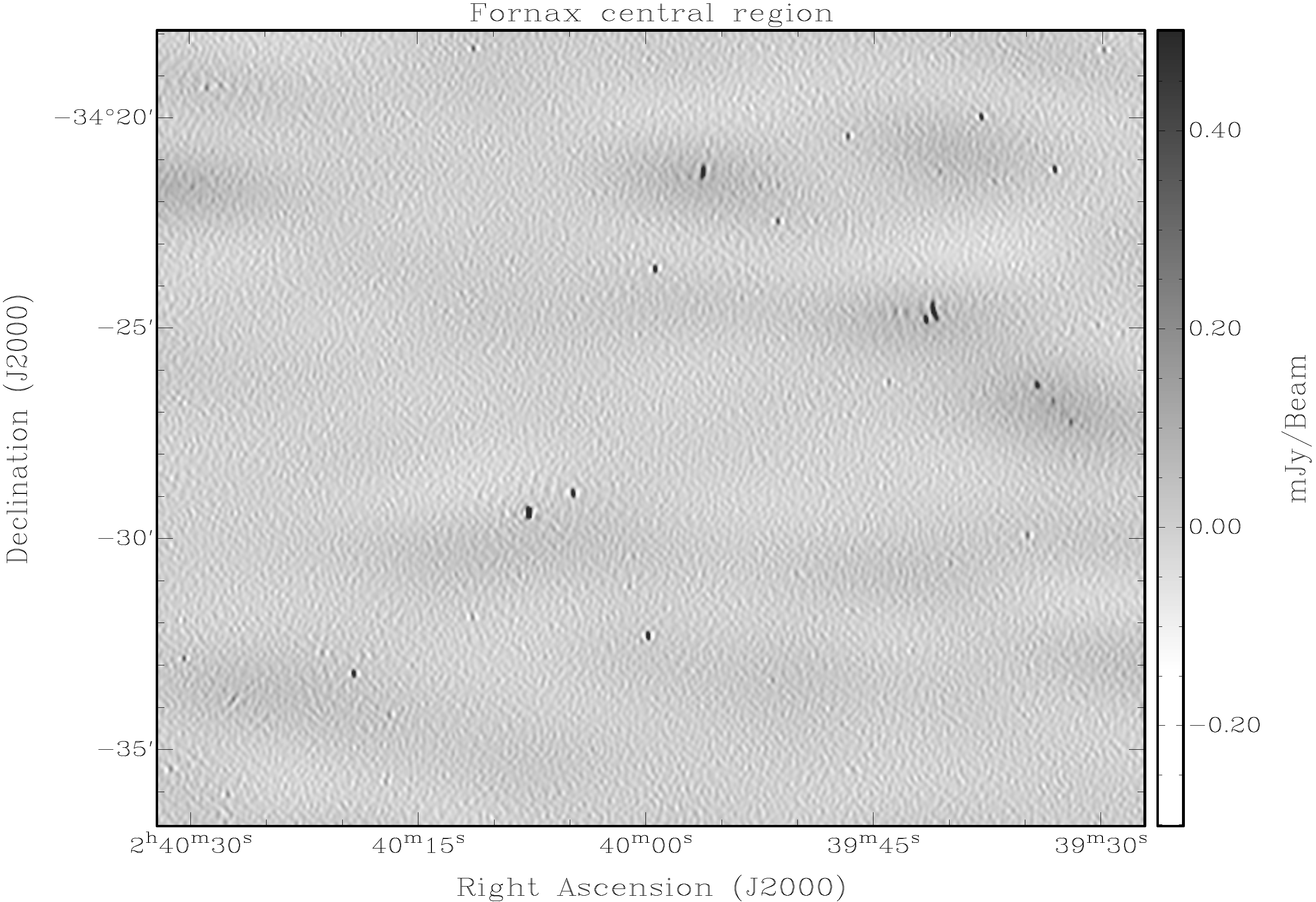}
 \end{minipage}\\ 
\vspace{5mm}
\hspace{-10mm}
 \begin{minipage}[htb]{7.5cm}
   \centering
   \includegraphics[width=0.85\textwidth,angle=-90]{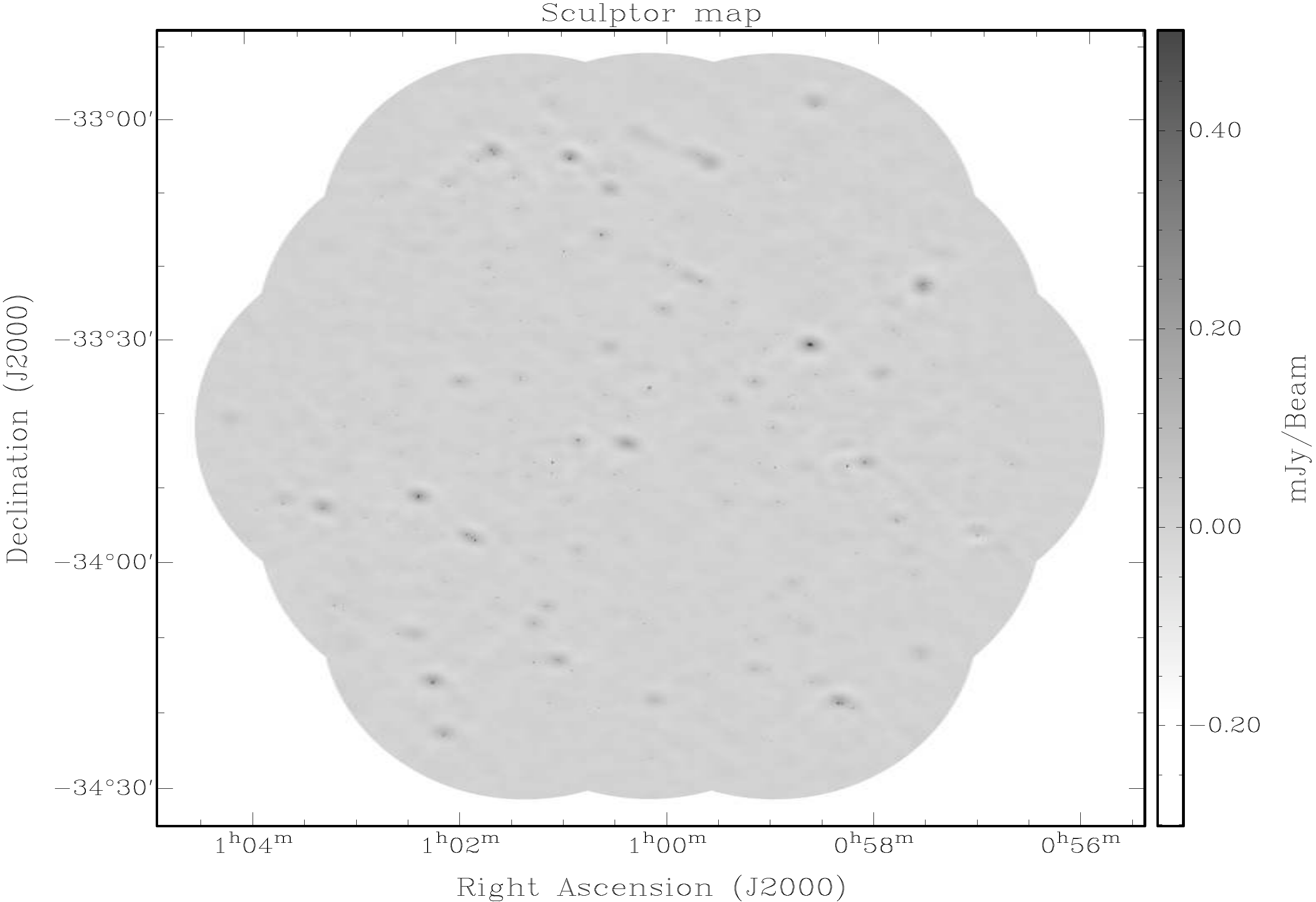}
 \end{minipage}
\hspace{+25mm}
 \begin{minipage}[htb]{7.5cm}
   \centering
   \includegraphics[width=0.85\textwidth,angle=-90]{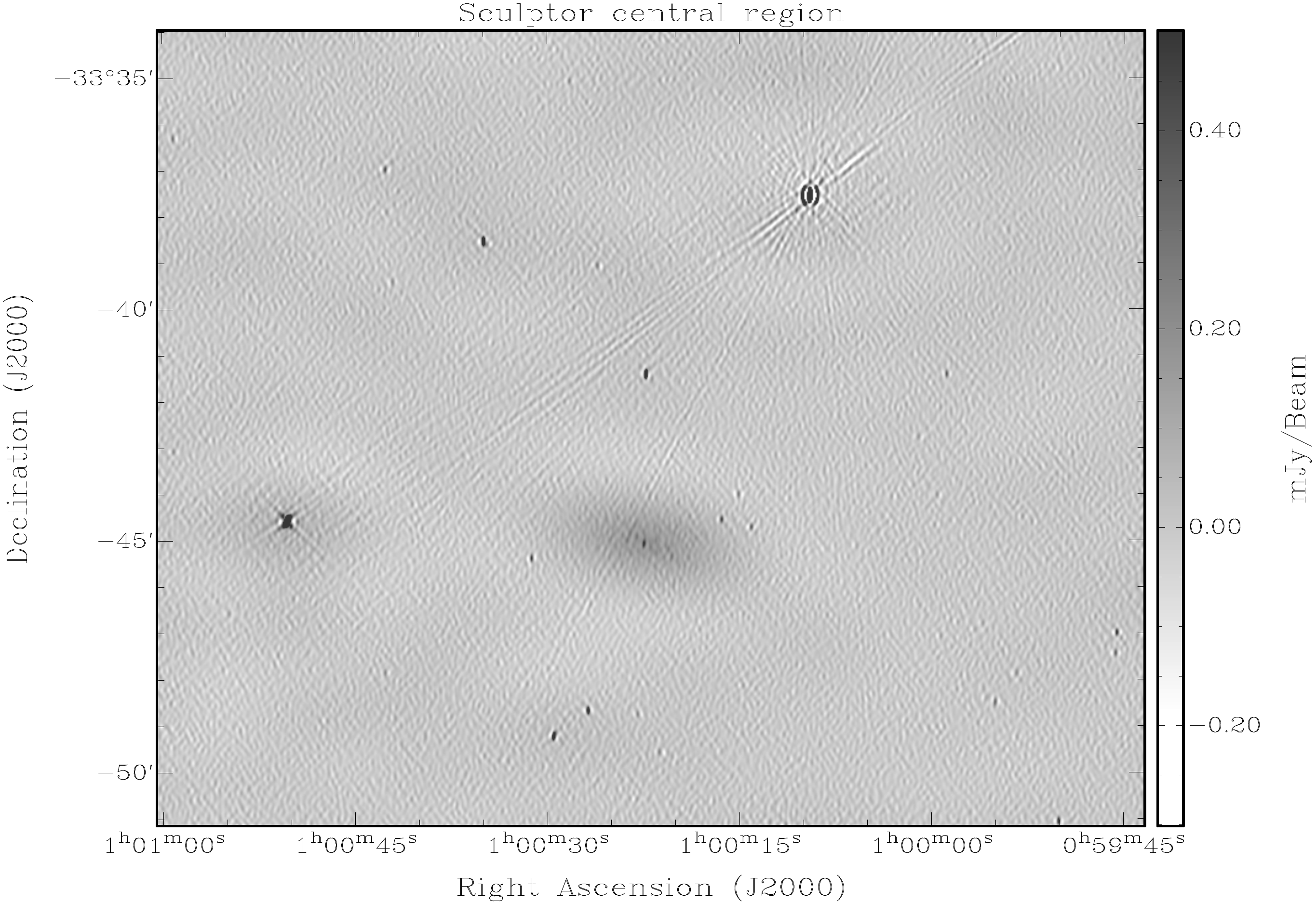}
 \end{minipage}
\caption{{\bf Maps.} Left: Grayscale of the observational mosaic maps after data reduction for the Carina, Fornax, and Sculptor FoV (from top to bottom). Right: Zoom in the central region.}
\label{fig:maps1}
\end{figure*}

\begin{figure*}
\centering
\hspace{-10mm}
 \begin{minipage}[htb]{7.5cm}
   \centering
   \includegraphics[width=0.85\textwidth,angle=-90]{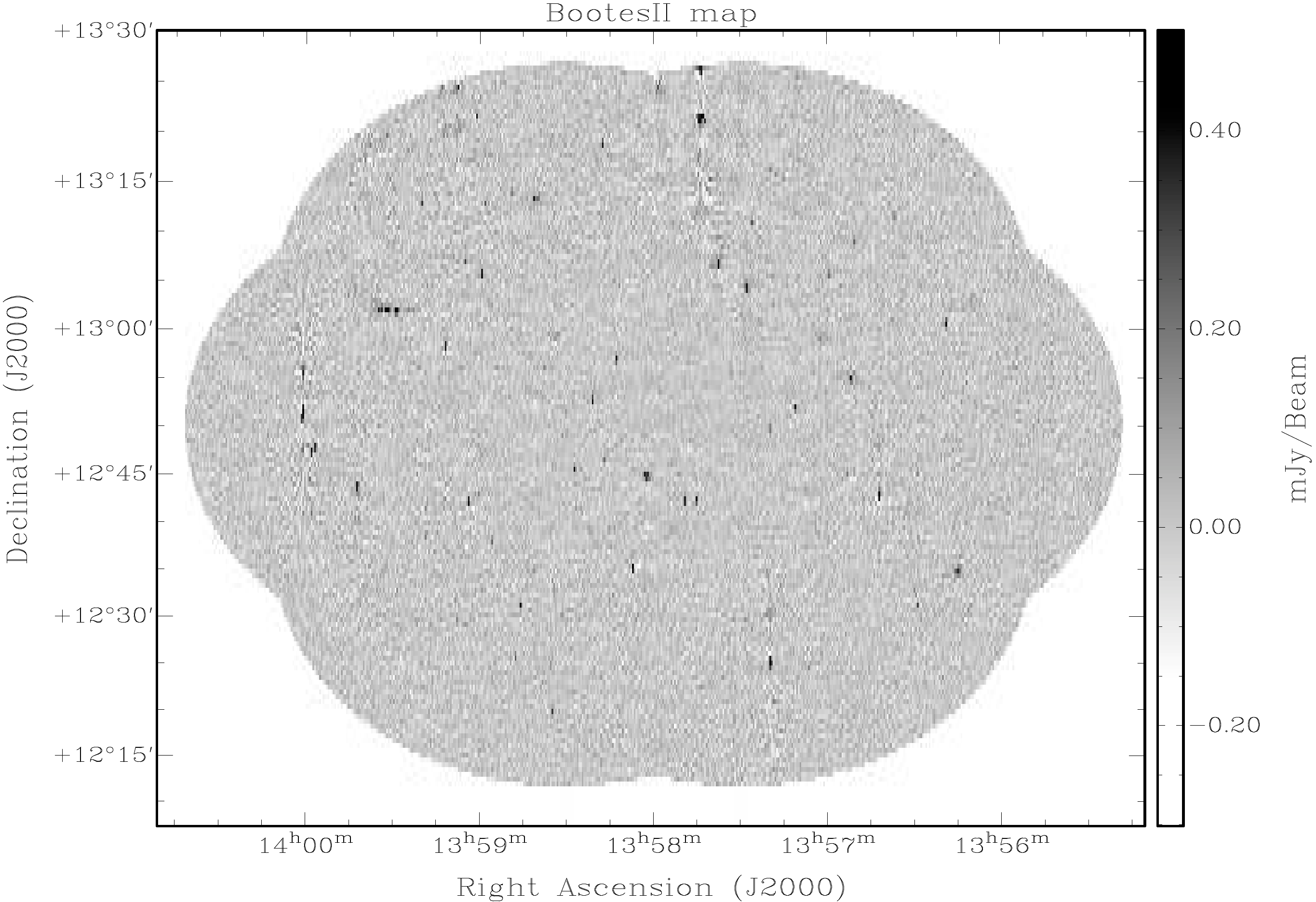}
 \end{minipage}
\hspace{+25mm}
 \begin{minipage}[htb]{7.5cm}
   \centering
   \includegraphics[width=0.85\textwidth,angle=-90]{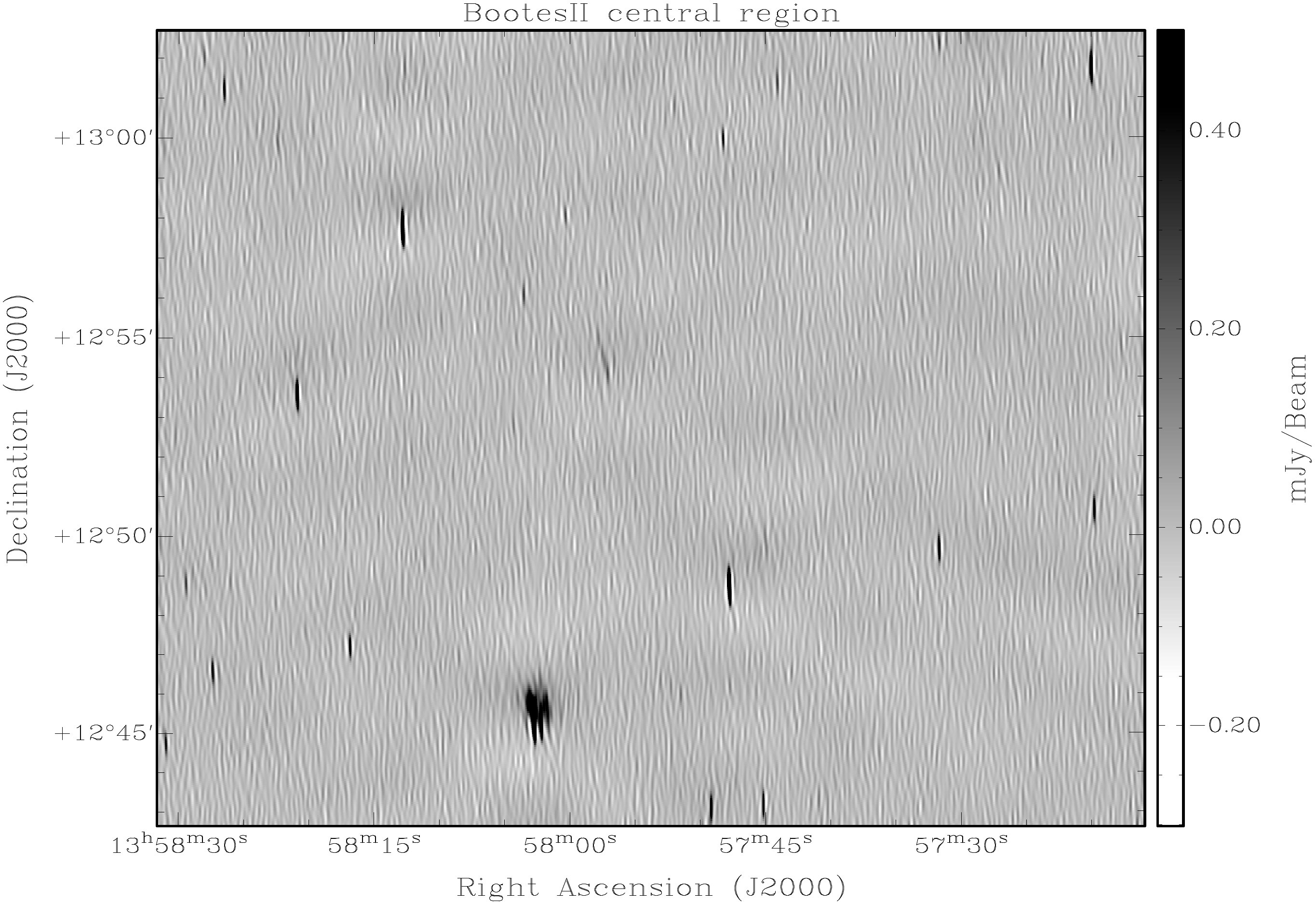}
 \end{minipage}\\ 
\vspace{5mm}
\hspace{-10mm}
 \begin{minipage}[htb]{7.5cm}
   \centering
   \includegraphics[width=0.85\textwidth,angle=-90]{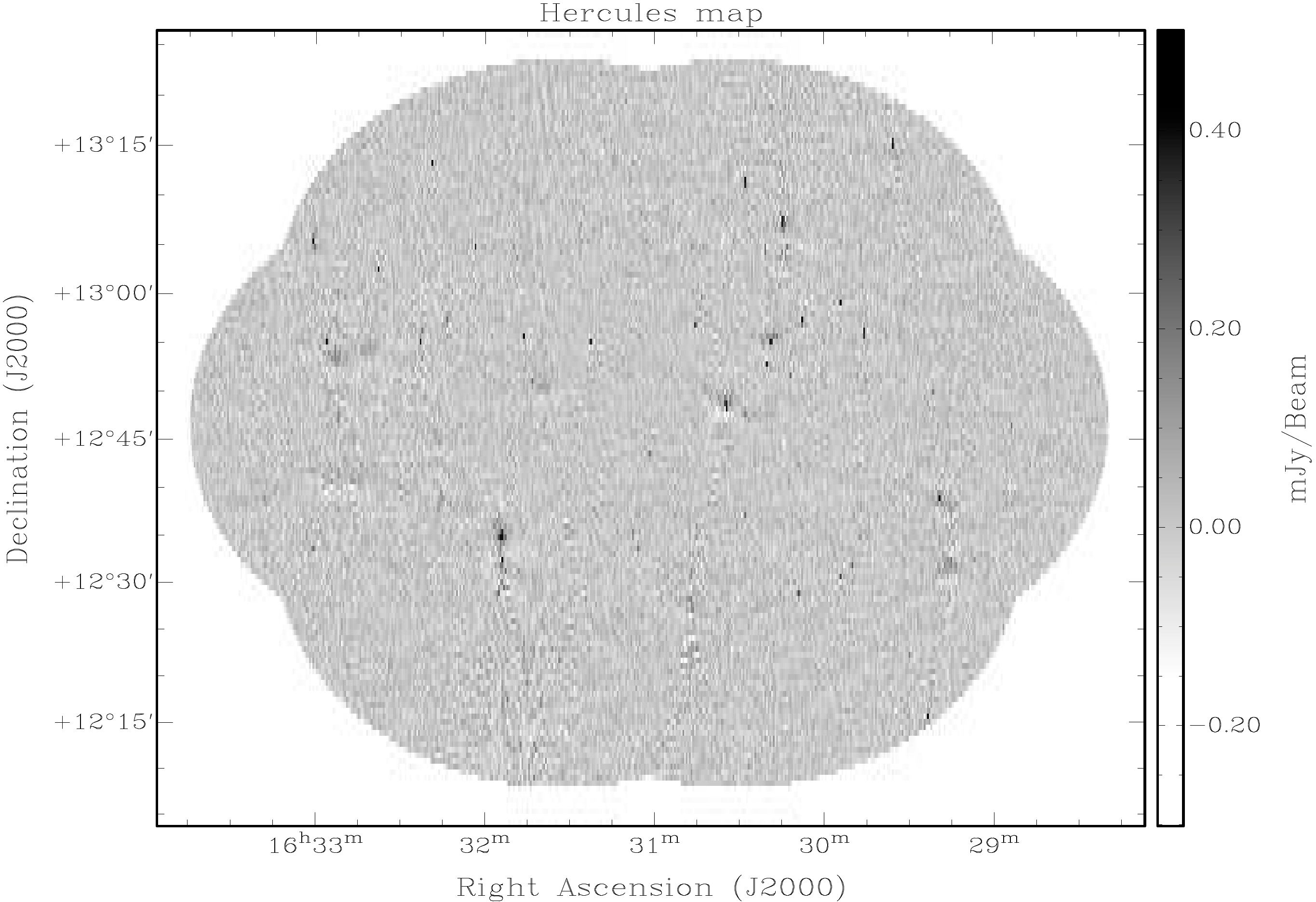}
 \end{minipage}
\hspace{25mm}
 \begin{minipage}[htb]{7.5cm}
   \centering
   \includegraphics[width=0.85\textwidth,angle=-90]{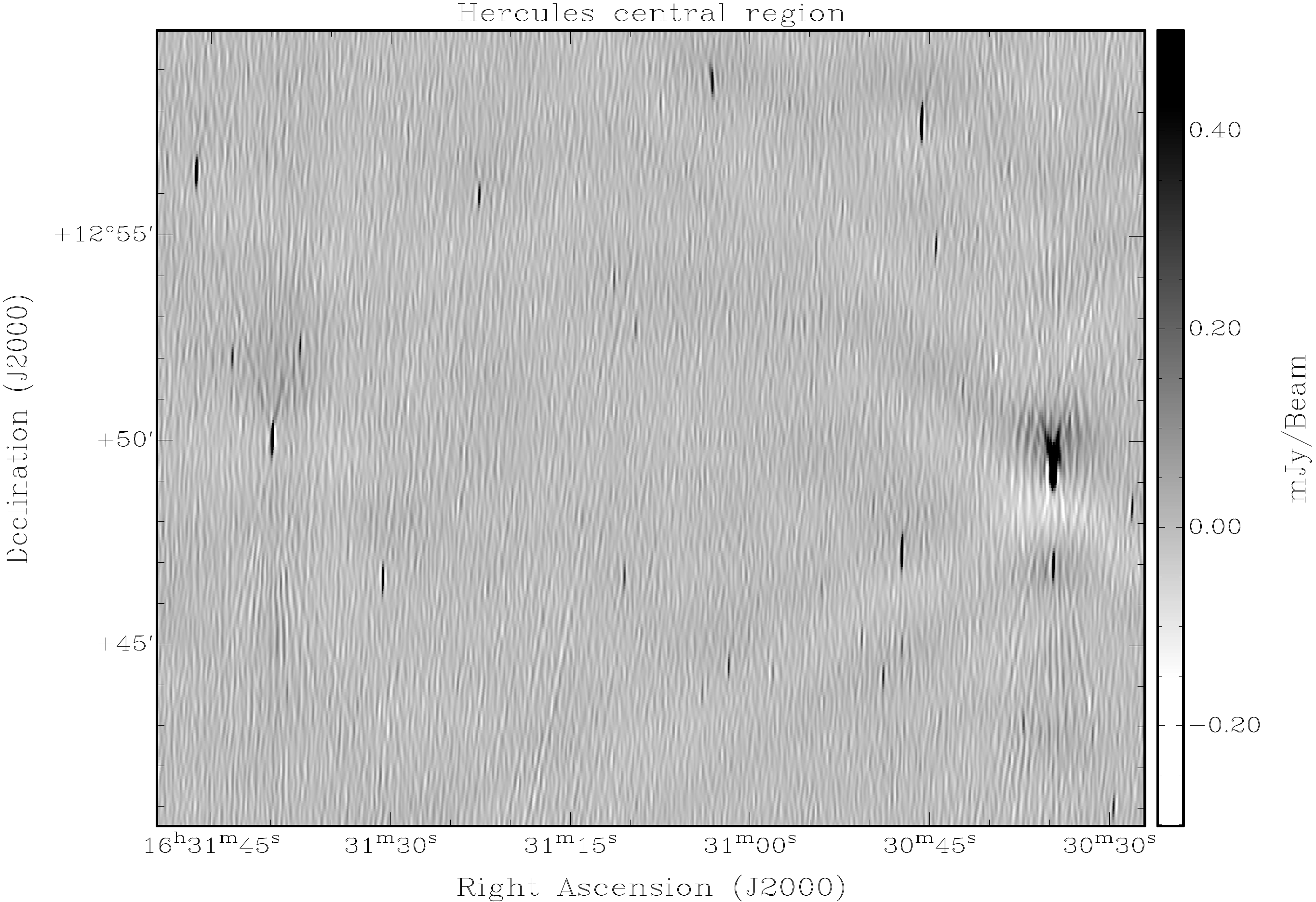}
 \end{minipage}\\ 
\vspace{5mm}
\hspace{-10mm}
 \begin{minipage}[htb]{7.5cm}
   \centering
   \includegraphics[width=0.85\textwidth,angle=-90]{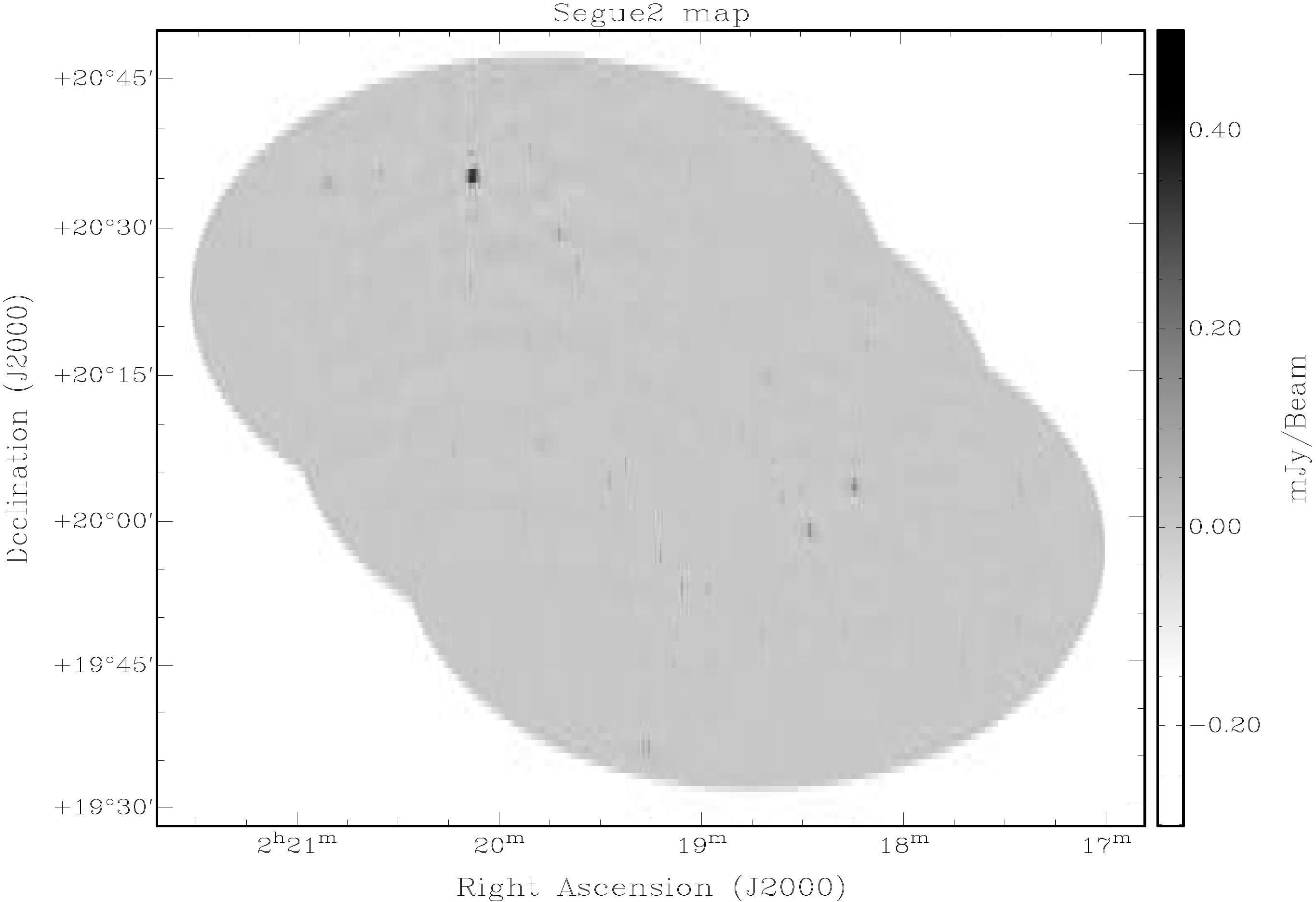}
 \end{minipage}
\hspace{25mm}
 \begin{minipage}[htb]{7.5cm}
   \centering
   \includegraphics[width=0.85\textwidth,angle=-90]{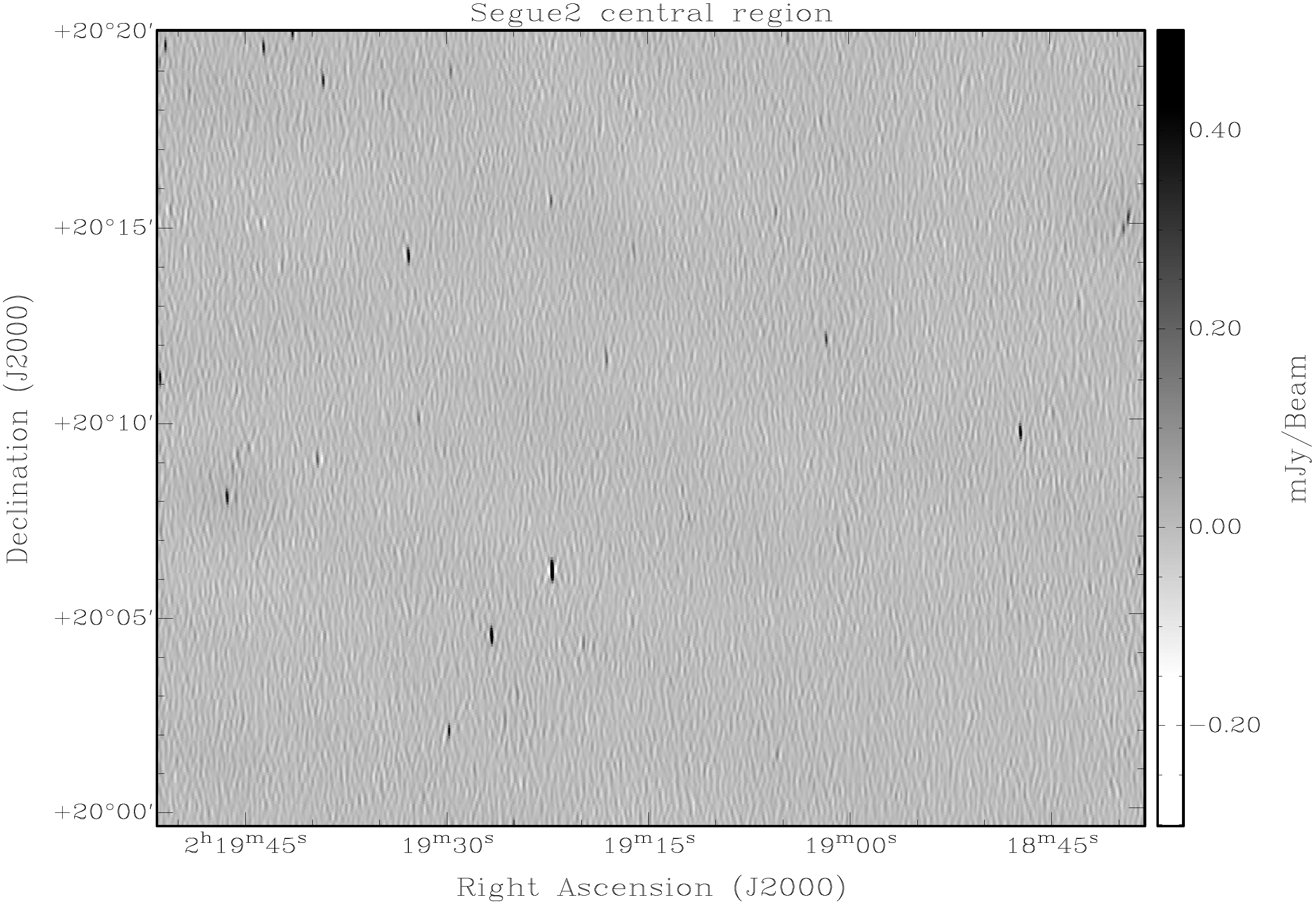}
 \end{minipage}
\caption{{\bf Maps.} Same of Fig.~\ref{fig:maps1} but for BootesII, Hercules, and Segue 2 FoV (from top to bottom).}
\label{fig:maps2}
\end{figure*}

\begin{figure*}
\centering
\hspace{-10mm}
 \begin{minipage}[htb]{7.5cm}
   \centering
   \includegraphics[width=0.85\textwidth,angle=-90]{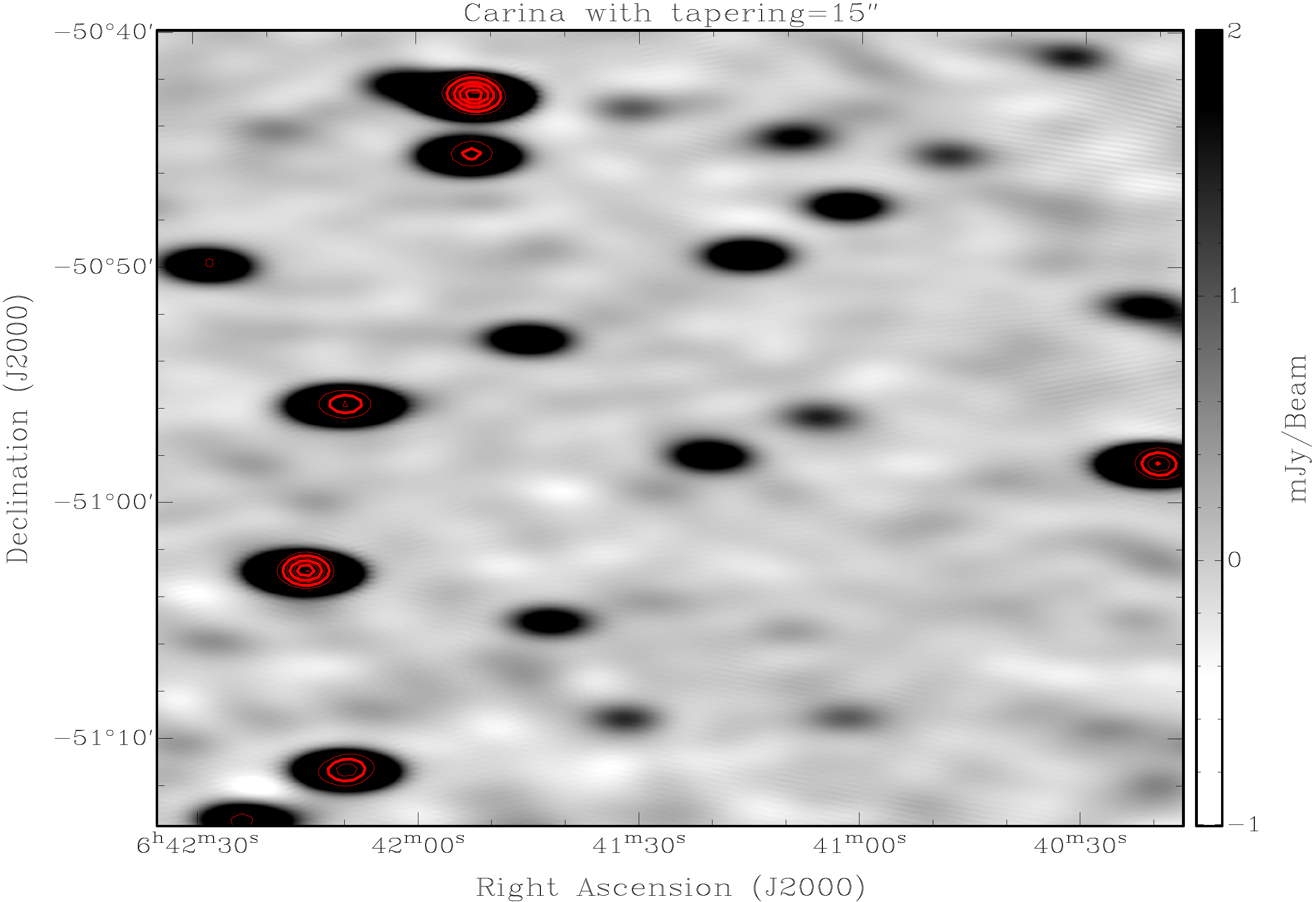}
 \end{minipage} 
\hspace{25mm}
 \begin{minipage}[htb]{7.5cm}
   \centering
   \includegraphics[width=0.85\textwidth,angle=-90]{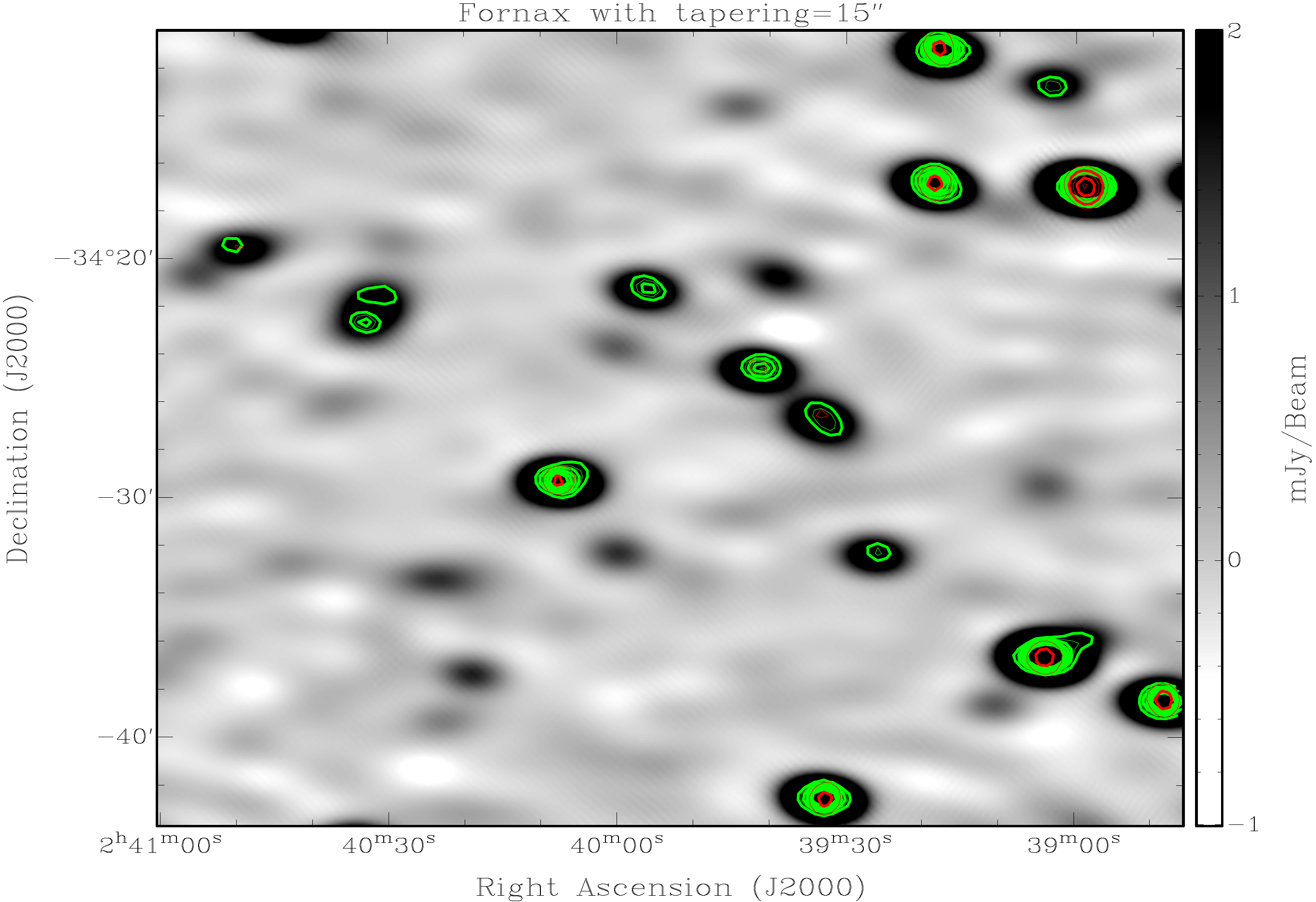}
 \end{minipage}\\ 
\vspace{5mm}
\hspace{-10mm}
 \begin{minipage}[htb]{7.5cm}
   \includegraphics[width=0.85\textwidth,angle=-90]{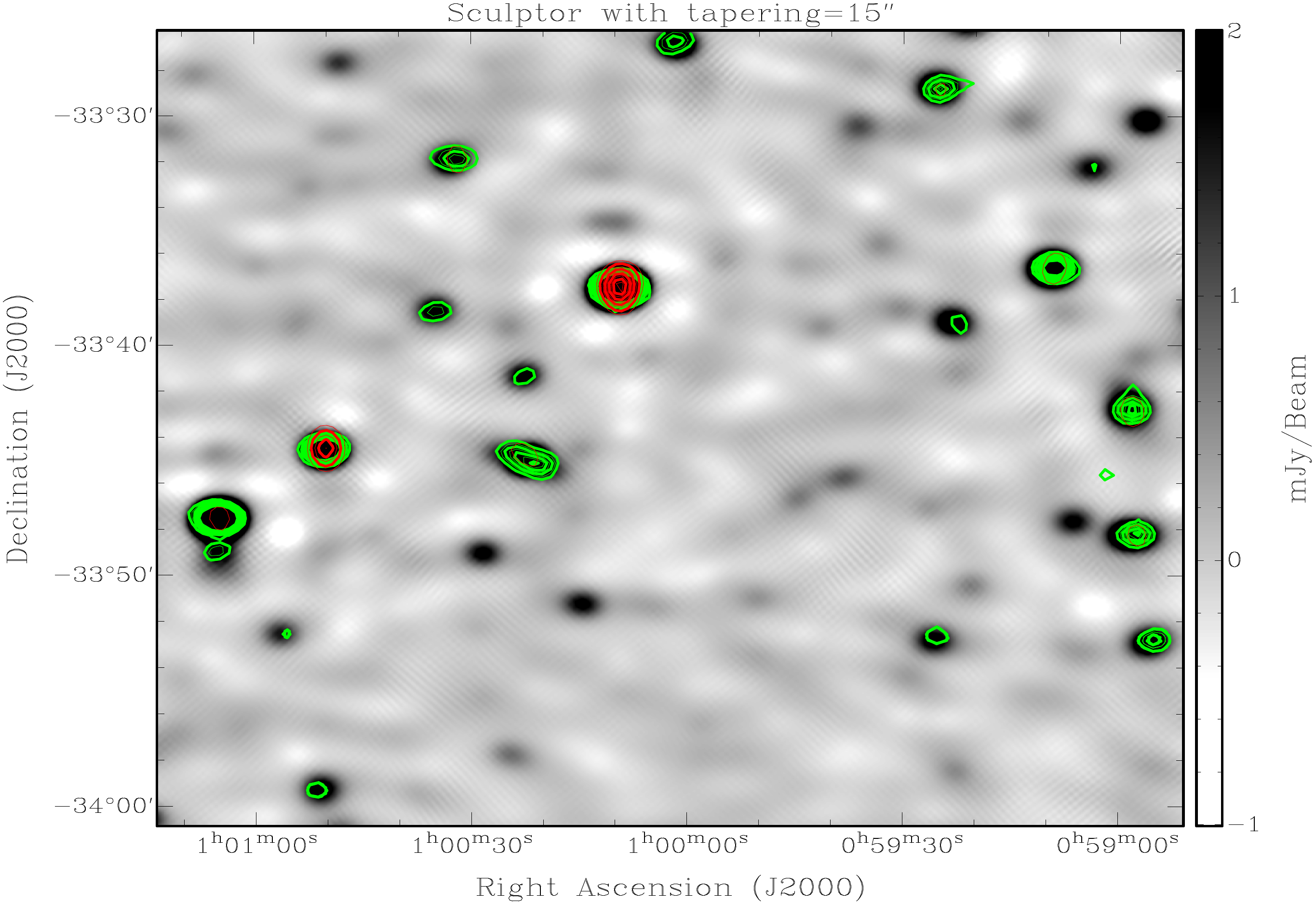}
 \end{minipage}
\hspace{25mm}
 \begin{minipage}[htb]{7.5cm}
   \centering
   \includegraphics[width=0.85\textwidth,angle=-90]{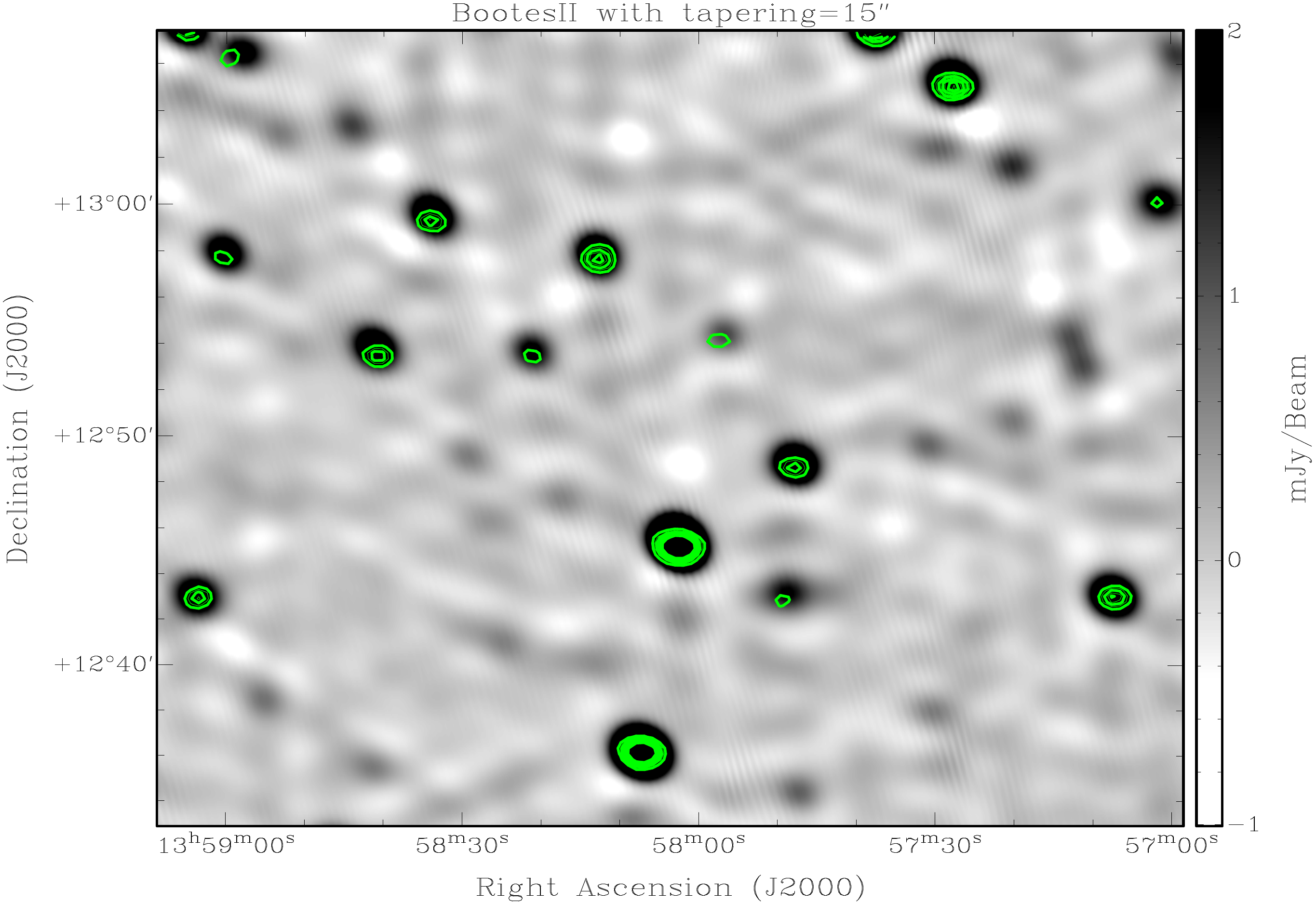}
 \end{minipage}\\ 
\vspace{5mm}
\hspace{-10mm}
 \begin{minipage}[htb]{7.5cm}
   \centering
   \includegraphics[width=0.85\textwidth,angle=-90]{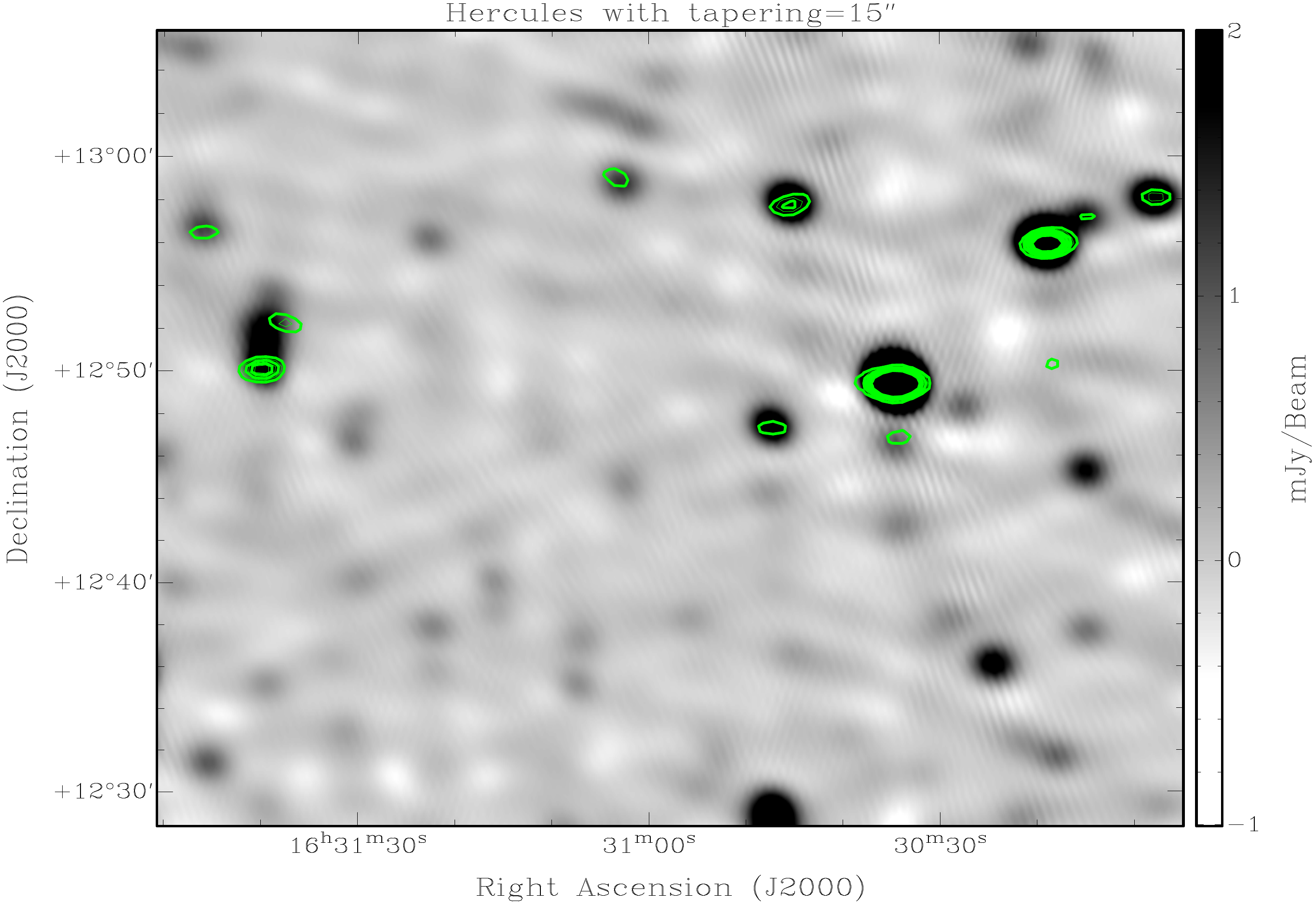}
 \end{minipage}
\hspace{25mm}
 \begin{minipage}[htb]{7.5cm}
   \centering
   \includegraphics[width=0.85\textwidth,angle=-90]{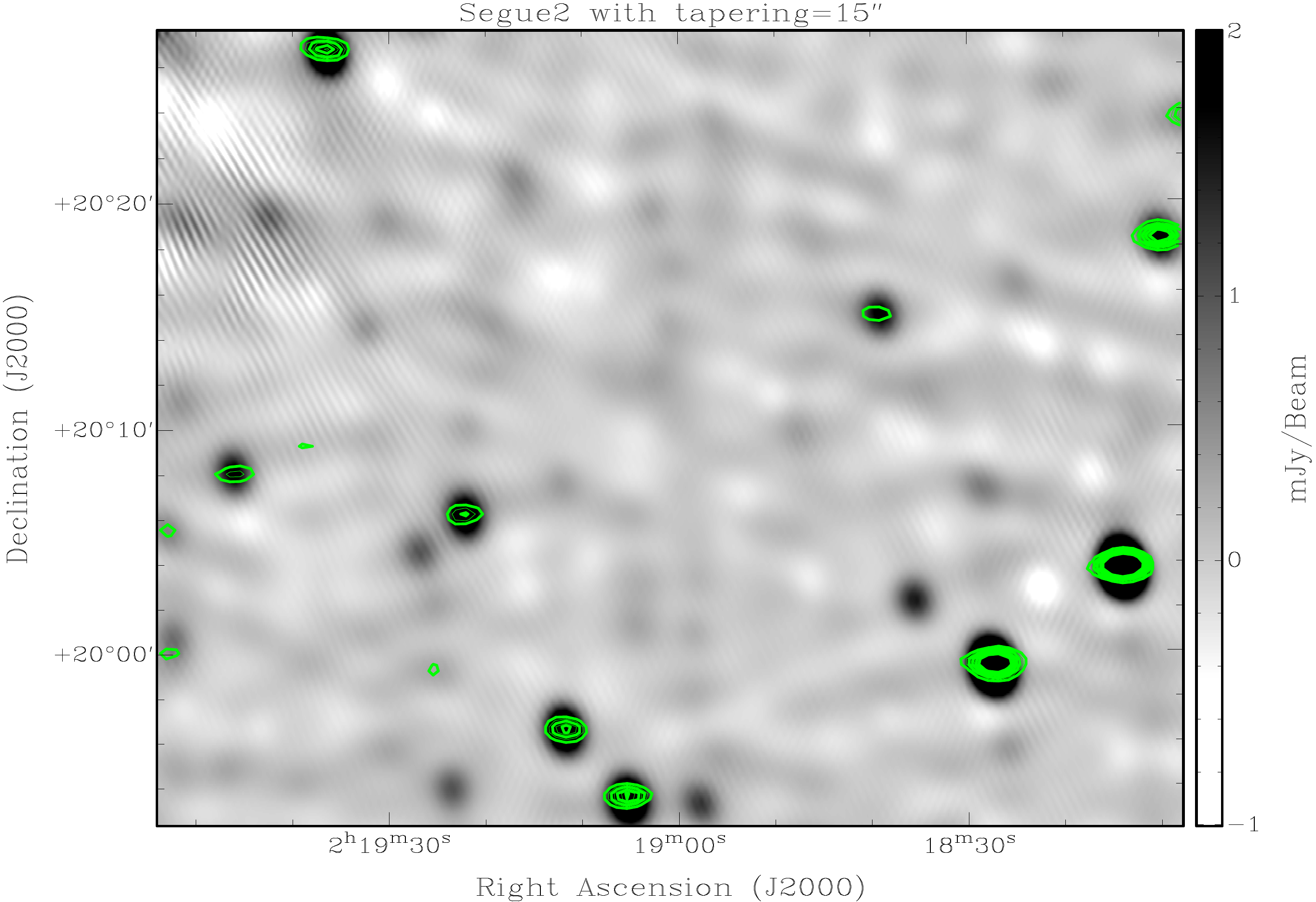}
 \end{minipage} 
\caption{{\bf Maps.} Central region of observational maps obtained with a Gaussian taper of 15 arcsec of FWHM for Carina, Fornax, Sculptor, BootesII, Hercules, and Segue2 (from top left to bottom right). Contours from NVSS (from 2 mJy, green) and SUMSS (from 10 mJy, red) sources are overlaid.}
\label{fig:maps3}
\end{figure*}

\begin{figure*}
\centering
 \begin{minipage}[htb]{7.5cm}
   \centering
   \includegraphics[width=0.6\textwidth,angle=-90]{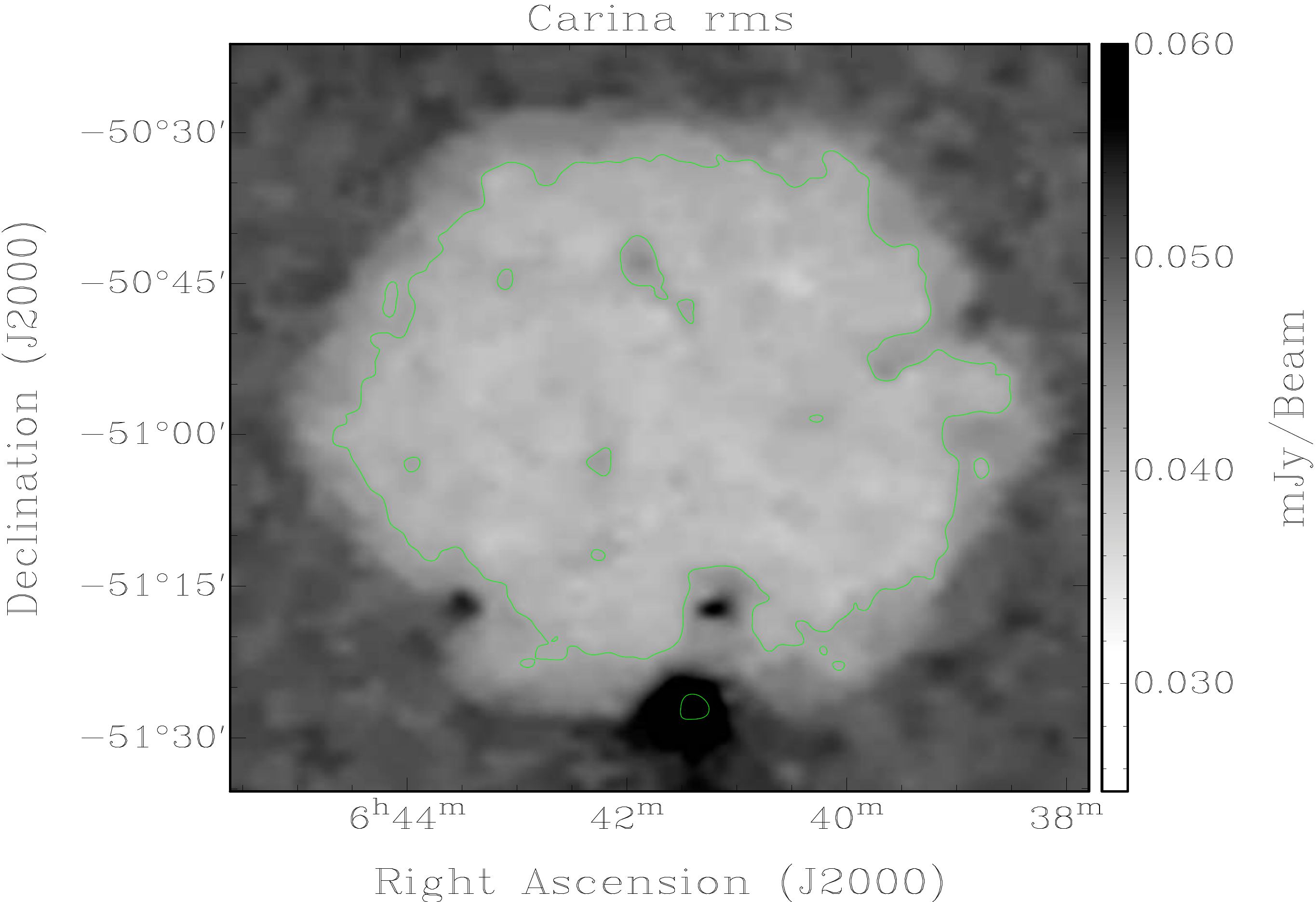}
 \end{minipage}
\hspace{5mm}
 \begin{minipage}[htb]{7.5cm}
   \centering
   \includegraphics[width=0.6\textwidth,angle=-90]{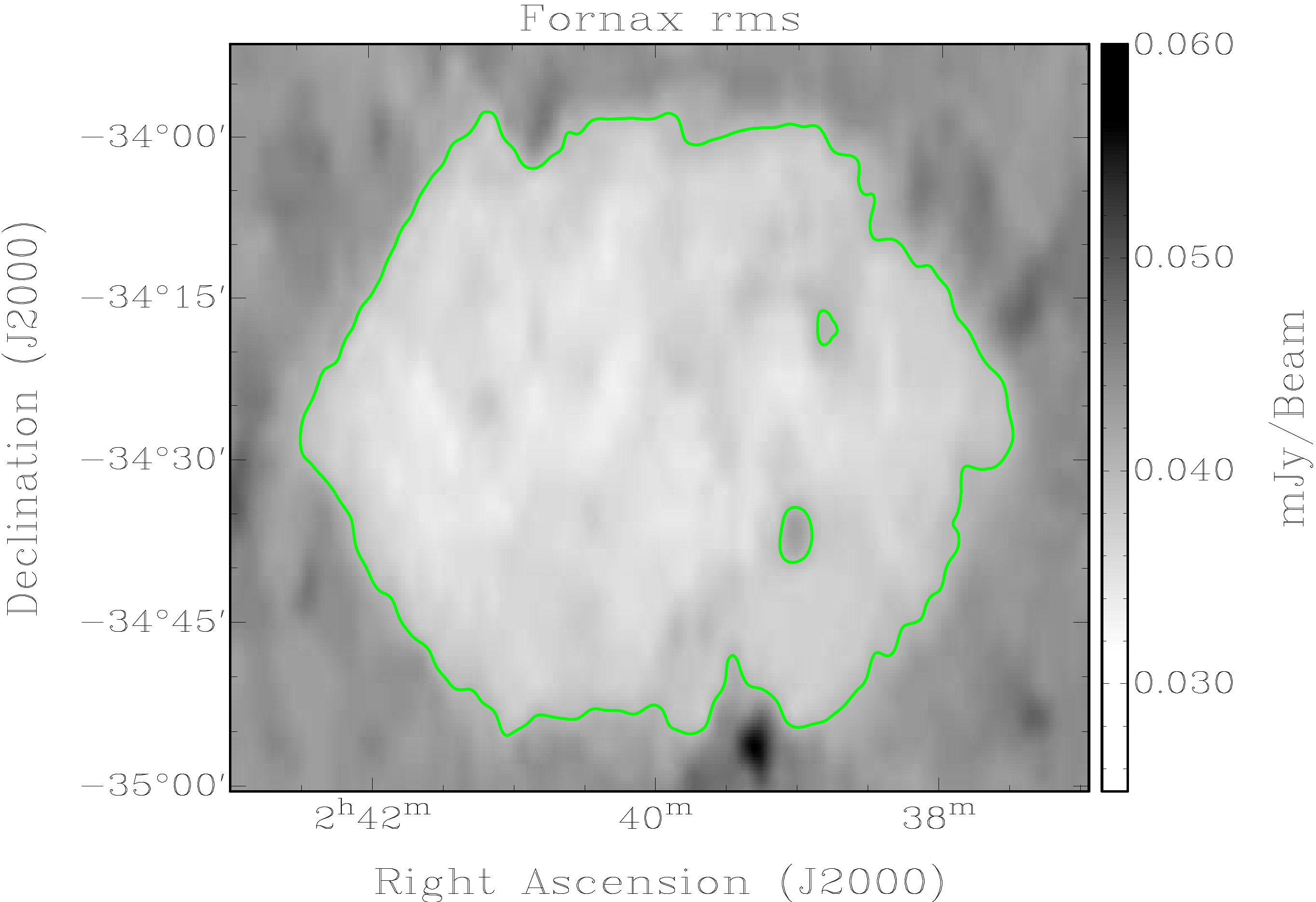}
 \end{minipage}\\
\vspace{5mm}
 \begin{minipage}[htb]{7.5cm}
   \centering
   \includegraphics[width=0.6\textwidth,angle=-90]{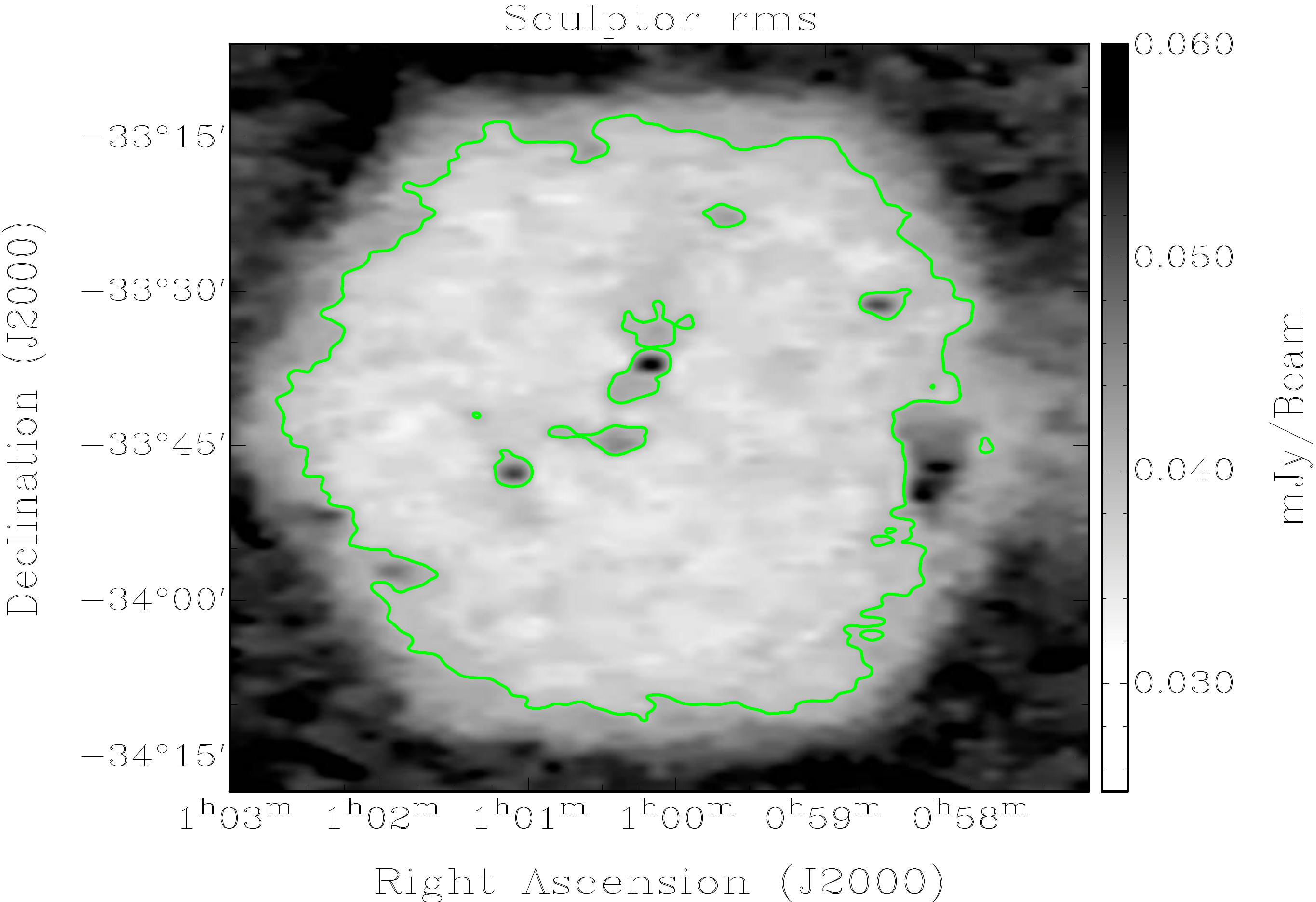}
 \end{minipage}
\hspace{5mm}
 \begin{minipage}[htb]{7.5cm}
   \centering
   \includegraphics[width=0.6\textwidth,angle=-90]{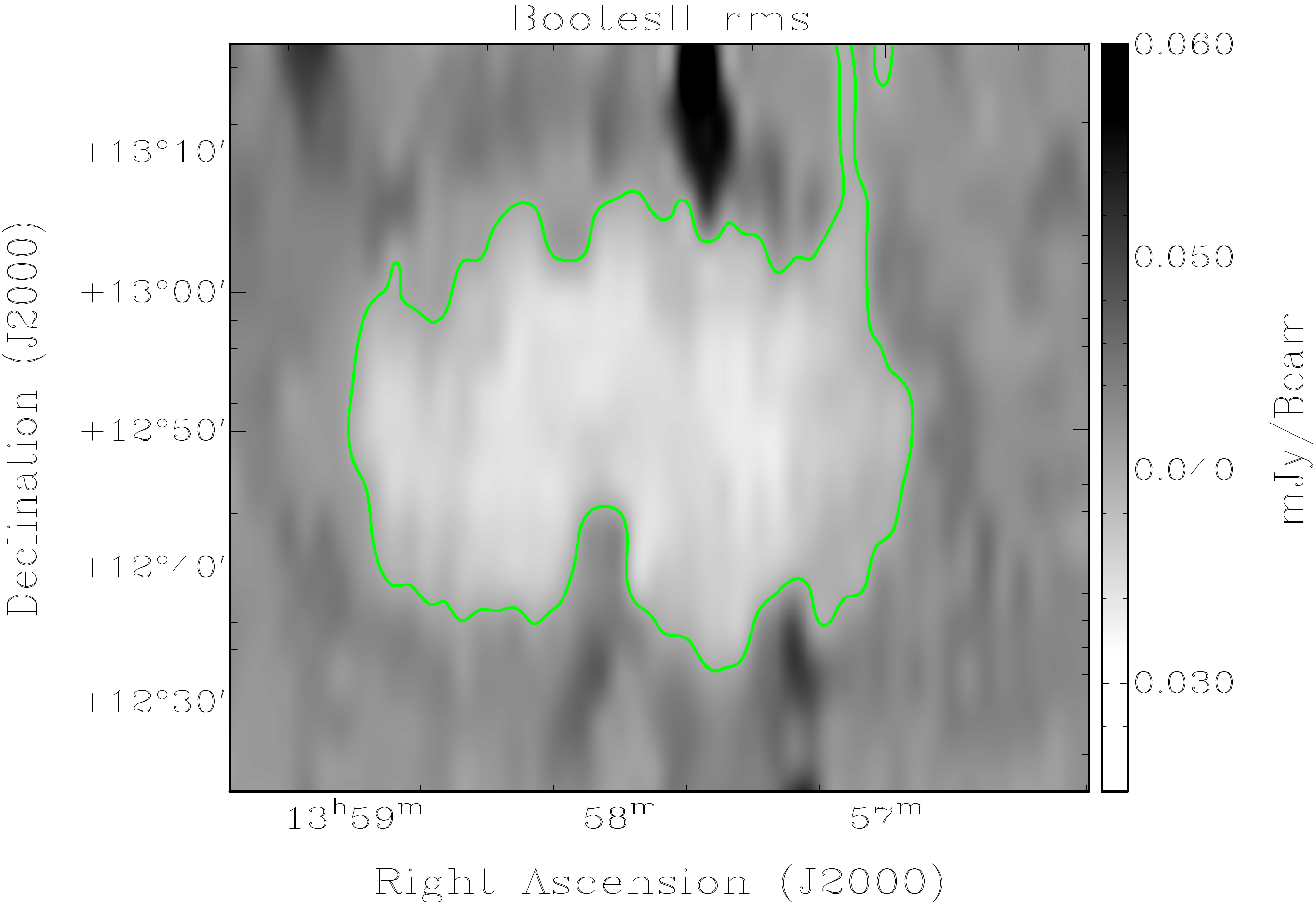}
 \end{minipage}\\
\vspace{5mm}
 \begin{minipage}[htb]{7.5cm}
   \centering
   \includegraphics[width=0.6\textwidth,angle=-90]{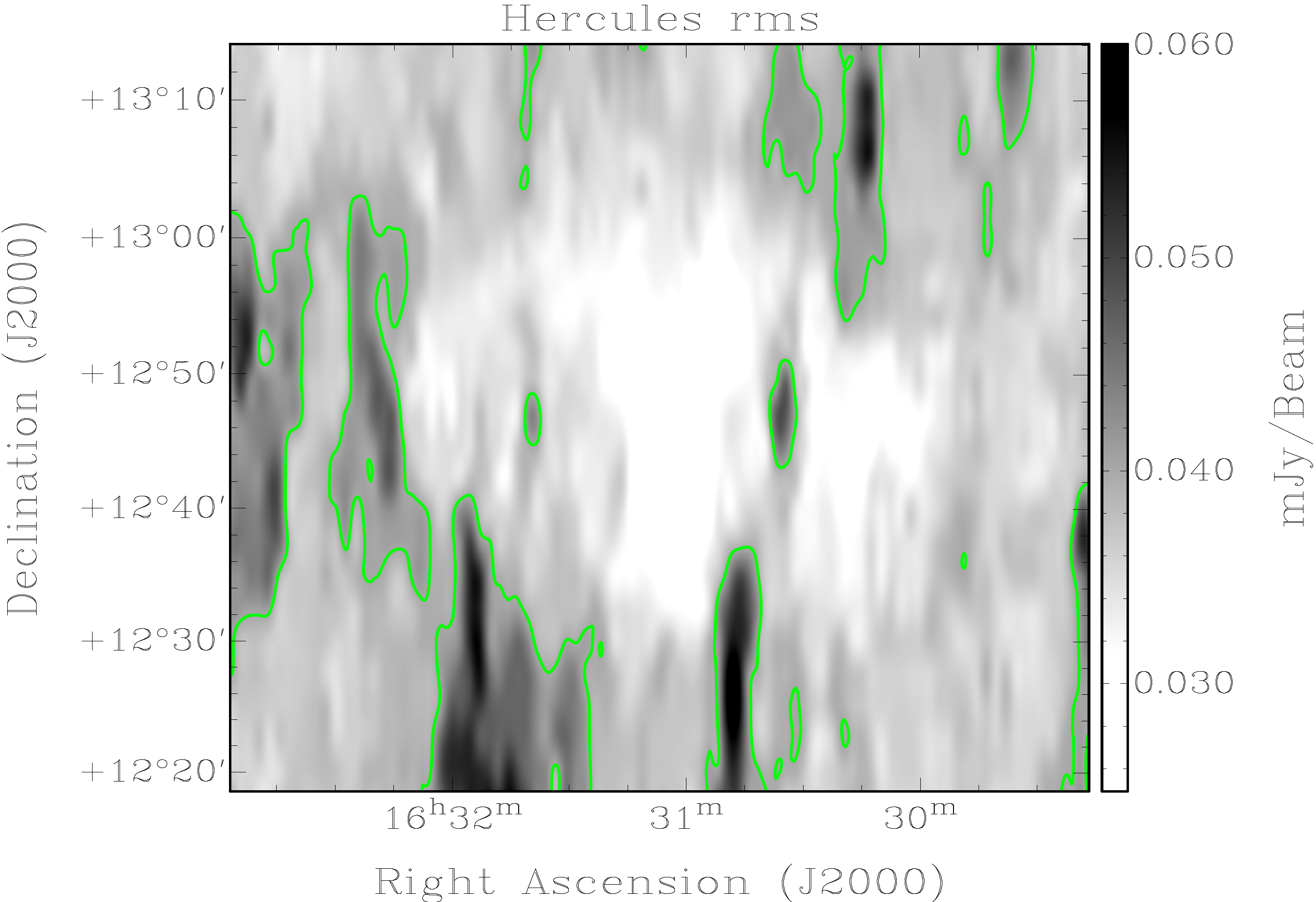}
 \end{minipage}
\hspace{5mm}
 \begin{minipage}[htb]{7.5cm}
   \centering
   \includegraphics[width=0.6\textwidth,angle=-90]{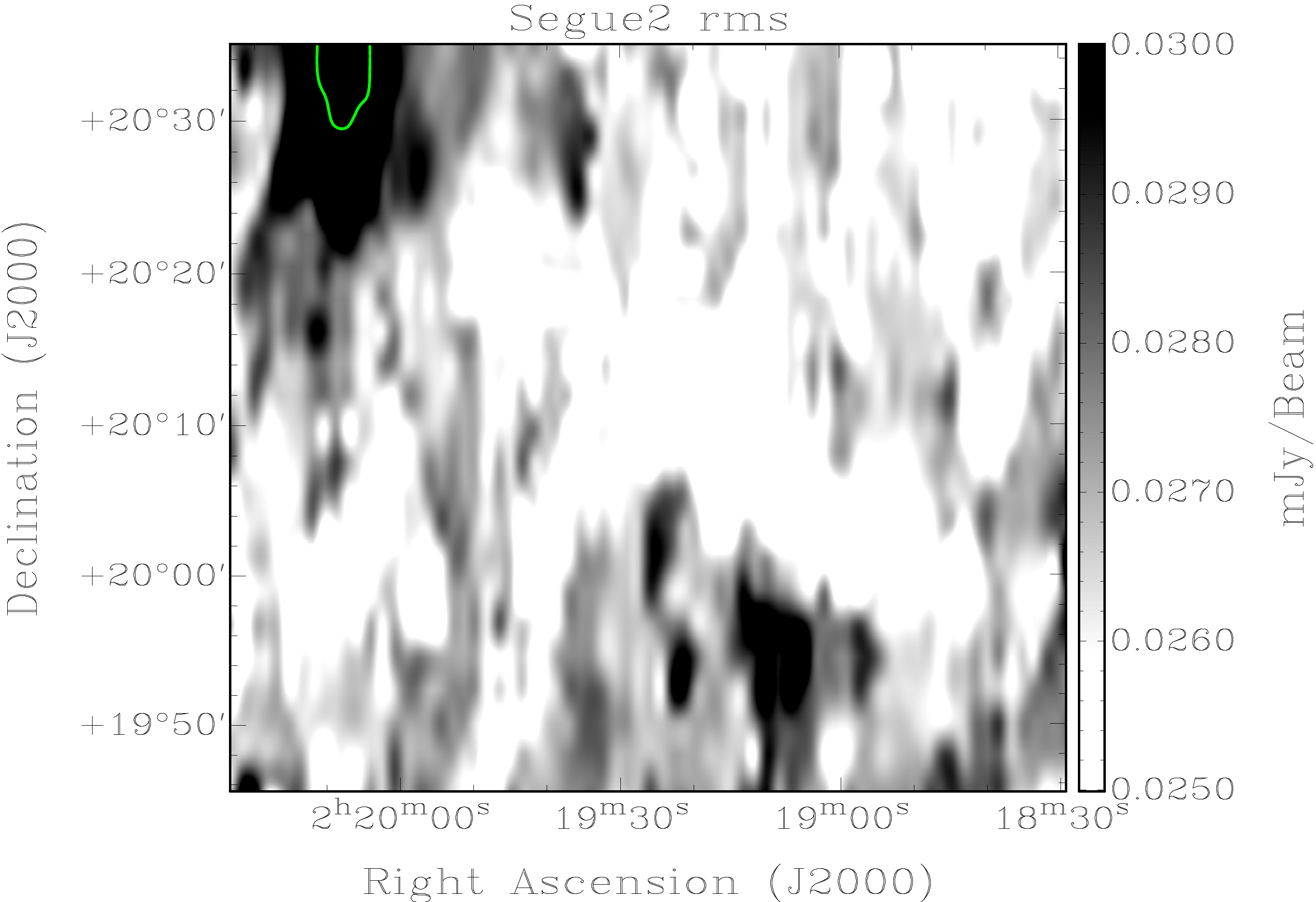}
 \end{minipage} \\
\vspace{-30mm}
 \begin{minipage}[htb]{8cm}
   \centering
   \includegraphics[width=\textwidth]{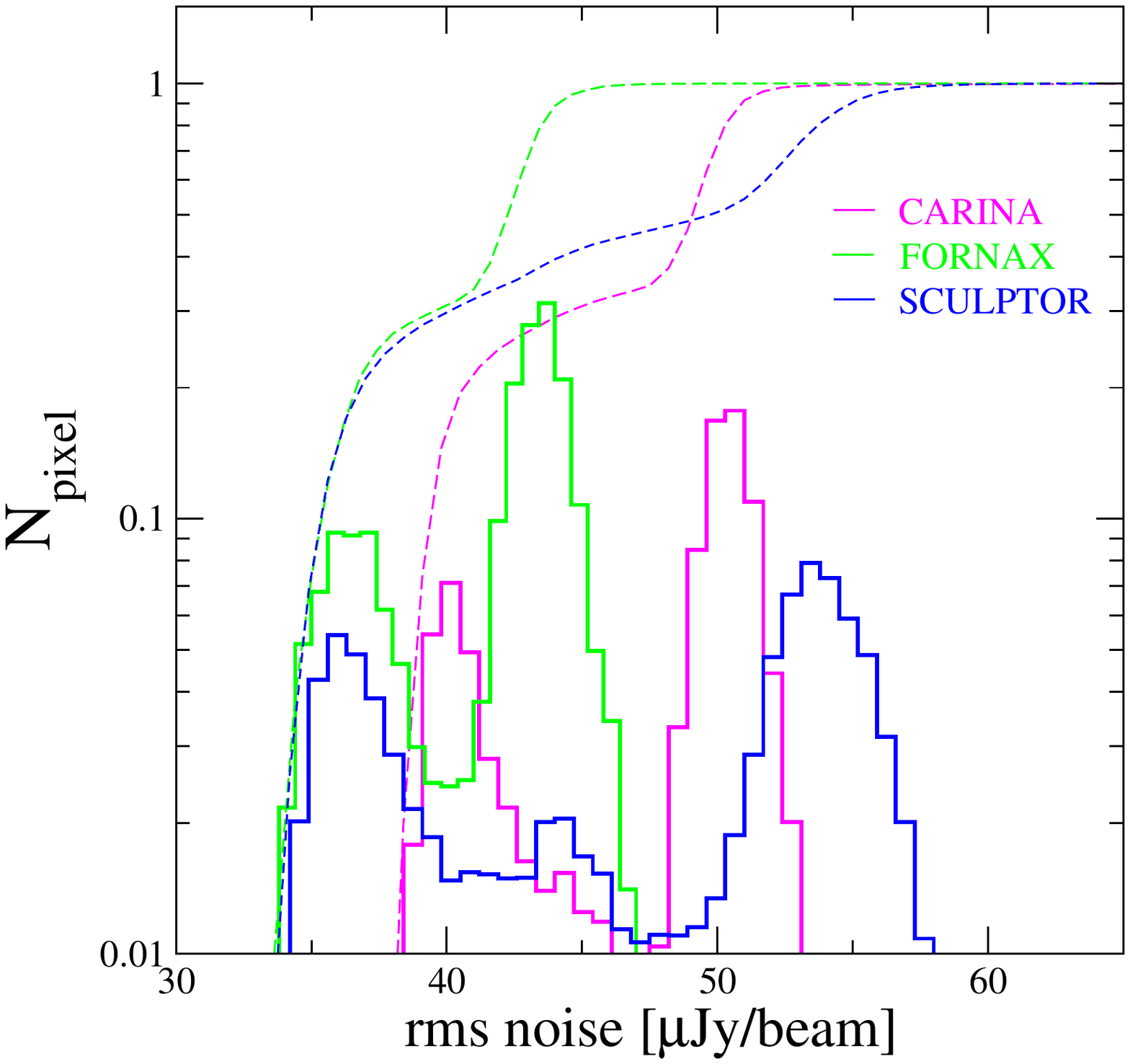}
 \end{minipage}
 \begin{minipage}[htb]{8cm}
   \centering
   \includegraphics[width=\textwidth]{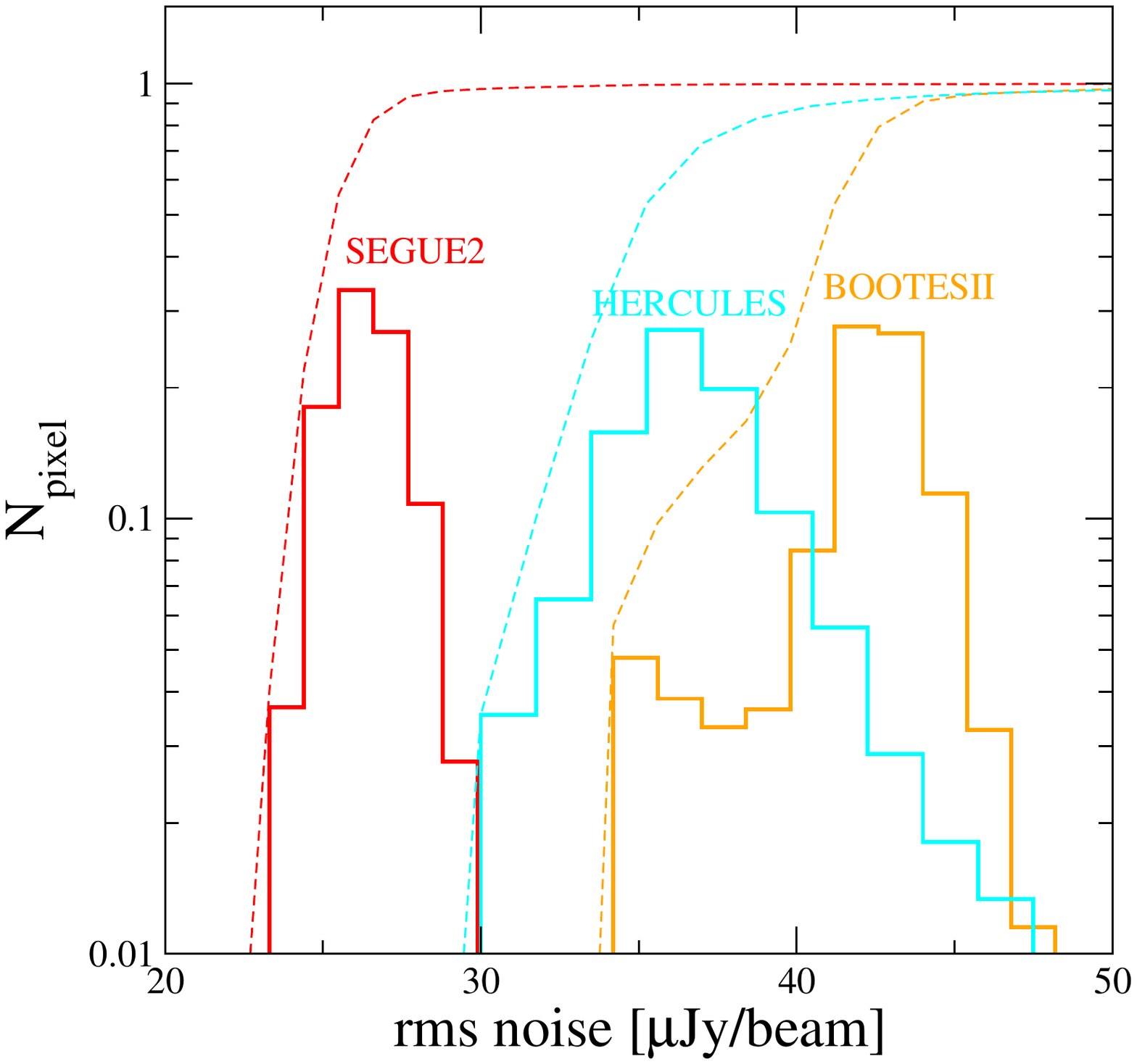}
 \end{minipage}
    \caption{{\bf Rms}. Upper and middle rows: RMS noise in the central part of the maps with contours delimiting the rms$<40 \,\mu$Jy region. The darkest spots are related to remnants from the cleaning of bright sources. Lower row: Number of pixels at a given noise and cumulative distribution (arbitrarily normalized). Note, especially in the CDS cases, the two peaks corresponding to inner and outer region. Left: Carina, Fornax, and Sculptor FoVs. Right: BootesII, Hercules, and Segue2 FoVs.}
\label{fig:rms}
 \end{figure*} 

\section{Source detection and catalogues}
\label{sec:cat}
The mosaiced robust~$-1$ images of each field are presented in Figs.~\ref{fig:maps1} and \ref{fig:maps2}, alongside zoomed in images of the central regions of each field. Fig.~\ref{fig:maps3} shows the central regions of the 15~arcsec Gaussian taper images, overlayed with NVSS and/or SUMSS contours for comparison.

We considered two automated routines for source-extraction and cataloging, which are provided by the SourceEXTRACTOR package~\citep{Bertin:1996fj} and the task SFIND in {\it Miriad}. In {\it Miriad}, the threshold for detection comes from the constraint of a maximal false detection rate (FDR), while in SExtractor it follows from a certain $\sigma-$level above the local background.

Previous analyses \citep[see, for example,][and references therein]{Huynh:2012hp} have found that mesh-sizes with widths of about 10 times the synthesised beam produce reliable noise estimation and completeness in deep radio continuum surveys. We estimated the local rms noise by splitting the map in regions corresponding to $\sim 10\times10$ of the synthesized beam. The rms maps obtained with SExtractor for the high-resolution maps are shown in Fig.~\ref{fig:rms}, upper panels.
The SExtractor algorithm consists of computing the mean and the standard deviation $\sigma$ of the distribution of pixel values within each sub-region. 
The most deviant values are then taken out and the means and standard deviations are re-computed. 
This is repeated until all the remaining pixel values are within 3-$\sigma$ from the mean.
The standard deviations in each sub-region form the noise map.

In the lower panel of Fig.~\ref{fig:rms}, we show the density and cumulative distributions of pixels in the noise map of each target.
The rms in the central (outer) region varies from a minimum of 25 $\mu$Jy/beam (29 $\mu$Jy/beam) in the Segue2 FoV to a maximum of 40 $\mu$Jy/beam (50 $\mu$Jy/beam) in the Carina FoV. 
It is of the same order of the nominal rms computed from the ATCA sensitivity calculator, as reported in Section~\ref{sec:obs}, adjusted to account for $33\%$ data loss due to RFI and for mosaic effects (and despite we used the more conservative robust -1 imaging rather than natural weighting).
The most successful imaging is for the cases of the Fornax and Sculptor FoVs, where there are no major issues. 
The high dynamic range in the Carina case and the high-DEC of the UDS FoVs (see above discussion on w-term effects) have made the deconvolution somewhat less successful for these targets.

For the CDS targets, we note a clear bi-modality in the noise distribution in Fig.~\ref{fig:rms}.
The peak at higher rms values is associated to the pixels in the outer ``ring'' of the maps, where fewer overlapping fields are present with respect to the central part of the mosaic.
A similar trend is seen also in the case of BootesII, but, in general, the rms distribution in UDS is more uniform due to the lower number of mosaic fields.

The source identification in SFIND was performed setting the parameter $\alpha$ to 0.1 (which means 99.9\% reliable catalogue for the case of a perfect image with pure Gaussian noise) and the rmsbox option to 10 synthesized beams.
On our maps, SFIND and SExtractor give nearly identical results for astrometry (number of sources and positions), once the threshold parameters in SExtractor are tuned (we found a threshold typically slightly above $5\sigma$).
The mismatch on positions is random, and about $1''$ on average for all FoVs. This value can be taken as an estimate of our positional accuracy.

Photometry can, on the other hand, give quite different results for some sources. We individually inspected a few of those cases, and concluded that it can be due to the fact that
SExtractor is optimised for optical images, which have significantly different signal and noise structures with respect to radio maps (e.g. correlation of noise on large scales is not present). In the following, we will use results from SFIND which is instead specifically written to analyse radio images, so accounting for artifacts and sidelobes. 
Photometric parameters are determined by a routine that selects contiguous monotonically decreasing adjacent pixels
from the FDR-selected ones, and fits them with a 2-D elliptical Gaussian.

The error on the estimated fluxes can be computed by adding in quadrature the local rms and the {\it Miriad} fit-error. In order to take into account possible inaccuracy in the calibration model and process (in particular associated to possible unaccounted RFI), we also conservatively added in quadrature an error equal to the 5\% of the flux to the error provided in the catalogue (see, for example, a similar approach in \citet{Massardi:2007kw}) .

A theoretical estimate of the positional error is $FWHM/2\cdot \sigma_{rms}/S_{peak}$, with $S_{peak}$ being the peak flux density. Since the threshold for detection is set to about $5\sigma_{rms}$ and the beam is $\sim 8''$, the error for faintest sources is at arcsec level. This is compatible with comparisons between SFIND and SExtractor.
In the Sculptor FoV we have a source from the ICRF catalogue: (ICRF J010009.3-333731)  0057-338  RA=01 00 09.39094184, DEC=-33 37 31.9360512.
The position obtained in our catalogue is RA=01 00 09.398, DEC=-33 37 31.98, thus in agreement with ICRF at 0.1 arcsec level.
The source is very bright ($\sim$100 mJy at 2 GHz), and we expect a degradation within one order of magnitude for faintest sources \citep[see, for example, simulations in][]{Huynh:2012hp}.

Putting all above arguments together, we can conservatively assume 1 arcsec as our positional uncertainty. 
In fact the estimate of the positional error provided by the SFIND algorithm is $\lesssim 1$ arcsec for all the detected sources.

\begin{table*}
{\scriptsize
\centering
\begin{tabular}{|c|c|c|c|ccc|ccc|ccc|}
\hline
FoV & r.m.s. & number of  & multiple & \multicolumn{3}{|c|} {NVSS} & \multicolumn{3}{|c|} {SUMSS} & \multicolumn{3}{|c|} {FIRST}  \\ 
    & $\mu$Jy & sources & sources & sources & $\langle \beta \rangle$ & $\langle \Delta \theta \rangle$ & sources & $\langle \beta \rangle$ & $\langle \Delta \theta \rangle$ & sources & $\langle \beta \rangle$ & $\langle \Delta \theta \rangle$ \\ 
\hline 
Carina    & 40 (50) & 225 & 32 & / & / & / & 39 (39) & $-0.9\pm0.1$ & $2.8''$& / & / & / \\
Fornax    & 36 (43) & 362 & 51 & 80 (79) & $-0.9\pm0.1$ & $3.5''$ & 46 (46) & $-0.8\pm0.1$ & $1.9''$ & / & / & / \\ 
Scultpor  & 31 (53) & 316 & 44 & 67 (59) & $-0.4\pm0.1$ & $4.1''$ & 40 (40) & $-0.6\pm0.1$ & $2.2''$ & / & / & / \\ 
BootesII  & 34 (41) & 173 & 20 & 39 (39) & $-1.0\pm0.1$ & $4.7''$ & / & / & / & 68 (65) & $-0.7\pm0.2$ & $1.3''$ \\ 
Hercules  & 30 (37) & 169 & 16 & 24 (23) & $-1.1\pm0.3$ & $4.5''$ & / & / & / & 58 (57) & $-1.0\pm0.1$ & $1.7''$ \\
Segue2    & 25 (29) & 147 & 15 & 18 (17) & $-2.0(1.1) \pm0.3$ & $5.1''$ & / & / & / & / & / & / \\ 
\hline
\end{tabular}
\caption{Summary of FoV properties. In the second column, the rms is reported for the inner (outer) region. Third column shows the number of detected sources in our survey, with the number of sources having more than one entry in the catalogue is in column 4. Comparisons with other surveys is performed only in the inner region (avoiding sources at distances smaller that 10 arcmin from the boundary of our image) of the FoV and shown from column 5. For each survey we show the total number of sources, the ones having a match in our catalogue (in bracket), average spectral index $\langle \beta \rangle$ (including multiple component sources), and average positional offset $\langle \Delta \theta \rangle$ (excluding multiple component sources).}
\label{tab:summary}
}
\end{table*}

In the high-resolution maps, we included the signal from long baselines involving the sixth antenna, and the synthesized beam is $\sim8''$. With the rule of thumb of 10 beam per source, one can estimate the confusion limit to be around 3 $\mu$Jy. Therefore, consistently to what is found, confusion does not represent an issue for these maps.
Including only shortest baselines (i.e., excluding the contribution from the sixth antenna), the beam grows to $\sim2'$ with a confusion limit at about 500 $\mu$Jy.

As discussed in Section~\ref{sec:red}, a good imaging is achieved setting the robustness parameter to -1. 
This however down-weights short baselines and the extended diffuse flux density is, in some cases, poorly reconstructed.
In order to recover it, we also consider a map where we apply a Gaussian taper of 15 arcsec (still with robustness parameter equal to -1), which basically strongly down-weights long baselines.
A combination of robustness parameter and Gaussian taper were used to explore imaging parameter space to 
produce optimal images, which was not possible by simply using natural weighting.
The Gaussian tapered image has a confusion noise significantly larger than the instrumental rms.

The flux densities obtained from this map have been compared to the flux densities of the un-tapered map.
When there is a one-to-one correspondence between sources (where in the un-tapered map, different components of multiple component sources have already been gathered), we use the flux density from the tapered map as the main estimate of the total flux density, since it recovers the diffuse part of the emission.
When instead different sources of the  un-tapered map are seen as a single source in the tapered one, we associate the extended flux density to the source of the un-tapered map which is closer to the tapered peak. 
The total flux density measured in the tapered map minus the flux density of the non-associated sources measured in the un-tapered map is then the estimate of the total flux density for the associated source.
For sources which do not have a counterpart in the tapered image (which has a larger noise), the total flux density is obviously not changed.
There are a small number of cases of sources detected in the tapered map but missed in the main map. We include them in the main catalogue for completeness.

Following such procedure, we found 1835 entries in the catalogue corresponding to a total of 1392 extracted sources with 178 cases being (possibly) multiple component sources.
The number of sources in each FoV is reported in Table~\ref{tab:summary}.

Radio sources can be made up of different components.
To decide whether nearby sources are separated sources or components of a single source, we visually inspected all the fields where either $\theta_d<1'$ (with $\theta_d$ being the distance between sources) or the criterium of \citet{Magliocchetti:1998hm} ($\theta_d<100''\sqrt{S_{peak}/10\,mJy}$), was satisfied. 
A more detailed study of the 178 possible multiple sources will be reported in a companion paper.

The extension of a source can be estimated through the ratio of the integrated flux $S_{tot}$ to the peak flux densities:
$S_{tot}/S_{peak} = \theta_{min}\theta_{maj}/(b_{min} b_{maj})$ with $b_i$ ($\theta_i$) being the synthesized beam (source) FWHM axes.
A criterium often adopted in the literature is to consider sources with $S_{tot}/S_{peak}<1.3$ to be unresolved \citep{White:2012se}.
However, since we combine two different maps and the total flux density can come from the tapered image, this kind of analysis is somewhat misleading. 
A robust deconvolution criterium is hard to be defined in this case.
In the catalogue, we always quote the fitted sizes of source axes from the un-tapered map, with the caveat that, when the total flux density estimate is significantly below the total flux density from the tapered image, they underestimate the real size of the source, since do not account for the diffuse components.
Bandwidth smearing can also, in principle, affect the source extension estimation in a complicated way.
Because of the small bandwidth, however, it is likely to have only a modest effect, as discussed in the next Section.

\subsection{Possible systematic effects on flux determination: Clean bias and Bandwidth smearing}
\label{sec:syst}

Due to the fact that the bandwidth is not infinitesimally small, the peak flux of a source can be reduced (but typically with corresponding increase in the source size, so conserving the total integrated flux density). This effect is known as radio bandwidth smearing and is analogous to optical chromatic aberration. 

For single pointings this effect has been often modeled with the relation~\citep{Condon:1998iy}:
\be
A=\frac{S_{peak}}{S_{peak}^0}=\frac{1}{\sqrt{1+\frac{2\,\ln 2}{3}\left(\frac{\Delta \nu}{\nu}\frac{d}{\theta_B}\right)}}\;,
\label{eq:BWS}
\ee
with $d$ being the distance from the center of the pointing and $\theta_B$ being the synthesized FWHM.
In the case of a mosaic, the bandwidth smearing can act in a complex way \citep[see, for example,][]{Bondi:2008uc}. A procedure to estimate the attenuation is to average Eq.~\ref{eq:BWS} over the primary beams of each pointing: 
\be
\bar A=\frac{\sum_{i=1}^{N_p}\,P(r-r^c_i)\,A(r-r^c_i)}{\sum_{i=1}^{N_p}\,P(r-r^c_i)}\;\;,
\label{eq:BWSmos}
\ee
where $P(x)=\exp(-4\,\ln 2\,(x/FWHM)^2)$ is the primary beam pattern and $r^c_i$ is the center of the pointing $i$.
However, due to a correlator bug, all of the mosaic panels were correlated at the position of the first panel, as already mentioned.
This means that Eq.~\ref{eq:BWS} applies to all panels with $d$ being the distance from the center of the central panel.

\begin{figure*}
\vspace{-30mm}
\centering
 \begin{minipage}[htb]{8cm}
   \centering
   \includegraphics[width=\textwidth]{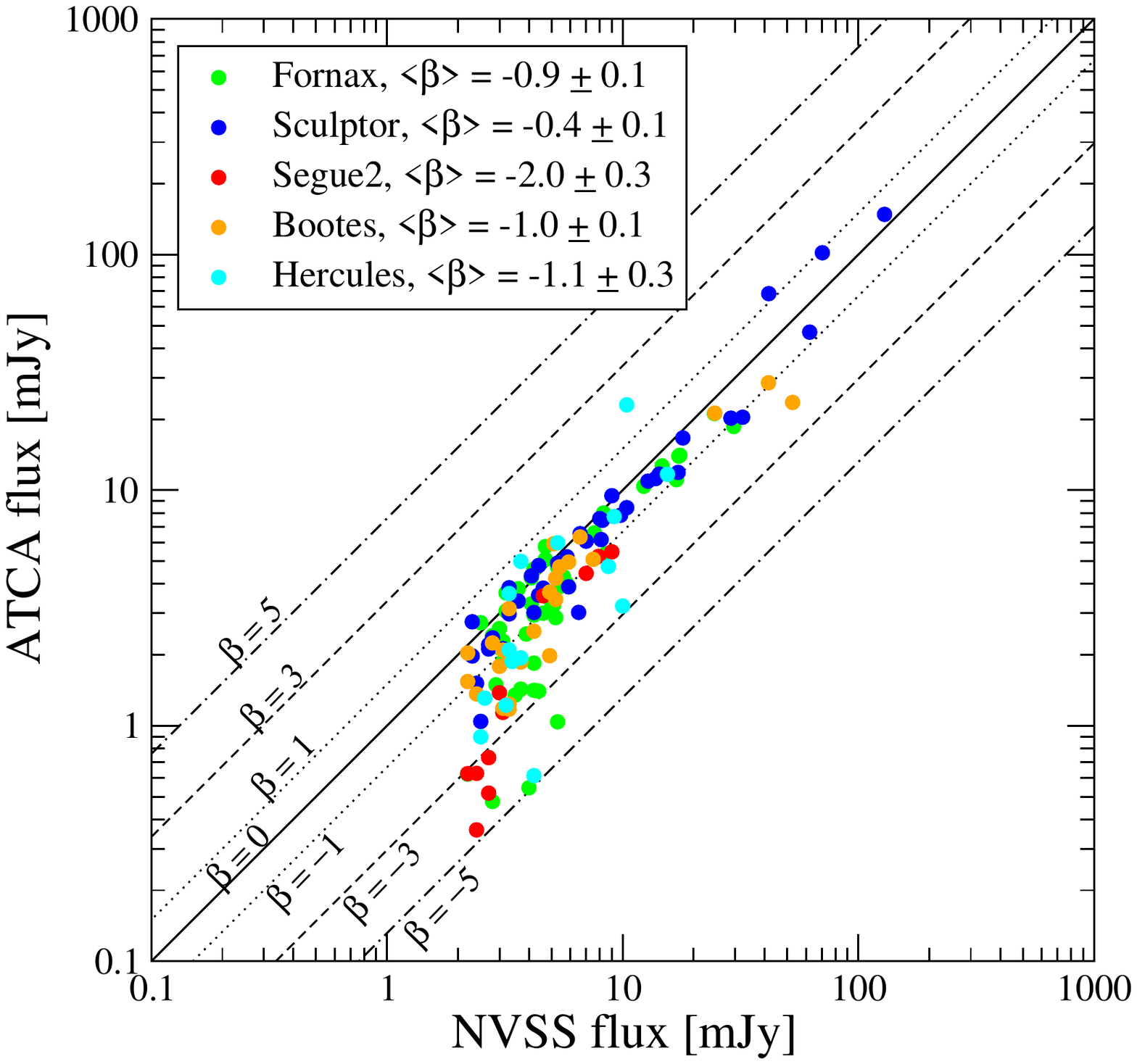}
 \end{minipage}
 \begin{minipage}[htb]{8cm}
   \centering
   \includegraphics[width=\textwidth]{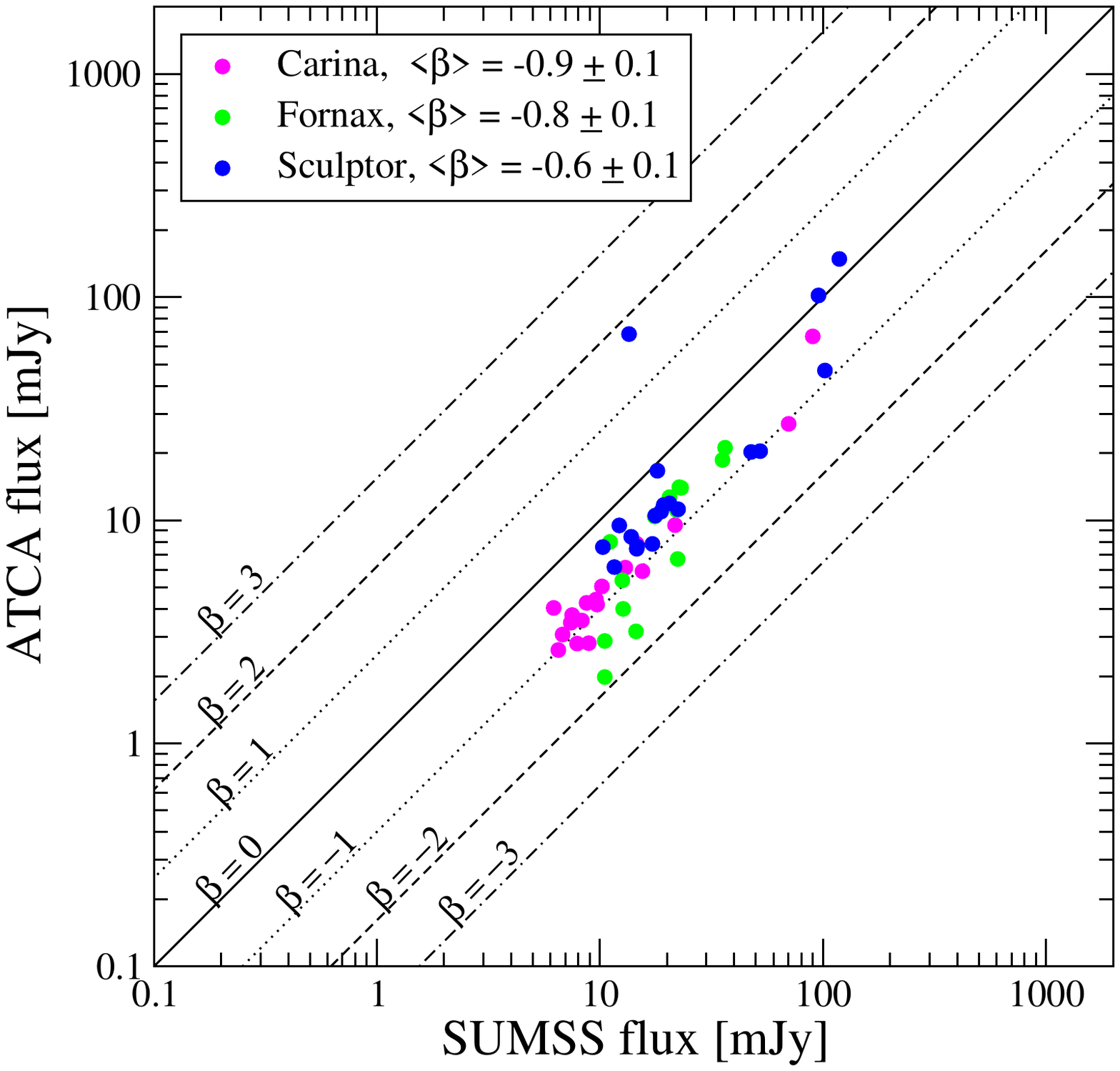}
 \end{minipage}\\
\vspace{-35mm}
 \begin{minipage}[htb]{8cm}
   \centering
   \includegraphics[width=\textwidth]{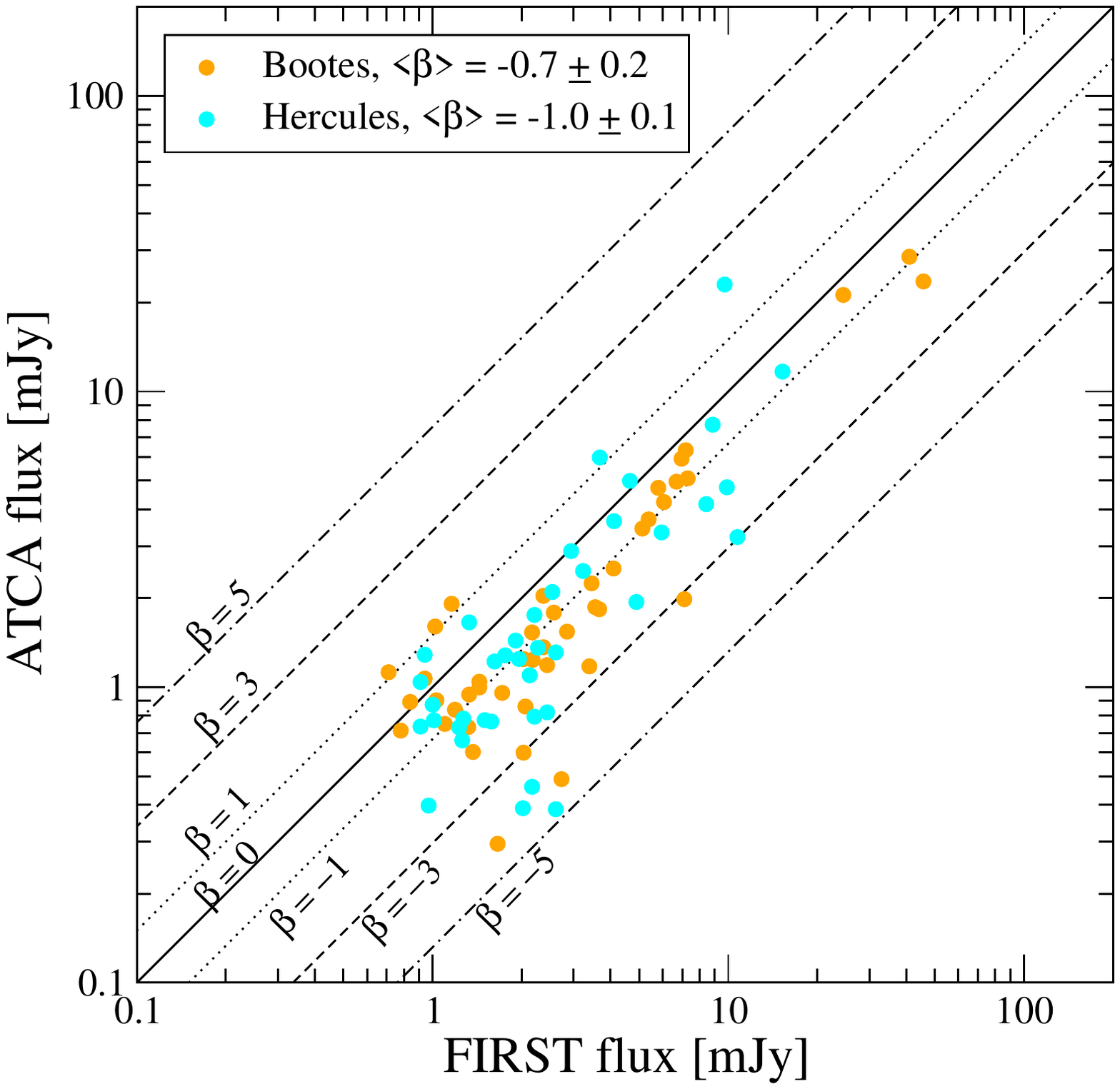}
 \end{minipage}
 \begin{minipage}[htb]{8cm}
   \centering
   \includegraphics[width=\textwidth]{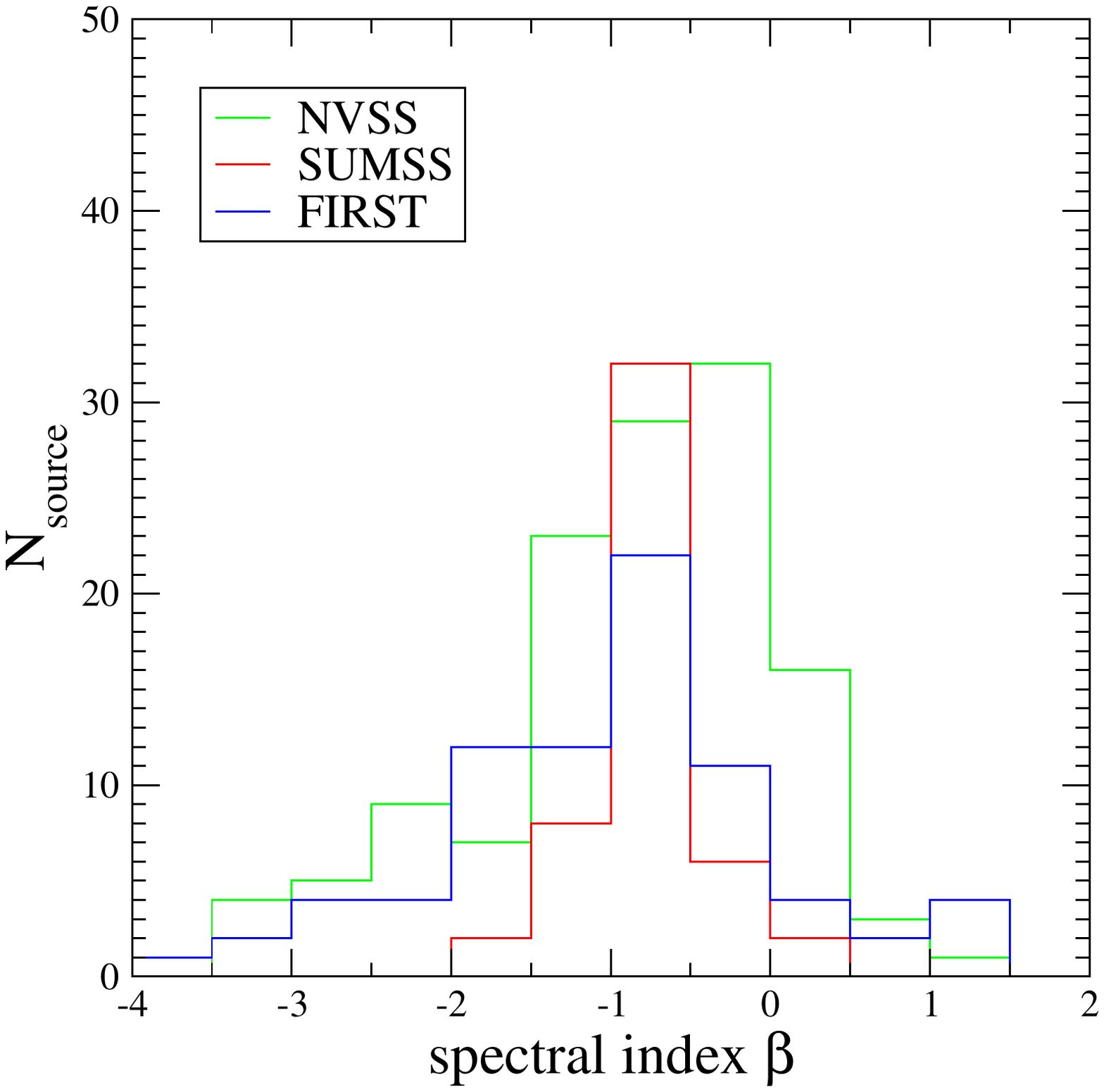}
 \end{minipage}
    \caption{{\bf Spectral index}. Comparison of derived source flux densities with the results of NVSS (top-left), SUMSS (top-right), and FIRST (bottom-left). Error bars are not reported to simplify the visualization.  Lines show spectral index level with $\beta$ being $\beta=\ln (S_{2}/S_{\nu_i})/\ln (2\,{\rm GHz}/\nu_i)$, where $\nu_i=0.843\,(1.4)$ GHz for SUMSS (NVSS, FIRST). Bottom-right panel: Number of sources as a function of spectral index. In all panels, the multiple component sources are not considered (the average spectral index $\langle \beta \rangle$ when they are included is reported for completeness). }
\label{fig:spind}
 \end{figure*} 

Due to our very small channel-width $\Delta \nu=1$ MHz, the bandwidth smearing does not represent a major issue.
Eq.~\ref{eq:BWS} gives corrections which are at most of the order of 5\% for sources at a distance of 1 degree (i.e., the boundary of CDS maps).

To empirically verify this conclusion, we performed the imaging of a few distant fields of CDS averaging over a large number of channels (so to increase $\Delta \nu$ in Eq.~\ref{eq:BWS} such that the effect grows to appreciable levels). The ratio of the obtained peak fluxes with respect to the peaks in the original maps is found to closely follow (within error bars) the relation of Eq.~\ref{eq:BWS}, with a conserved total flux. This confirms that Eq.~\ref{eq:BWS} is indeed a reliable estimate of the effect. In our original observing setup, sources are therefore not significantly smeared radially from the first field centre, and the bandwidth smearing correction can be neglected.

Due to incomplete UV coverage, the cleaning process can redistribute the flux from sources to noise peaks. This effect is known as clean bias, and, in general, it is a significant problem only for snapshot observations (where UV coverage is indeed poor). 
Previous analyses \citep[e.g.][]{Prandoni:2000tx} showed that a possible way to mitigate clean bias effects is to stop the cleaning process at a maximum residual flux well above the theoretical noise. We followed this approach stopping at about 3 times the nominal rms.

This procedure together with the relatively good UV coverage of our observations protect against clean bias.
Therefore we do not apply any correction to the fluxes reported in the catalogue.

\section{Comparison with other radio surveys}
\label{sec:comp}

In this Section, we compare our findings with existing radio catalogues, in particular with the large surveys FIRST~\citep{FIRST}, NVSS~\citep{Condon:1998iy}, and SUMSS~\citep{Mauch:2003zh}.
The FIRST FoV overlaps with BootesII and Hercules regions.
The NVSS FoV covers all our fields except for Carina.
The SUMSS survey includes the three CDS: Carina, Fornax, and Sculptor.

The spectral indices of sources in our catalogue are obtained through the comparison with above surveys.
We restricted the comparison to sources at distances larger than 10 arcmin from the boundary of our image, in order to avoid effects from primary beam or highly non-uniform rms.
In case of multiple components, we add up the flux densities of the various components.
They are however conservatively excluded in the plots of Fig.~\ref{fig:spind}.

All the cases which present some significant level of mismatch have been individually inspected and reported below.
The only systematic issue we found is a possible loss of diffuse flux for sources which are close to each other without forming a multiple component source.
In this case, the map with tapering might see them as a single source and, as mentioned above, we chose to associate the corresponding diffuse flux to the source which is closer to the peak of the latter. 
With this approximation, we might be missing the extended diffuse emission of the farther sources. However, this is only a factor for less than 3\% of the total number of sources. 

Counterparts in other frequency bands and the corresponding identifications will be discussed in a companion paper.

In the following, we will define the spectral index $\beta$ with $\beta=\ln (S_{2}/S_{\nu_i})/\ln (2\,{\rm GHz}/\nu_i)$. The frequency of observation of the above surveys is denoted by $\nu_i$ and is 843 MHz for SUMSS and 1.4 GHz for NVSS and FIRST. $S_{\nu_i}$ is the corresponding flux density, while $S_2$ is the flux density measured in this work at, approximately, 2 GHz. 

\subsection{NVSS}
\label{sec:nvss}
The comparison with the NVSS catalogue is summarized in Table \ref{tab:summary}.
The spectral indices are consistent with a prevalence of synchrotron radio continuum sources.
The average offset for the source positions is of the order of a few arcsec. This is consistent with NVSS errors, which are likely to be larger than our positional errors given their larger synthesized beam (about $45''$).

In the Fornax FoV, one NVSS source (J024238-341710) is unassociated. There is no corresponding low C.L. peak, and it might suggest a strongly variable source.
We find a significant mismatch in the fluxes of 5 sources. Three of them (J024031-342132, J024219-335933, J024253-342345) are close to brighter sources and part of the diffuse flux might be missing in our map (for the reason mentioned above), see low $\beta$ cases in Fig.~\ref{fig:spind}. The mismatch in the remaining two (J023924-335632, J023737-335920) has no apparent reason, so they could be moderately variable sources.

In the Sculptor FoV, eight NVSS sources are unassociated. One source (J005900-331411) is however present at low C.L. so possibly pointing towards some variability. Four of them (J005842-330735, J005847-333400, J005900-334552, J010017-333843) are close to bright sources, so could be part of multiple component sources or sidelobes in NVSS.  One source (J010324-333545) is not far from a boundary of our image in a noisy region. The remaining two (J005806-330934, J005817-330654) have instead no apparent reason for the mismatch, so could be truly variable sources. Finally, we note a bright source (J010105-334732) with a quite strong inverted spectrum $\beta=+1.4$ (the latter agreeing with other archival data, see \citet{Healey:2007by}). 

In the BootesII FoV, there is no unassociated source. The average spectral index is close to expectations ($\beta=-1.0\pm0.1$), with no extreme cases.

The catalogue in the Hercules FoV also matches quite well with NVSS, although the spectral index is somewhat lower ($\beta=-1.1\pm0.3$). One NVSS source (J163018+125016) is unassociated. It is close to a bright sources and just above the NVSS detection threshold, which suggests to be either a sidelobe in NVSS or missed in our map because of the noisy region. Since it is not present in the FIRST catalogue as well, the first option might look more plausible. The source J163047+122711 shows a strong inverted spectrum ($\beta=+2$). The source J163137+125217 is close to a brighter source and in the association process loses most of its diffuse flux, showing a very low spectral index ($\beta=-4.7$). Another source (J163254+124034) shows a notably low spectral index ($\beta=-2.8$), with no observational problems apparent in this case.

As already mentioned above, the Segue2 image presents some issues related to the high DEC and the presence of a very bright source $4C\, +20.10$ (which poses dynamic range issues).
This is particularly relevant for the tapered image leading to a loss of diffuse flux, especially for sources in the surrounding of the $4C$ source J022007+203540.
This is the reason for the very low spectral index reported in Table~\ref{tab:summary} and in Fig.~\ref{fig:spind}. We checked that a way to significantly alleviate the issue would be to consider the total flux from the maximum between the flux in the long and short baseline maps (which highlights the fact that the short-baseline map has a poorer flux reconstruction than for the other FoVs). However, for the sake of consistency, we stuck to the method adopted so far.
On the other hand, taking only sources at distances larger than $30'$ from the $4C$ source, the spectral index grows to a more canonical value of $-1.1\pm0.3$ and this is reported in the Table~\ref{tab:summary} (with parenthesis).
Of the six NVSS sources with very low spectral index in Fig.~\ref{fig:spind}a, four (J021938+200918, J021952+200534, J022014+202406, J022016+201729) are close to the $4C$ source, and one (J021925+195925) is in a crowded region (and is not very well reconstructed in our tapered image). The reason behind the low index of the remaining source (J021805+200543) is not straightforward, and it could be a truly variable source.
Finally, one NVSS source (J021750+200330) is unassociated. The area of the source has no apparent issues in both our and NVSS maps, so the source could be a strongly variable source or with an intrinsically very low spectral index (NVSS flux$=3.3$ mJy).

\subsection{SUMSS}
\label{sec:sumss}
The comparison with the SUMSS catalogue is summarized in Table \ref{tab:summary}. 
Again, the spectral indices are consistent with a prevalence of synchrotron sources.
The average offset for the source positions is of the order of a few arcsec consistent with SUMSS positional uncertainties (which are likely to be larger than our positional errors given their larger beam, of the order of $45''$).

In the Fornax, Sculptor, and Carina FoVs, all the sources have been matched, with fairly standard values for the flux ratios and with positional differences within the expected errors.
We only highlight a bright source (J010105-334736) in the Sculptor FoV with a quite strong inverted spectrum $\beta=+1.8$. A similar spectral index is found in the comparison with NVSS.

\subsection{FIRST}
\label{sec:first}
The FIRST survey can resolve structures on scales from 2 to $30''$. Therefore it is ideal to compare with the results of our long-baseline maps. For both BootesII and Hercules FoV, we find a good agreement. The average positional mismatch is about 1 arcsec, consistent with our estimated positional error (see above). This also indicates that the systematic offset in the source position mentioned in Section~\ref{sec:red} has been successfully fixed by means of the NCP projection also for the high-DEC FoVs.
The spectral indices are only mildly reduced with the respect to the comparison with the full catalogue (i.e., their average is -1.2 in the BootesII FoV and -1.4 in the Hercules FoV), which can be ascribed to the lack of diffuse flux on scales of a few tens of arcsec in our long-baseline maps.

The comparison of the FIRST catalogue with our full catalogue is instead summarized in Table \ref{tab:summary}. 
The spectral indices indicate, as expected, a prevalence of synchrotron sources.

In the BootesII FoV, there are three unassociated sources. Two of them, J135946.5+134549 and J135956.4+125550, are at the FIRST detection limit, with the second one also being in a noisy region of our map. The third case (J135818.7+131340) has instead no explanation in terms of possible observational issues and can be a truly variable source.
Eight sources have a spectral index below $-2$ and three above $+1$. However, after cross-checking with the NVSS catalogue, and taking into account possible observational issues mentioned above, no puzzling cases are left.

In the Hercules FoV, one source (J163026.7+125832) is unassociated. It is close to a bright source and at the detection limit for FIRST. Therefore, if it is not variable, either we miss the detection because of the noisy region (no peak at low C.L. is present) or it is a sidelobe of the FIRST map.
Four faint sources J163221.5+130021, J163139.6+131114, J162902.1+124522, and J162939.2+130727 present notably low spectral indices because of no detection in the tapered image (and lack of diffuse emission in the original map). The sources J163047.1+122711 and J163254.6+124035 show a high and low spectral index, respectively, with no apparent observational issue, as already mentioned in the comparison with NVSS. 

\section{Number counts}
\label{sec:num}

Fig.~\ref{fig:rms} shows that the rms is $\lesssim 50 \mu$Jy in all the maps. Therefore our radio sample can be considered complete (at 5--$\sigma$) up to $250 \,\mu$Jy  in terms of peak flux density.
As discussed in Section~\ref{sec:syst}, bandwidth smearing and clean bias are negligible and do not lead to sizable incompleteness. 
However, source counts are a function of the total integrated flux and it is not straightforward to set an a priori completeness threshold (also because of the observational issues on detecting diffuse emissions mentioned above). Different observational biases can cause incompleteness and therefore affect the source counts.

\begin{figure*}
\vspace{-30mm}
\centering
 \begin{minipage}[htb]{8cm}
   \centering
   \includegraphics[width=\textwidth]{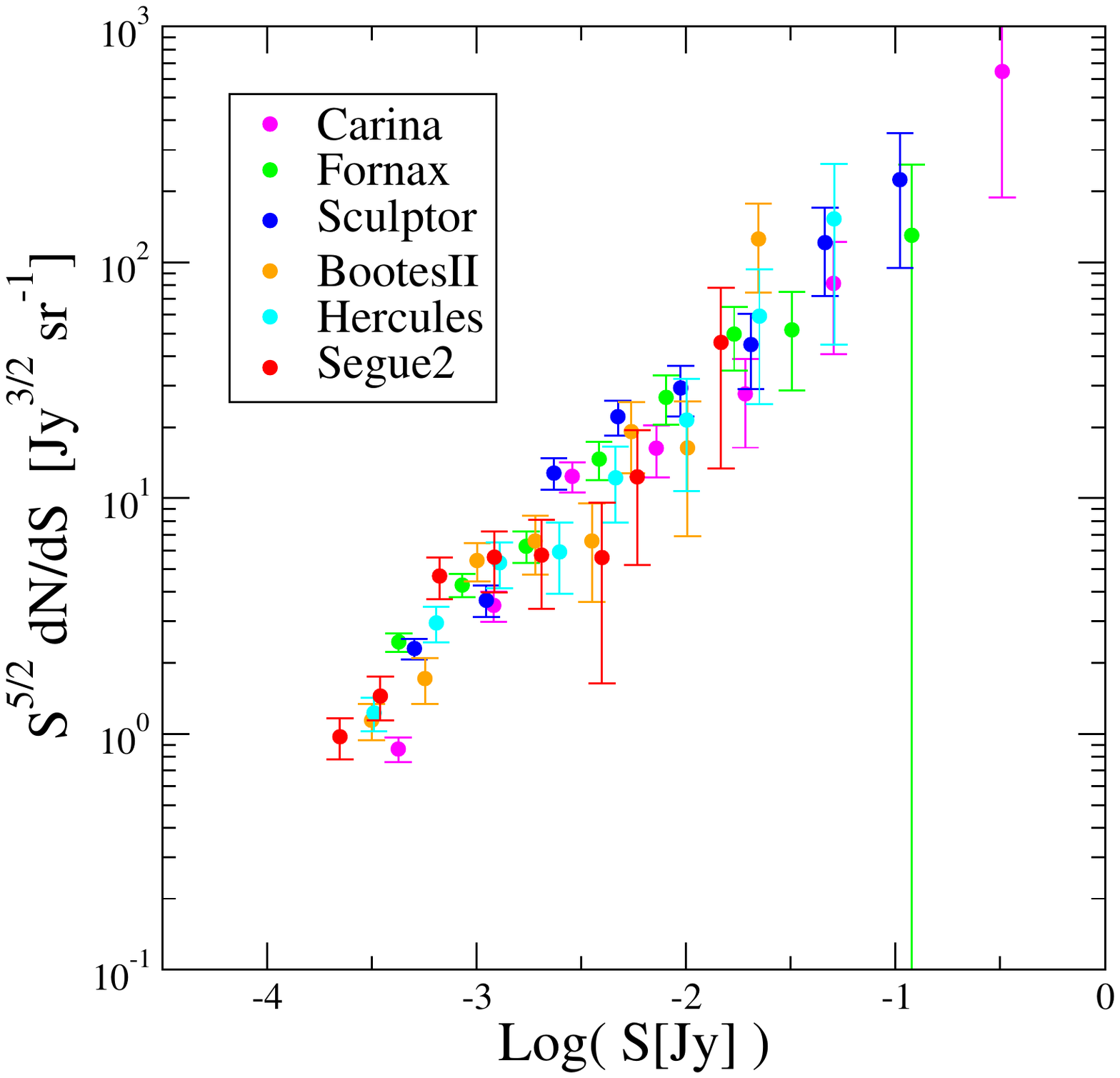}
 \end{minipage}
 \begin{minipage}[htb]{8cm}
   \centering
   \includegraphics[width=\textwidth]{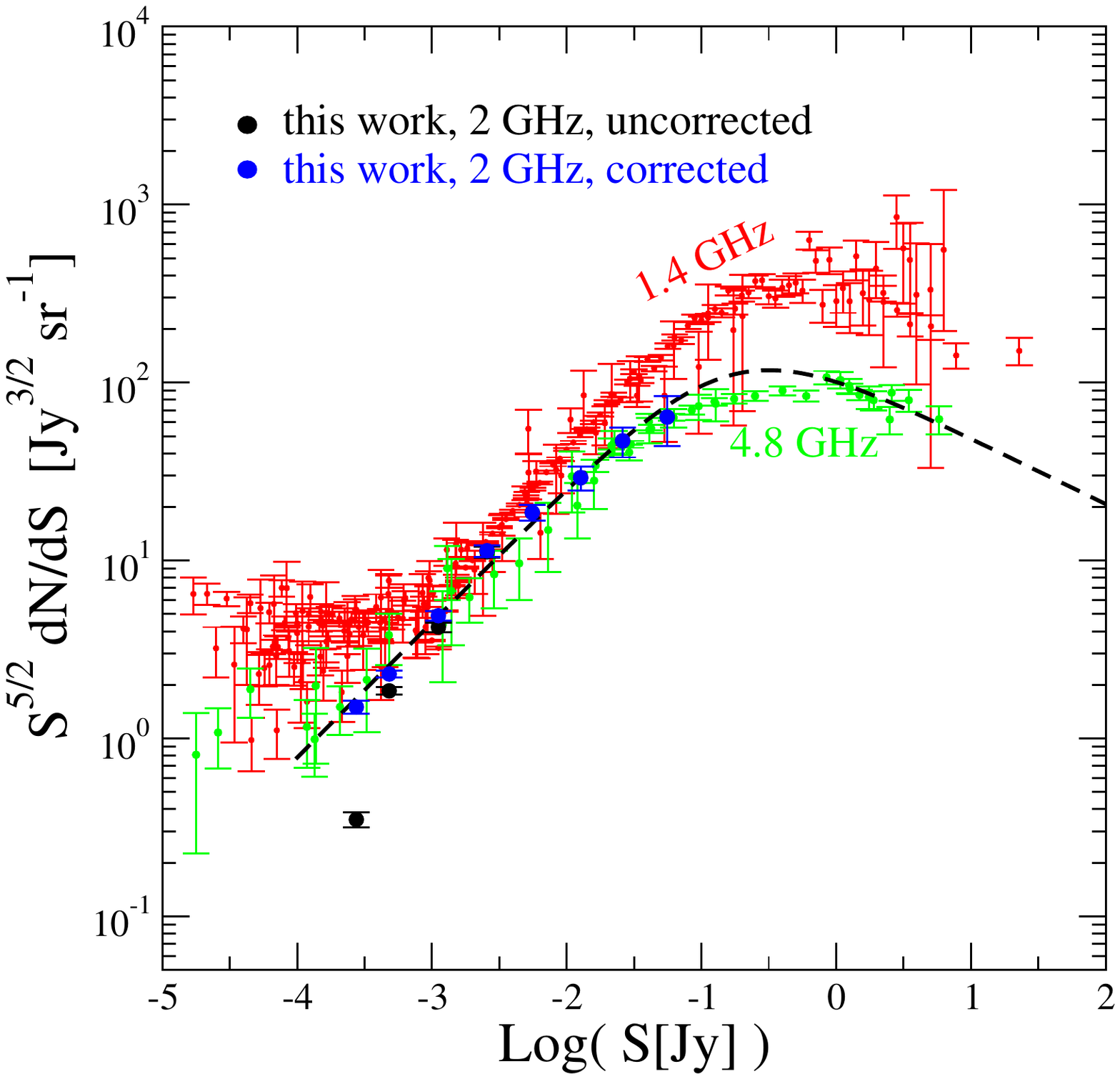}
 \end{minipage}
    \caption{{\bf Counts}. Left: Source number counts in the different fields of view of our survey. Right: Source number counts combining the different fields of view. Blue (black) dots show corrected (uncorrected) number counts where the correction factors have been described in the text. The model of \citet{Gervasi:2008rr} as well as a compilation of observational data at 1.4 GHz and 4.8 GHz from \citet{DeZotti:2009an} are shown for comparison. }
\label{fig:counts}
 \end{figure*} 

First, the actual visibility area needs to be computed.
Indeed, the effective area over which a source of a certain flux density can be detected depends on the noise distribution. The latter can be made inhomogeneous by different effects, such as primary beam response and the presence of bright sources.
We accounted for the varying sensitivity of the survey by assigning to each source a weight equal to the inverse of the area over which the source could have been detected. In practice, the latter is given by the dashed curve in Fig.~\ref{fig:rms}a replacing the rms with $S_{peak}/5$.
The visibility area reaches 100\% for peak flux density between $150 \,\mu$Jy (Segue2) and $300 \,\mu$Jy (Sculptor). 

Two other potentially important biases are instead flux density boosting and resolution bias.
Since the source detection algorithm searches for peaks above an average local noise background, sources on noise peaks have a higher probability of being detected (with boosted measured fluxes), while sources on noise dips might be excluded. This effect is known as Eddington bias or flux density boosting and can affect the number counts in the faintest flux density bins.
In order to precisely estimate the size of the effect, a simulation involving a radio population with similar chatacteristics as those of the catalogue would be in order \citep[see, for example,][]{Biggs:2006}.
However, unless the aim is to look for number counts at very low flux densities, including sources with signal to noise ratio close to the observational limit, the flux density boost is typically found to be below a few percent \citep[see, for example,][]{Huynh:2012hp,White:2010gz}. 
We computed number counts only for $>5$-$\sigma$ detected sources, assuming this correction to be negligible.

The most important correction to make source counts complete in terms of the total flux density is the resolution bias. 
Weak extended sources with large total integrated flux densities might have peak flux densities below the detection threshold.
The resolution bias is a function of the intrinsic angular size distribution of the sources versus the maximum detectable angular
size of the observations, the latter depending on the flux density and on the observational beam.
To compute the correction, we followed \citet{Prandoni:2000ua}, \citet{Huynh:2005fs} and \citet{Huynh:2012hp}.

The maximum angular size $\theta_{max}$ of a source with total flux density $S_{tot}$ before it falls below the detection threshold is given by:
\be
 \theta_{max}=\sqrt{\frac{S_{tot}\,b_{min}\,b_{maj}}{5\,\sigma_{rms}}}\;,
\label{eq:thetamax}
\ee
with $b_{min}$ and $b_{maj}$ being the synthesized beams, and we consider the rms noise $\sigma_{rms}$ averaged over the map.
$\theta_{max}$ is computed from the tapered maps down to fluxes which are above their detection thresholds (for details about synthesized beam and noise of the tapered maps, see Paper II), since they are the maps from which we derived the estimate of $S_{tot}$.
Below approximately $700 \,\mu$Jy, $\theta_{max}$ is instead derived from the un-tapered map.

Depending on the deconvolution efficiency, there is a minimum angular size $\theta_{min}$ below which sources cannot be successfully resolved. While it corresponds to the synthesized beam at high flux density, $\theta_{min}$ might be significantly larger at low signal to noise ratio. 
To derive $\theta_{min}$, we assumed that values of $S_{tot} < S_{peak}$ are due to noise fluctuations, and defined an envelope $S_{tot}/S_{peak} =1-\mathcal{A}/(1+S_{peak}/\sigma_{rms})^{1.5}$ containing 90\% of the sources with $S_{tot}/S_{peak} < 1$ by fitting $\mathcal{A}$ in each FoV.
Assuming that similar statistical errors are also present for sources with $S_{tot} > S_{peak}$, and 
reflecting the envelope on the positive side, one can consider sources lying above the upper envelope to be successfully deconvolved.
The upper envelope provides an estimate for $\theta_{min}$:
\be
\frac{\theta_{min}^2}{b_{min}\,b_{maj}} =1+\mathcal{A}/(1+S_{peak}/\sigma_{rms})^{1.5}\;.
\label{eq:envelope}
\ee
In order to have a proper gauging of $\mathcal{A}$ and of deconvolution efficiency, we performed the fit on a list of sources determined setting the parameter $\alpha=10$ in the task SFIND of {\it Miriad} (instead of the more conservative $\alpha=0.1$ used for the catalogue) so to have a more significant amount of sources with $S_{tot} < S_{peak}$.

For what concerns the true integral angular size distribution of the ``faint'' radio population we are interested in,
we considered the estimate given in \citet{Windhorst:1990}: $h(\theta)= \exp[-\ln 2\, (\theta/\theta_{med})^{0.62}]$, where $\theta_{med}=2'' S_{1.4\,{\rm GHz}}^{0.3}$ and  $S_{1.4\,{\rm GHz}}$ is the flux at 1.4 GHz in mJy (we take $S_{1.4\,{\rm GHz}}=S_{2\,{\rm GHz}}\,(2/1.4)^{0.75}$).

The resolution bias correction can be then computed as $1/(1-h(\theta_{lim}))$, with $\theta_{lim}={\rm max}(\theta_{min},\theta_{max})$.
Clearly, the procedure related to Eq.~\ref{eq:envelope} is well-defined only if applied to the tapered and un-tapered maps separately.
On the other hand, for the tapered map, the resolution bias is negligible, as can be easily understood by noting that $\theta_{med}$ is significantly smaller than its synthesized beam (of order of 1 arcmin).
This is not the case for the un-tapered map, and we did compute the correction for counts at flux density below the detection thresholds of the tapered maps.
Typically, $\theta_{min}$ can become important at low flux density levels, where $\theta_{max}$ can fall below the synthesized beam size. 
However, since we derived the catalogue with a quite conservative threshold (namely, $\alpha=0.1$ in SFIND), the level of flux density is such that we found $\theta_{lim}= \theta_{max}$ except for the lowest flux bin.

Following such procedure, we apply the correction $1/(1-h(\theta_{lim}))$ to the observed counts for sources in our catalogue.
The largest resolution bias correction is in the lowest flux bin, and amount to about 30\%.

The resulting number counts are reported in Table~\ref{tab:count} and shown in Fig.~\ref{fig:counts}.
Fluxes of different components of multiple sources (identified as mentioned in Section~\ref{sec:cat}) have been summed together.

The presence of highly elongated beams can complicate the estimate the resolution bias discussed above.
Indeed, in this case, the major (minor) axis can be significantly larger (smaller) than the typical source size, while above expressions adopt an average value.
On the other hand, we do not see appreciable differences between CDS and UDS in the uncorrected number counts in Fig.~\ref{fig:counts} (left), and, since UDS have a much more elongated beam shape than CDS (see Table~\ref{tab:beams}), we believe the derived correction to still be a fair approximation.

In the right panel of Fig.~\ref{fig:counts}, we split the flux range into 8 logarithmic bins from $150\,\mu$Jy to 0.1 Jy. The correction factor is very large in the first bin because of the choice of the extrema $[150-338]\,\mu$Jy. Indeed, only in the Segue2 FoV the detection threshold is everywhere below $150\,\mu$Jy, while the visibility area correction is large for the other FoVs.
In the left panel, we instead set 8 logarithmic bins starting from the lowest observed flux (and up to the highest flux) of each map. Therefore, since the rms noise is quite uniform, the corrected counts (which are not shown) would be more similar to the uncorrected case also in the low-flux tail (with a maximum increase of 30-40\%, mostly due to resolution bias, for the first bin). 
We found a good agreement between the different fields of view. 

In the right panel of Fig.~\ref{fig:counts}, we compare our results with source counts derived from other surveys at nearby frequencies (data are taken from \citet{DeZotti:2009an}; see also references therein). The corrected counts derived in this work lie in between estimates at 1.4 and 4.8 GHz, as expected.
The uncorrected counts deviate from this trend below about 0.5 mJy.
By such comparison, and by looking at the estimated correction factors, we can conclude that the catalogue starts becoming incomplete at fluxes below the mJy level.

The dashed line describes theoretical expectations for AGNs, which have been computed from \citet{Gervasi:2008rr}, by rescaling the best fit of the high flux population at 2.7 GHz in their Table 2, assuming a spectral index of -0.7.
Data are in fair agreement. A flattening of the counts due to star forming galaxies is expected at flux level of a few tens of $\mu$Jy (see, for example, data at 1.4 GHz in Fig.~\ref{fig:counts}), but we probably need an improvement of a factor of a few in sensitivity to detect it.
Another way to explore counts below the detection threshold of the presented observations can be achieved by computing the so-called $P(D)$ distribution.
see, for example, \citet{Vernstrom:2014} for a state-of-the art analysis, and \citet{Scheuer:1957} for the original idea.
This is however a complex analysis, which is beyond the goal of the present work.

\begin{table}
\centering
\begin{tabular}{ccccc}
\hline
flux bin & $\langle S\rangle$ & uncorr. & $S^{2.5}\,\frac{dN_{uncorr}}{dS}$ & $S^{2.5}\,\frac{dN}{dS}$  \\ 
mJy & mJy & counts & Jy$^{1.5}$ sr$^{-1}$ & Jy$^{1.5}$ sr$^{-1}$  \\ 
\hline
0.15 - 0.34 & 0.28 & 98 & 0.35 $\pm$ 0.04 & 1.5 $\pm$ 0.1 \\
0.34 - 0.76 & 0.48 & 384  & 1.8 $\pm$ 0.1 & 2.3 $\pm$ 0.1 \\ 
0.76 - 1.7  & 1.1  & 242  & 4.2 $\pm$ 0.3  & 4.9 $\pm$ 0.3 \\ 
1.7  - 3.9  & 2.6  & 181  & 11.2 $\pm$ 0.8  & 11.3 $\pm$ 0.8\\ 
3.9  - 8.7  & 5.6  & 98   & 19 $\pm$ 2 & 19 $\pm$ 2 \\
8.7  - 20   & 13   & 44  & 29 $\pm$ 4  & 29 $\pm$ 4 \\ 
20   - 44   & 26   & 25  & 47 $\pm$ 9  & 47 $\pm$ 9\\ 
44   - 100  & 56   & 10   & 64 $\pm$ 20 & 64 $\pm$ 20 \\
\hline
\end{tabular}
\caption{Source counts obtained adding up the sources of the observed six fields of view. For each flux density bin, we report
the mean flux density $\langle S\rangle$, the number of detected sources, the differential radio
source count $dN/dS$ ($dN_{uncorr}/dS$) with (without) applying the corrections described in the text.
}
\label{tab:count}
\end{table}

\section{Conclusions}
\label{sec:concl}
In this work, we provided a detailed analysis of the presence of radio point-like sources in the dSph fields that is essential for the study of the large-scale diffuse radio emission.
The main goal of the presented project is the search for a diffuse radio emission from six MW satellites, Carina, Fornax, Sculptor, BootesII, Hercules, and Segue2.
The analysis of the extended signal is described in Paper II and Paper III, with specific reference to non-thermal radio emission and DM-induced signal, respectively.

The study of the diffuse radio emission in dSph galaxies is crucial to address various astrophysical and cosmological questions.
Indeed, dwarf spheroidals are key probes for near-field cosmology and for galaxy formation and evolution at small-scales.
However, little is known about them, and no thermal or non-thermal emission has been so far detected in association to a dSph.
In addition, MW satellites are also one of the key probes for indirect searches of particle DM signals.

One of the major problems in identifying a diffuse radio emission in these systems is the contamination of maps by background source contributions.
Large beams are needed to detect a diffuse signal. On the other hand, by improving the sensitivity, the confusion limit is rapidly reached.
Subtraction of point-sources is thus mandatory.
This work moves along this direction by providing a deep survey which aims at precisely mapping the background sources present in the selected dSph fields.

We presented observations at 16 cm wavelength of the fields of the mentioned six dSph galaxies in the Local Group.
A total of about 8 square degrees of the sky were observed with the Australia Telescope Compact Array by means of a mosaic strategy (for a total of 74 pointings).
We produced images with rms noise levels between 25 and 50 $\mu$Jy, depending on the specific region, and resolution of a few arcseconds. 
We extracted a total number of 1392 sources (1835 source components) which form the released catalogue. The first few lines are reported in Table~\ref{tab:cat}.

We produced two types of maps: high resolution maps, encoding the signal from the ATCA long-baselines, and low resolution maps, given by the emission measured in the compact core of the array.
The first provided the astrometric information, while the total flux density of the sources was mainly derived from the latter.

We compared our source catalogue with existing GHz radio observations of the dSph fields. In particular, we considered the FIRST~\citep{FIRST} (1.4 GHz), NVSS~\citep{Condon:1998iy} (1.4 GHz), and SUMSS~\citep{Mauch:2003zh} (843 MHz) surveys.
In the fields involved in the comparison, our catalogue contains 217 of the 228 NVSS sources, 122 of the 126 FIRST sources, and 125 of the 125 SUMSS sources. The few percent mismatch with FIRST and NVSS in the number of detected sources can be ascribed in many cases to truly variable sources, and to a limited amount of artifacts (in FIRST and NVSS) or minor observational issues in our setup (see Section~\ref{sec:comp} for more details).
The average spectral index was found to be $\langle \beta \rangle=-0.8$ for all the three cases, suggesting that our source catalog is dominated by synchrotron sources, as expected.

The number of extracted sources is significant and allowed us to derive source counts with very low statistical errors down to about 0.25 mJy.
After correcting the counts for incompleteness at low flux density, as described in Section~\ref{sec:num}, our results are in agreement with models of counts for a source population dominated by AGNs.

In a future work of the series associated with this project, we will also explore the multi-wavelength cross-matching of the catalogue sources, in order to determine the spectral energy distribution, redshift, and type identification, as well as to perform a detailed study of the multiple component sources.
This analysis will determine whether the catalogue contains possible candidates for being the first radio source belonging to a dSph ever discovered.

\section*{Acknowledgements}
We thank Laura Bonavera for assisting with the C2499 ATCA observations of this project.
We wish also to thank Mark Wieringa for the support on the {\it Miriad} data reduction and Piero Ullio for insightful discussions during the early stages of the project.
S.C. acknowledges support by the South African Research Chairs Initiative of the Department of Science and Technology and National Research Foundation and by the Square Kilometre Array (SKA).
S.P. is partly supported by the US Department of Energy, Contract DE-FG02-04ER41268.
M.R. acknowledges support by the research grant {\sl TAsP (Theoretical Astroparticle Physics)} funded by the Istituto Nazionale di Fisica Nucleare (INFN).
The Australia Telescope Compact Array is part of the Australia Telescope National Facility which is funded by the Commonwealth of Australia for operation as a National Facility managed by CSIRO. 

\begin{table*}
{\footnotesize
\centering
\caption{First few lines of the catalogue.
Columns (1) and (2) show the right ascension and declination of sources in J2000.
The peak flux density at 2 GHz (un-tapered map), in mJy, is in column (3). The values are not corrected for possible systematic effects, which are estimated to be negligible (see discussion in the text).  
The integrated flux density at 2 GHz (un-tapered map), in mJy, is in column (4). Errors are obtained summing in quadrature the local rms, the fit error, and a 5\% of the flux density (to account for possible inaccuracy in the calibration model and process, especially due to RFI). 
In columns (5), (6) and (7), we report the FWHM major axis ($b_{maj}$ in arcmin), the minor axis ($b_{min}$ in arcmin), and the position angle (P.A., measured north to east, in degrees) of the source. For sources with $S_{tot}/S_{peak}<1.3$, the source is not successfully deconvolved and these values should not be considered.
Column (8) shows the integrated flux density at 2 GHz in the tapered map, in mJy.
Column (9) reports a flag for multiple component sources: S = single component source, M = multiple component source (followed by a number identifying the multiple source to which the component belongs).
The full catalogue is available in ASCII format in the online material.
}
\label{tablecar}
\begin{tabular}{rrrrrrccccccc}
\hline
\multicolumn{6}{c}{J2000}  & & & \multicolumn{2}{c}{Angular size} & P.A. & &Multiple\\
\multicolumn{3}{c}{RA}  & \multicolumn{3}{c}{Dec} &$F_{r_{-1}}^{peak}$ [mJy]&$F_{r_{-1}} \pm \delta F_{r_{-1}}$ [mJy]&$b_{maj}$ [$'$]&$b_{min}$ [$'$]&$\theta$ [deg]&$F_{gta}$ [mJy]& flag \\
\hline

6 & 46 & 23.6 & -51 & 07 & 4.0 & 0.88 & 0.97 $\pm$ 0.07 & 0.07 & 0.04 & -7.90 & 1.58 & S\\ 
6 & 46 & 18.8 & -50 & 55 & 26.1 & 1.12 & 1.44 $\pm$ 0.09 & 0.07 & 0.05 & -12.60 & 7.00 & M1\\
6 & 46 & 16.7 & -50 & 55 & 23.9 & 0.96 & 2.41 $\pm$ 0.16 & 0.10 & 0.07 & 65.60 & 0.00 & M1\\ 
6 & 46 & 12.0 & -51 & 11 & 52.1 & 0.94 & 1.04 $\pm$ 0.08 & 0.07 & 0.04 & -9.40 & 1.52 & S\\ 
6 & 46 & 07.4 & -51 & 15 & 11.2 & 0.41 & 0.53 $\pm$ 0.06 & 0.08 & 0.05 & 1.40 & 1.16 & S\\ 
6 & 45 & 50.5 & -50 & 35 & 14.4 & 0.30 & 0.34 $\pm$ 0.05 & 0.07 & 0.04 & -5.70 & 1.21 & S\\ 
6 & 45 & 48.3 & -51 & 18 & 4.9 & 0.60 & 0.72 $\pm$ 0.06 & 0.08 & 0.05 & -13.10 & 1.38 & S\\
6 & 45 & 47.9 & -50 & 30 & 12.2 & 1.97 & 2.50 $\pm$ 0.19 & 1.52 & 1.06 & 75.70 & 0.00 & S\\ 
6 & 45 & 42.0 & -51 & 00 & 48.3 & 16.02 & 22.63 $\pm$ 1.23 & 0.09 & 0.05 & -13.30 & 46.23 & M2\\ 
6 & 45 & 41.8 & -51 & 00 & 41.2 & 7.99 & 11.59 $\pm$ 0.66 & 0.09 & 0.05 & -10.10 & 0.00 & M2\\ 
6 & 45 & 29.1 & -51 & 06 & 14.7 & 2.11 & 3.03 $\pm$ 0.17 & 0.08 & 0.05 & -8.30 & 9.54 & M3\\ 
6 & 45 & 28.2 & -51 & 06 & 16.3 & 2.04 & 2.65 $\pm$ 0.15 & 0.08 & 0.05 & -7.40 & 0.00 & M3\\ 
6 & 45 & 27.8 & -51 & 13 & 20.1 & 0.30 & 0.49 $\pm$ 0.06 & 0.08 & 0.06 & 5.50 & 1.78 & S\\ 
6 & 45 & 26.8 & -51 & 11 & 16.1 & 0.41 & 0.51 $\pm$ 0.06 & 0.08 & 0.05 & 7.30 & 0.00 & S\\ 
6 & 45 & 28.0 & -51 & 28 & 55.9 & 0.38 & 0.47 $\pm$ 0.06 & 0.07 & 0.05 & -5.80 & 0.00 & S\\ 
6 & 45 & 19.7 & -50 & 48 & 57.1 & 1.03 & 1.56 $\pm$ 0.10 & 0.09 & 0.05 & 19.50 & 3.09 & S\\ 
6 & 45 & 17.9 & -50 & 35 & 48.6 & 0.51 & 0.66 $\pm$ 0.06 & 0.08 & 0.05 & -16.10 & 1.38 & S\\ 
6 & 45 & 17.7 & -50 & 50 & 54.7 & 2.45 & 2.54 $\pm$ 0.14 & 0.07 & 0.04 & -6.40 & 3.31 & S\\ 
6 & 45 & 05.4 & -50 & 39 & 34.6 & 1.57 & 1.77 $\pm$ 0.11 & 0.07 & 0.05 & -5.30 & 3.07 & S\\ 
\end{tabular}
}
\label{tab:cat}
\end{table*}


\bibliographystyle{mn2e}
\bibliography{paper1_refs}


\end{document}